\documentclass[letterpaper, showkeys,onecolumn,11pt,unpublished]{quantumarticle}
\pdfoutput=1
\usepackage[utf8]{inputenc}
\usepackage[english]{babel}
\usepackage[T1]{fontenc}
\usepackage{amsmath}

\usepackage{tikz}
\usepackage{lipsum}

\usepackage[
colorlinks=true,
linkcolor=blue,
urlcolor=blue,
citecolor=blue]{hyperref}

\usepackage{lineno}
\usepackage{dcolumn}
\usepackage{bm}
\usepackage{mathtools}
\usepackage{braket}
\usepackage{physics}
\usepackage{afterpage}
\usepackage{nicefrac}
\usepackage{xcolor}
\usepackage{amsfonts}
\usepackage{amssymb}
\usepackage{comment}
\usepackage{stfloats}
\usepackage{float}

\usepackage[numbers,sort&compress]{natbib}

\begin{document}
	
	\title{Chirped Pulse Control of Raman Coherence in Atoms and Molecules}

	\author{~Jabir~Chathanathil and Svetlana~A.~Malinovskaya}
	\affil{Department of Physics, Stevens Institute of Technology, Hoboken, New Jersey 07030, USA}
		
	\maketitle
	
	\begin{abstract}
		
		A novel chirped pulse control scheme is presented based on Coherent Anti-Stokes Raman Spectroscopy (C-CARS) aiming at maximizing the vibrational coherence in atoms and molecules. The scheme utilizes chirping of the three incoming pulses, the pump, the Stokes and the probe, in the four-wave mixing process of C-CARS to fulfill the adiabatic passage conditions. The derivation of the scheme is based on simplifying the four-level system into a ‘super-effective’ two level system via rotating wave approximation and adiabatic elimination of the excited state manifold. The robustness, spectral selectivity and adiabatic nature of C-CARS method may prove useful for sensing, imaging, and detection. It is demonstrated that the selectivity in excitation of vibrational degrees of freedom can be controlled by carefully choosing the spectral chirp rate of the pulses. The C-CARS control scheme is applied to a surrogate methanol molecule to generate an optimal anti-Stokes signal backscattered from a cloud of molecules a kilometer away. The theory is based on the solution of the coupled Maxwell-Liouville von Neumann equations and focuses on the quantum effects induced in the target molecules by the control pulse trains. The propagation effects of pulses through the medium are evaluated and the buildup of the molecular-specific anti-Stokes signal is demonstrated numerically. A deep learning technique, using Convolutional Neural Networks (CNN), is implemented to characterize the control pulses and evaluate time-dependent phase characteristics from them. The effects of decoherence induced by spontaneous decay and collisional dephasing are also examined. Additionally, we present the technique of Fractional Stimulated Raman Adiabatic Passage (F-STIRAP) and demonstrate that it can be utilized for remote detection in a multi-level system by creation of a maximally coherent superposition state.
		
	\end{abstract}
	
	
	\newpage
	\tableofcontents
	
	\section{INTRODUCTION}
	
	
	The quest for understanding and mastering the world has been an inherent part of human life ever since it existed. As a student of physics, whenever I explained to my friends and family members of my research, it always fascinated them to know that we can predict the results of a microscopic process using a mathematical derivation. The fact that a pen and a paper are all that theoretical physicists require in order to describe and predict the profound realities of the universe still inspires me every day. The perfect sync between the results of mathematical calculations and physical processes is the manifestation of an elegant reality we live in.
	
	The development of quantum theory in the early twentieth century revolutionized the way we understood physical realities of the universe. The phenomena of wave-matter duality and the uncertainty principle could only be perceived as ‘miracles’ in the classical world. The implications of quantum mechanics posed deep philosophical questions and troubled scientists like Albert Einstein. It took several more decades to settle the question of whether ‘God plays dice or not’. After the introduction of Bell’s theorem and its experimental realization, quantum theory was finally established as a fundamental physical reality of the universe. As we improved our understanding about the physical world, we mastered the ways in which this knowledge could be used to improve human lives. Just as the development of mechanical engines and industrial revolution can be attributed to Newton’s laws, the modern technological advancements are, in part, the result of our understanding of the quantum mechanical processes. Today, after a century of progress, we may be close to realizing a quantum computer that will outperform classical computers.
	
	In the early twentieth century, along with the development of quantum theory came the discovery of Raman scattering process. A quantum mechanical description of physical systems was necessary to understand the Raman scattering. Since then, modern Raman Spectroscopy paved new ways to learn about the structure of atoms and molecules. The advancement in laser technology since 1960s accelerated this trend and opened wide range of possibilities to control microscopic systems for various applications including sensing, imaging and detection.
	
	Before the discovery of Raman scattering, Rayleigh scattering could successfully describe classical phenomena like the color of sky.
	This is a classical process of elastic scattering in which an electromagnetic field scatters off of a particle without changing the incident frequency. This happens when the particle has much smaller wavelength compared to the incident field. But as the particle gets smaller, the scattering process becomes inelastic and the classical description of the Rayleigh scattering proves to be insufficient to explain the process. This is because the output fields in this case do not have a continuous spectrum of frequencies, rather it possesses a frequency spectrum that corresponds to the quantum energy levels of the sample. This inelastic scattering of the electromagnetic field with microscopic systems is called Raman scattering, or Spontaneous Raman scattering. 
	In this process, a pump field, which excites the molecule into a virtual state, scatters off of the molecule, generating a field with a lower (red-shifted) or higher (blue-shifted) frequency than the pump, and bringing the molecule back to one of its vibrational states. The red-shifted field, known as Stokes, brings the molecule to a higher vibrational state, while a blue-shifted field, known as anti-Stokes brings the molecule into a lower vibrational state.
	
	Even though the discovery of Modern Raman spectroscopy delivered tremendous progress in our probing and understanding of the microscopic structures, it needed improvements as the output signal was very weak, incoherent and non-directional. This led to the development of other improved versions of Raman spectroscopy such as Stimulated Raman Spectroscopy (SRS) \cite{SRS_main}, 
	and Coherent anti-Stokes Raman Spectroscopy (CARS)\cite{CARS_first}. 
	SRS is a third order nonlinear process in which two fields, a pump and a Stokes, are incident on the sample, stimulating the vibrational transition. In contrast to the spontaneous Raman scattering, this is a coherent process and the resonant enhancement of the transition happens when the frequency difference of pump and Stokes ($\omega_p - \omega_s$) is matched with the vibrational frequency. CARS is a four-wave mixing process in which the pump and Stokes pulses, having frequencies $\omega_p$ and $\omega_s$ respectively, excite the molecular vibrations to create a coherent superposition that a probe pulse, having frequency $\omega_{pr}$, interacts with to generate an anti-Stokes signal. The output signal is blue-shifted and has a frequency of $\omega_{as} = \omega_p - \omega_s + \omega_{pr}$.
	Owing to addressing the inherent vibrational properties of matter, CARS is one of the best suited and most frequently used methods for imaging, sensing and detection without labeling or staining \cite{XieCRSMicroscopy, CARS_tutorial_2012, Xie_CARS_2008, Xie_CARS_2004, Quantitative_CARS_2011, Polymer_2022}. The high sensitivity, high resolution and non-invasiveness of CARS has been exploited for imaging of chemical and biological samples \cite{Xie_1999, Xie_imaging_2005, Xie_laser_2002, Xie_imaging_2001, Potma_hydro_2001, Potma_imaging_2006, Potma_lipid_2003, Potma_imaging_2004, Potma_imaging_2010, Saykally_imaging_2002, nature_imaging_2014}, standoff detection \cite{Dantus_standoff_2008, Dantus_standoff_2013, scully-cars, Sokolov_detection_2005} and combustion thermometry \cite{Gord_CARS_2010, Gord_comparison_2017}. Recent developments in the applications of CARS in biology include imaging and classification of cancer cells that help early diagnosis \cite{cancer1, cancer2, cancer3} and rapid and label-free detection of the SARS-CoV-2 pathogens \cite{COVID}. CARS has also been used recently for observing real-time vibrations of chemical bonds within molecules \cite{Potma_real-time_2014}, direct imaging of molecular symmetries \cite{nature_symmetry_2016}, graphene
	imaging \cite{nature_graphere_2019}, and femtosecond spectroscopy 
	\cite{CARS_multiplex_2014, flame_wall_2015}.
	
	Both SRS and CARS, being coherent processes, provide signals many orders of magnitude higher in amplitude compared to the spontaneous Raman process. But the detection in SRS requires a scheme that is more complicated than CARS as the output and input signals in SRS have the same frequencies. In CARS, the signal can be easily separated by an optical filter due to the anti-Stokes field having a  frequency blue-shifted compared to the incoming pulses \cite{CARS_tutorial_2012}. The direction of signal in CARS is determined by the phase matching conditions \cite{review_skeletal_2016, Zhetlikov_2000, Levis_BOXCARS_2007, Scully_coherent_vs_incoherent_2007}. SRS has an advantage over CARS, because of the absence of a nonresonant background \cite{Xie_SRS_2008}, as the signal is stimulated only when the frequency difference matches with that of the vibrational mode. The presence of the non-resonant background appearing in the spectra, which limits image contrast and sensitivity, has been one of the main challeges in CARS. To overcome the limitations of CARS, there has been tremendous work on removing the background from nonresonant processes and enhancing the signal amplitude \cite{Potma_background-free_2006, Potma_background-free_2004, background-free_2003, background-free_2008, background-free_2010, background-free_2010-2, nature_high-resolution_2018, chirped-probe-pulse_2011, Pestov_Optim_2007, Kumar_background-free_2011}.
	
	Quantum control is a way to manipulate the dynamics in a quantum system in order to make a more useful outcome. This is usually done with a tailored external field possessing controllable parameters. The amplitude of the anti-Stokes field in CARS is related to coherence between the electronic vibrational states within the Maxwell's equations framework. Thus, maximizing coherence is the key to optimizing the intensity of the anti-Stokes signal \cite{Ch21, scully-cars}. The primary aim of this work is to develop a quantum control scheme applicable to CARS that maximizes coherence by manipulating the field parameters involved in the process.
	
	In this chapter, we take a semiclassical approach to analyze the dynamics of light-matter interactions, which means a classical electromagnetic field is interacting with a quantum system of atoms or molecules is considered. The evolution of any quantum system is determined by the Schr\"odinger equation:
	\begin{equation}\label{Schrodinger}
		i\hbar \frac{\partial }{\partial t}\ket{\boldsymbol{\Psi}(t)} = \mathbf{H}(t)\ket{\boldsymbol{\Psi}(t)}
	\end{equation}  
	where $\ket{\boldsymbol{\Psi}(t)}$ is the wave function that describe the quantum system and is a superposition of its $n $ eigenstates $\ket{n}$.
	\begin{equation}
		\ket{\boldsymbol{\Psi}(t)} = \sum_{n}^{}a_n\ket{n}\,.
	\end{equation}
	The co-efficients $a_n(t)$ are known as the probability amplitudes, the square of which give the probability of observing the nth eigenstate when making a measurement of an observable in the eigenstate basis. In most cases of light-matter interactions, the Hamiltonian $\mathbf{H}(t)$ can be decomposed as:
	\begin{equation}\label{Hamil_basic}
		\mathbf{H}(t) = \mathbf{H}_0 + \mathbf{V}(t)
	\end{equation}
	where $\mathbf{H}_0$ is the time-independent Hamiltonian of the system and $\mathbf{V}(t)$ is the time-dependent interaction Hamiltonian. For a field $\mathbf{E}(t)$ interacting with the quantum system, the interaction Hamiltonian is given by: $\mathbf{V}(t) = -\boldsymbol{\mu}\cdot\mathbf{E}(t)$, where $\boldsymbol{\mu}$ is the transition dipole moment. Using Hamiltonian \eqref{Hamil_basic} in the Schrodinger equation \eqref{Schrodinger} yields the dynamics of the quantum system during the interaction. As the simplest example, we can consider the quantum system with two energy levels $\ket{g}$ and $\ket{e}$, having frequency $\omega_{g}$ and $\omega_{e}$ respectively, interacting with a monochromatic light field with frequency $\omega_{L}$. The quantum state of this system at any time can be written as the Schr\"odinger wave function: $\ket{\boldsymbol{\Psi}(t)} = a_g\ket{g} + a_e\ket{e} $.
	
	When the system involves multiple quantum states interacting with many fields, it convenient to move to a different reference frame to simplify the numerical simulation of the problem. One way is to move to a frame that rotates with the
	frequencies of the given quantum states. This way of representing the evolution of the system is called the ``interaction representation". To do this, the probability amplitudes are transformed by the equations:
	\begin{equation}
		a_n(t) = \tilde{a}_n(t)e^{-i\omega_n t}\,, \ \ \ \ \ n = g, e
	\end{equation}
	This transformation of the Hamiltonian into this representation removes the time independent diagonal elements and absorb them into the interaction part of the Hamiltonian.
	
	Another representation, in which the system can be transformed to a reference frame rotating at the frequency of incident fields, is called the ``field-interaction representation". For the two level system described above, this can be done using the below equations:
	\begin{equation}
		a_g(t) = \tilde{\tilde{a}}_g(t)e^{i\omega_L(t)t/2} \,, \ \ \ \ a_e(t) = \tilde{\tilde{a}}_e(t)e^{-i\omega_L(t)t/2}
	\end{equation}
	where $\omega_L(t)$ represents the time dependent frequency in the case of a chirped field. This transformation removes the exponential terms in the Hamiltonian and represents the energy levels in terms of the detunings, which is the difference between laser frequency and and frequency of energy splitting in the system, $(\omega_e-\omega_g)-\omega_L(t)$. This representation is very convenient when analyzing dynamics of energy levels during the light-matter interaction.
	
	An alternative way to of expressing quantum states is using density matrix formalism. When dealing with the ensemble of quantum states, the density matrix representation is more convenient especially when the effects of decoherence are taken into account. The evolution of quantum states in density matrix representation are described by the Liouville-von Neumann equations:
	
	\begin{equation}\label{Liouville}
		i\hbar \dot{\boldsymbol{\rho}}(t) = [\mathbf{H}(t), \boldsymbol{\rho}(t)]
	\end{equation}
	where the density matrix elements can be expressed in terms of the probability amplitudes as: $\boldsymbol{\rho}_{ij}=a_ia_j^*$.
	
	There has been a considerable number of studies published on different control methods aiming to improve the technique of coherent anti-Stokes Raman spectroscopy. Chirped pulses have been used in CARS-based imaging techniques to achieve high spectroscopic resolution \cite{Saykally_chirped_2006, Saykally_chirped_2007} and maximum coherence\cite{Pandya20, Malinovsky_2009, Ma07}. A method for selective excitation in a  multimode system using a transform limited pump pulse and a lineraly chirped Stokes pulse in stimulated Raman scattering was proposed in \cite{Ma2006}. The effects of a chirped pump and Stokes pulses on the nonadiabatic coupling between vibrational modes are discussed in \cite{Patel-2011}. A `roof' method of chirping to maximize coherence was introduced in \cite{Ma01} based on adiabatic passage in an effective two-level system. In this method, the Stokes pulse was linearly chirped at the same rate of the chirped pump pulse in the first half of the pulse duration and was chirped with a negative rate afterwards. In the current work, we develop a scheme in which all the pulses in CARS are chirped in the a way that creates maximum vibrational coherence in a robust and selective way.
	
	The primary aim of this chapter is to develop theoretical frameworks that improve the existing methods of imaging, sensing and detection using quantum control methods. We seek to improve the techniques of coherent anti-Stokes Raman Spectroscopy (CARS) and Stimulated Raman adiabatic passage (STIRAP) by controlling the field parameters involved in the process. In section 2, we present a theoretical description of a general and robust technique of creating maximal-coherence superpositions of quantum states that can be used to optimize the signals in CARS-based applications. This technique is named as C-CARS and is based on the idea of manipulating the amplitudes and phases of incoming pulses in order to optimize the output signal and suppress the background. In the third section, we present a semiclassical theory based on this control scheme to simulate the output from a molecular system at kilometer scale aiming at remote detection. This theory combines the C-CARS scheme with the coupled Maxwell-Liouville-von Neumann equations, augmented with relaxation terms, and makes use of a machine learning technique to analyze the phase values of the scattered signal. In section 4, we give details of this machine learning technique, which uses deep Convolutional Neural Networks (CNN), and outline the application of this technique in quantum control methods. In section 5, we describe the process of Stimulated Raman Adiabatic Passage (STIRAP), the conditions for adiabatic passage, and the effects of chirping pulses in STIRAP. We show that a variation of STIRAP, namely fractional STIRAP (F-STIRAP), can be used to maximize coherence in a multi-level system. This provides another robust way for optimizing the output signal in detection methods. The chapter is concluded with a summary of results in section 6.
	
	\section{QUANTUM CONTROL IN CARS USING CHIRPED PULSES}
	
	\subsection{Coherent Anti-Stokes Raman Spectroscopy}
	
	In the previous section, we discussed the limitations of Coherent Anti-Stokes Raman Spectroscopy (CARS) and noted that it is important to maximize the vibrational coherence and suppress the background signal in CARS. Here, we present a chirping scheme in CARS, in which all the incoming pulses are chirped to achieve this goal. The selectivity, robustness and adiabatic nature of this control scheme make it a viable candidate for improving the current methods for imaging, sensing and detection using CARS.
	
	\begin{figure}
		\includegraphics[scale=0.6]{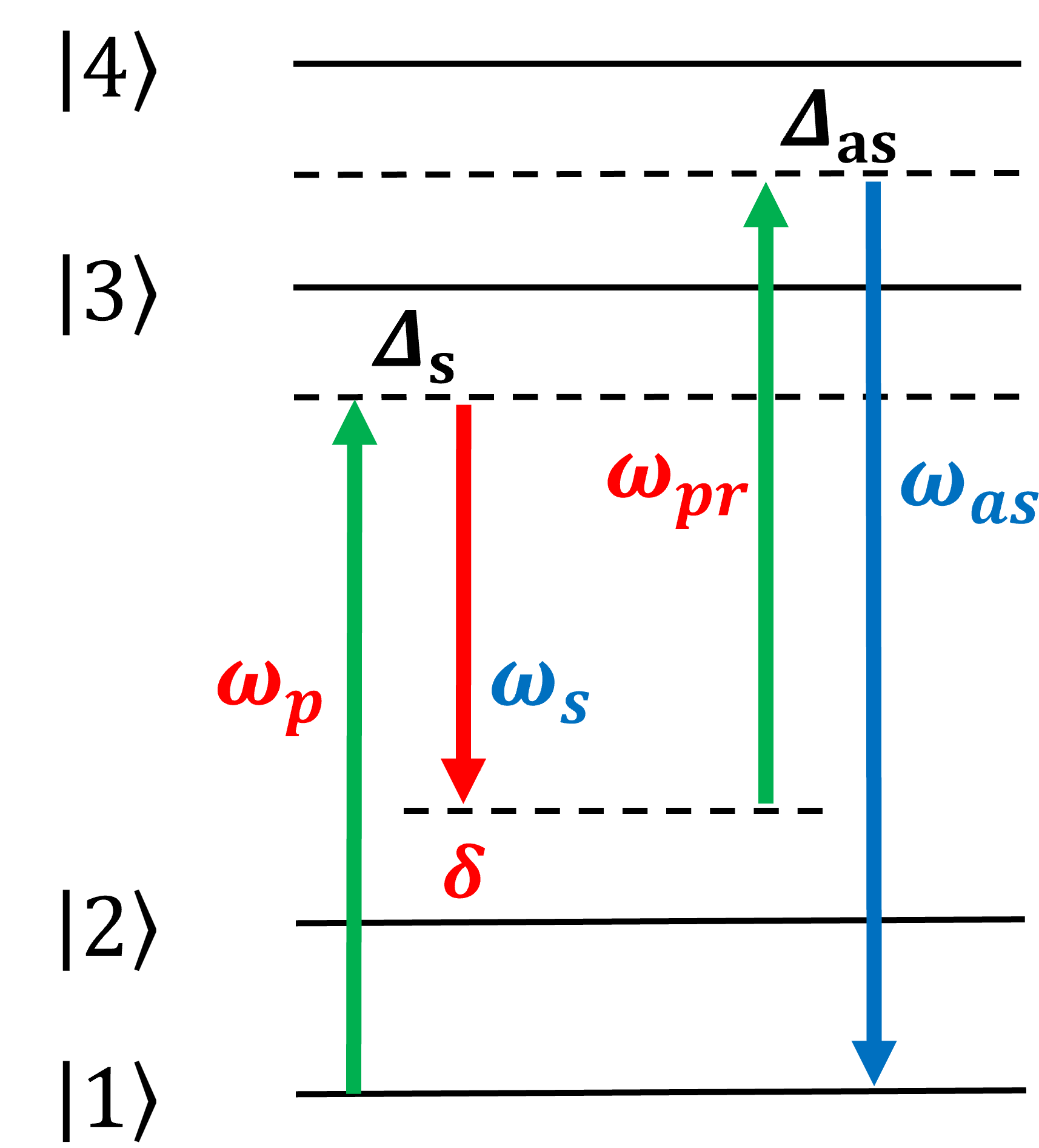}\centering
		\caption{ Schematic of Coherent Anti-Stokes Raman Spectroscopy (CARS): the pump ($\omega_{p}$) and the Stokes ($\omega_{s}$) fields interact with the ground vibrational state $\ket{1}$ and the excited vibrational $\ket{2}$ state of the ground electronic state in the target molecule to create a superposition  state with coherence $\rho_{12}$. The probe ($\omega_{pr}$) field interacts with this superposition state to generate the anti-Stokes field at frequency $\omega_{as}$. Parameters $\Delta_s$ and $\Delta_{as}$ are the one-photon detunings, and $\delta$ is the two-photon detuning.} \label{CARS_scheme}
	\end{figure}
	
	A schematic diagram of the CARS process is given in Fig.\,\ref{CARS_scheme}. The pump and Stokes fields create a coherent superposition between states $\ket{1}$ and $\ket{2}$ which a probe field scatters off of to generate an anti-Stokes signal. Consider chirped pump, Stokes and probe pulses with temporal chirp rates $\alpha_q$, $q=p,s,pr$ as
	\begin{equation}
		E_q(t) = E_{q_0}(t)\cos\left [\omega_q(t-t_c)+\frac{\alpha_q}{2}(t-t_c)^2\right ]\label{pulse_equations}
	\end{equation}
	and having Gaussian envelopes
	\begin{equation}
		E_{q_0}(t) = \frac{\tilde{E}_{q_0}}{\left (1+\frac{\alpha_q'^2}{\tau_0^4}\right )^{1/4}}e^{-\frac{(t-t_c)^2}{2\tau^2}}, 
	\end{equation}
	where $\tau_0$ is the tranform-limited pulse duration, $\tau$ is the chirp-dependent pulse duration given by: $\tau = \tau_0[1+\alpha_q'^2/\tau_0^4]^{1/2}$ and $\alpha'_q$ is the spectral chirp rate which is related to the temporal chirp rate by: $\alpha_q = \alpha_q'/\tau_0^2(1+\alpha_q'^2/\tau_0^4)$.
	The interaction Hamiltonian of the four-level system, after defining the one photon detunings, $\Delta_s = \omega_p - \omega_{31}$ and $\Delta_{as} = \omega_{as} - \omega_{41}$,
	reads
	
	\begin{equation}\label{HAM_4_level_int}
			H = \tfrac{\hbar}{2}
			\begingroup 
			\setlength\arraycolsep{1pt}
			\begin{pmatrix}
				0	&	0	&	\Omega_{p_0}(t)e^{i\Delta_{s}t+i\tfrac{\alpha_p}{2}t^{2}}	&	\Omega_{as_0}(t)e^{i\Delta_{as}t}\\
				0	&	0	&	\Omega_{s_0}(t)e^{i\Delta_{s}t+i\tfrac{\alpha_s}{2}t^{2}}	&	\Omega_{pr_0}(t)e^{i\Delta_{as}t+i\tfrac{\alpha_{pr}}{2}t^{2}}\\
				\Omega_{p_0}^*(t)e^{-i\Delta_{s}t-i\tfrac{\alpha_p}{2}t^{2}}	&	\Omega_{s_0}^*(t)e^{-i\Delta_{s}t-i\tfrac{\alpha_s}{2}t^{2}}	&	0	&	0\\
				\Omega_{as_0}^*(t)e^{-i\Delta_{as}t}	&	\Omega_{pr_0}^*(t)e^{-i\Delta_{as}t-i\tfrac{\alpha_{pr}}{2}t^{2}}	&	0	&	0\\
			\end{pmatrix}
		\endgroup
	\end{equation}
	
	where Rabi frequencies are given by $\Omega_{q_0} = -\mu_{ij}E_{q_0}/\hbar$. 
	
	\subsection{Adiabatic Elimination of excited states}
	This Hamiltonian can be simplified to a two-level super-effective Hamiltonian by eliminating the states $\ket{3}$ and $\ket{4}$ adiabatically under the assumption of large one-photon detunings. The dynamics of the four-level system interacting with the fields in Eq.(\ref{pulse_equations}) is described by the Liouville-von Neumann equation $ i\hbar \dot{\boldsymbol{\rho}}(t) = [\mathbf{H}_{int}(t), \boldsymbol{\rho}(t)].$
	We define the two-photon detuning $\delta = \omega_p - \omega_s - \omega_{21} = \omega_{as} - \omega_{pr} - \omega_{21}$,  make the transformations in the interaction frame:
	\begin{align} 
		\begin{aligned}
			\rho_{11} =& \tilde{\rho}_{11} \\ 
			\rho_{12} =& \tilde{\rho}_{12}e^{-i(\omega_1-\omega_2)(t-t_c)}\\ 
			\rho_{13} =& \tilde{\rho}_{13}e^{i\omega_p(t-t_c)}\\ 
			\rho_{14} =& \tilde{\rho}_{14}e^{i\omega_{as}(t-t_c)}\\ 
			\rho_{22} =& \tilde{\rho}_{22}\\
			\rho_{23} =& \tilde{\rho}_{23}e^{-i(\omega_2-\omega_1-\omega_p)(t-t_c)}\\
			\rho_{24} =& \tilde{\rho}_{24}e^{-i(\omega_2-\omega_{1}-\omega_{as})(t-t_c)} \\
			\rho_{33} =& \tilde{\rho}_{33} \\
			\rho_{34} =& \tilde{\rho}_{34}e^{-i(\omega_p-\omega_{as})(t-t_c)} \\
			\rho_{44} =& \tilde{\rho}_{44}
		\end{aligned}
	\end{align}
	and obtain a system of differential equations for the density matrix elements after dropping the {\em tilde} on both sides:
	\begin{align} 
		\begin{aligned}
			i\dot{\rho}_{11} =& \tfrac{1}{2}\Omega_{p0}(t)e^{\frac{i}{2}\alpha_{p}(t-t_c)^2}\rho_{31} + \tfrac{1}{2}\Omega_{as0}(t)\rho_{41} - c.c \,, \\
			i\dot{\rho}_{22} =&    \tfrac{1}{2}\Omega_{s0}(t)e^{-i\delta(t-t_c)+\frac{i}{2}\alpha_s(t-t_c)^2}\rho_{32} + \tfrac{1}{2}\Omega_{pr0}(t)e^{-i\delta(t-t_c)+\frac{i}{2}\alpha_{pr}(t-t_c)^2}\rho_{42} - c.c \,,\\
			i\dot{\rho}_{33} =&  \tfrac{1}{2}\Omega^*_{p0}(t)e^{-\frac{i}{2}\alpha_{p}(t-t_c)^2}\rho_{13} + \tfrac{1}{2}\Omega^*_{s0}(t)e^{i\delta(t-t_c)-\frac{i}{2}\alpha_s(t-t_c)^2}\rho_{23} - c.c \,,\\
			i\dot{\rho}_{44} =&  \tfrac{1}{2}\Omega^*_{as0}(t)\rho_{14} + \tfrac{1}{2}\Omega^*_{pr0}(t)e^{i\delta(t-t_c)-\frac{i}{2}\alpha_s(t-t_c)^2}\rho_{24} - c.c \,,\\
			i\dot{\rho}_{12} =& \tfrac{1}{2}\Omega_{p0}(t)e^{\frac{i}{2}\alpha_{p}(t-t_c)^2}\rho_{32} + \tfrac{1}{2}\Omega_{as0}(t)\rho_{42} -  \tfrac{1}{2}\Omega^*_{s0}(t)e^{i\delta(t-t_c)-\frac{i}{2}\alpha_s(t-t_c)^2}\rho_{13} \\ &-\tfrac{1}{2}\Omega^*_{pr0}(t)e^{i\delta(t-t_c)-\frac{i}{2}\alpha_{pr}(t-t_c)^2}\rho_{14} \,,\\
			i\dot{\rho}_{13} =&   \Delta_{s}\rho_{13} + \tfrac{1}{2}\Omega_{p0}(t)e^{\frac{i}{2}\alpha_{p}(t-t_c)^2}\rho_{33} + \tfrac{1}{2}\Omega_{as0}(t)\rho_{43} -  \tfrac{1}{2}\Omega_{p0}(t)e^{\frac{i}{2}\alpha_{p}(t-t_c)^2}\rho_{11} \\ &-\tfrac{1}{2}\Omega_{s0}(t)e^{-i\delta(t-t_c)+\frac{i}{2}\alpha_{pr}(t-t_c)^2}\rho_{12} \,,\\
			i\dot{\rho}_{14} =&   \Delta_{as}\rho_{14} + \tfrac{1}{2}\Omega_{p0}(t)e^{\frac{i}{2}\alpha_{p}(t-t_c)^2}\rho_{34} + \tfrac{1}{2}\Omega_{as0}(t)\rho_{44} -  \tfrac{1}{2}\Omega_{a0}(t)\rho_{11} \\ &-\tfrac{1}{2}\Omega_{pr0}(t)e^{-i\delta(t-t_c)+\frac{i}{2}\alpha_{pr}(t-t_c)^2}\rho_{12} \,,\\
			i\dot{\rho}_{23} =& \Delta_{s}\rho_{23} +  \tfrac{1}{2}\Omega_{s0}(t)e^{-i\delta(t-t_c)+\frac{i}{2}\alpha_s(t-t_c)^2}\rho_{33} + \tfrac{1}{2}\Omega_{pr0}(t)e^{-i\delta(t-t_c)+\frac{i}{2}\alpha_{pr}(t-t_c)^2}\rho_{43} \\ &-\tfrac{1}{2}\Omega_{p0}(t)e^{\frac{i}{2}\alpha_{p}(t-t_c)^2}\rho_{21} - \tfrac{1}{2}\Omega_{s0}(t)e^{-i\delta(t-t_c)-\frac{i}{2}\alpha_s(t-t_c)^2}\rho_{22} \,,\\
			i\dot{\rho}_{24} =& \Delta_{as}\rho_{24} +  \tfrac{1}{2}\Omega_{s0}(t)e^{-i\delta(t-t_c)+\frac{i}{2}\alpha_s(t-t_c)^2}\rho_{34} + \tfrac{1}{2}\Omega_{pr0}(t)e^{-i\delta(t-t_c)+\frac{i}{2}\alpha_{pr}(t-t_c)^2}\rho_{44} \\ &-\tfrac{1}{2}\Omega_{as0}(t)\rho_{21} - \tfrac{1}{2}\Omega_{pr0}(t)e^{-i\delta(t-t_c)+\frac{i}{2}\alpha_{pr}(t-t_c)^2}\rho_{22} \,,\\
			i\dot{\rho}_{34} =&  (\Delta_{as}-\Delta_{s})\rho_{34} + \tfrac{1}{2}\Omega^*_{p0}(t)e^{-\frac{i}{2}\alpha_{p}(t-t_c)^2}\rho_{14} + \tfrac{1}{2}\Omega^*_{s0}(t)e^{i\delta(t-t_c)-\frac{i}{2}\alpha_s(t-t_c)^2}\rho_{24} \\ &-\tfrac{1}{2}\Omega_{as0}(t)\rho_{31} - \tfrac{1}{2}\Omega_{pr0}(t)e^{-i\delta(t-t_c)+\frac{i}{2}\alpha_{pr}(t-t_c)^2}\rho_{32} \,.\\
		\end{aligned}
	\end{align}
	The condition for chirping of the probe pulse, $\alpha_{pr} = \alpha_s - \alpha_p$, is then imposed which is necessary to equate the exponentials. The above set of equations can be simplified considering the conditions for adiabatic elimination, $\dot{\rho}_{33}=\dot{\rho}_{44}=\dot{\rho}_{34} = 0$ and substituting for $\rho_{14}, \rho_{24}, \rho_{23}$ and $\rho_{24}$ in the equations of $\dot{\rho}_{11}, \dot{\rho}_{22}$ and $\dot{\rho}_{12}$. After defining the new Rabi frequencies:
	
	\begin{equation}
		\Omega_{1}(t)=\frac{|\Omega_{p0}(t)|^{2}}{4\Delta_s}+\frac{|\Omega_{as0}(t)|^{2}}{4\Delta_{as}}, \ \ \ \ \
		\Omega_{2}(t)=\frac{|\Omega_{s0}(t)|^{2}}{4\Delta_s}+\frac{|\Omega_{pr0}(t)|^{2}}{4\Delta_{as}}\,,
	\end{equation}	
	and 
	\begin{equation}
		\Omega_{3}(t)=\frac{\Omega_{p0}(t)\Omega^*_{s0}(t)}{4\Delta_s}+\frac{\Omega^*_{pr0}(t)\Omega_{as0}(t)}{4\Delta_{as}}\,,
	\end{equation}
	the density matrix equations are reduced to:
	\begin{align}
		\begin{aligned}
			i\dot{\rho}_{11} =& \Omega_3(t)e^{i\delta(t-t_c)-\frac{i}{2}(\alpha_s - \alpha_{p})(t-t_c)^2}\rho_{21} - c.c \,, \\
			i\dot{\rho}_{22} =& \Omega^*_3(t)e^{-i\delta(t-t_c)+\frac{i}{2}(\alpha_s - \alpha_{p})(t-t_c)^2}\rho_{12} - c.c \,, \\
			i\dot{\rho}_{12} =& \left[\Omega_1(t)-\Omega_{2}(t)\right]\rho_{12} + \Omega_3(t)e^{i\delta(t-t_c)-\frac{i}{2}(\alpha_s - \alpha_{p})(t-t_c)^2}(\rho_{22}-\rho_{11})\,.
		\end{aligned}
	\end{align} 
	
	Further transformations: $\rho_{11}=\tilde{\rho}_{11}$, $\rho_{12}=\tilde{\rho}_{12} e^{i\delta(t-t_c)-\frac{i}{2}(\alpha_s - \alpha_{p})(t-t_c)^2}$, $\rho_{22} = \tilde{\rho}_{22}$ and shifting of diagonal elements lead to the following Hamiltonian in the field-interaction representation for the ``super-effective'' two-level system
	\begin{equation}\label{HAM_supereff}
		\mathbf{H}_{se}(t) = 
		\frac{\hbar}{2}
		\begingroup 
		\setlength\arraycolsep{-15pt}
		\begin{pmatrix}
			\small
			\delta-(\alpha_s-\alpha_p)(t-t_{c})+\Omega_{1}(t)-\Omega_{2}(t)  & 2\Omega_{3}(t)\\ 
			2\Omega^*_{3}(t) & -\delta+(\alpha_s-\alpha_p)(t-t_{c})-\Omega_{1}(t)+\Omega_{2}(t)  \\
		\end{pmatrix}\
		\endgroup
	\end{equation}
	
	The amplitudes of incoming fields can be manipulated to make the AC stark shifts equal,
	$\Omega_1(t) = \Omega_2(t)$, which can be satisfied by taking $\Omega_{s0} = \Omega_{pr0} = \Omega_{p0}/\sqrt2$, considering the fact that the anti-Stokes field is absent before the interaction, $\Omega_{as0} = 0$. The effective Rabi frequency,  which is the relevant quantity in the dynamics, can then be written as: 
	\begin{equation}
		\Omega_3(t) = \frac{\Omega_{3(0)}}{\left[(1+\frac{\alpha_p'^2}{\tau_0^4})(1+\frac{\alpha_s'^2}{\tau_0^4})\right]^{1/4}}e^{-\frac{(t-t_c)^2}{\tau^2}}\,.
	\end{equation}
	
	\begin{figure}
		\includegraphics[scale=1.0]{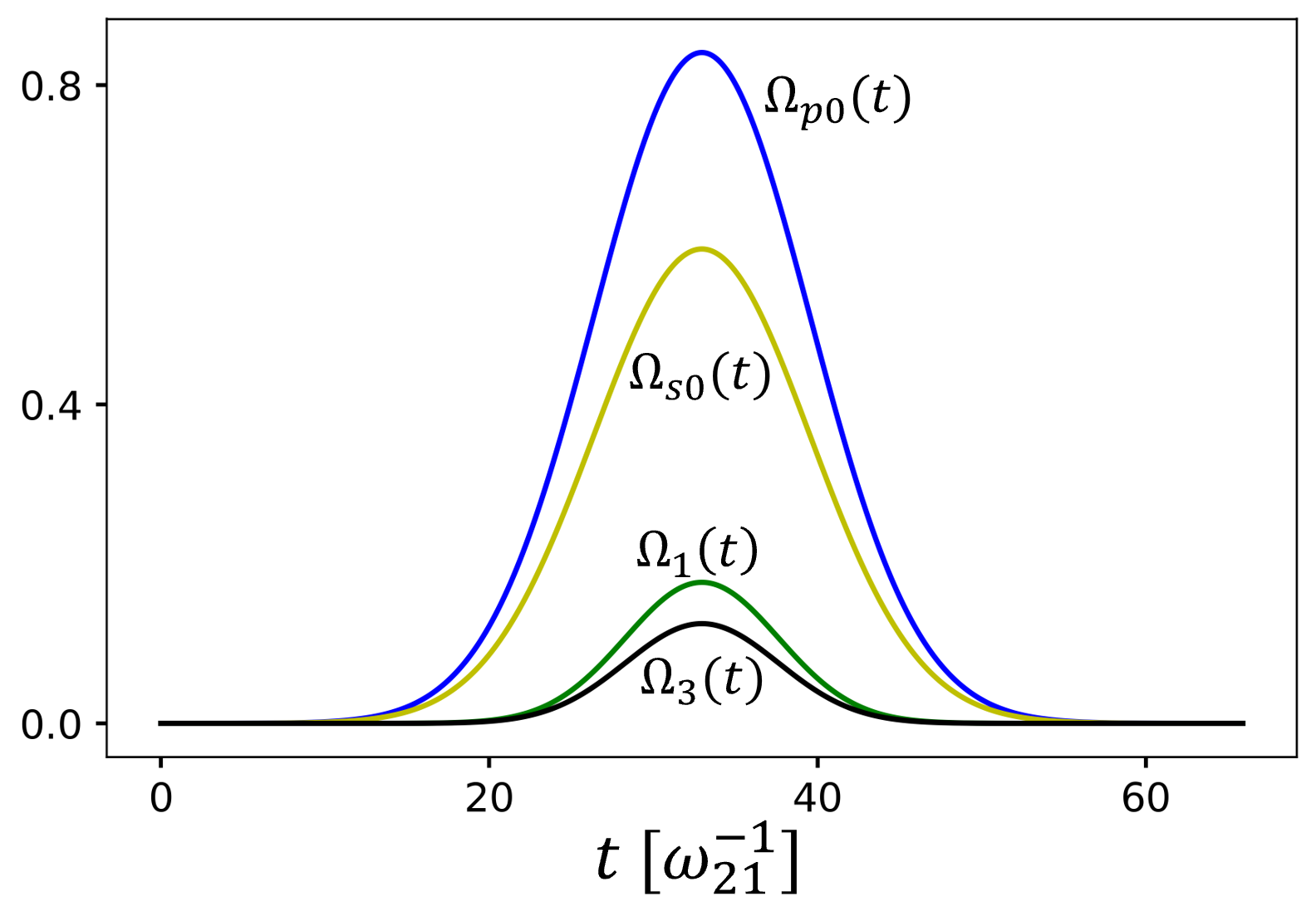}\centering
		\caption{The evolution of different Rabi frequencies in C-CARS scheme. The Stokes and probe Rabi frequencies have the same amplitude which is less than the amplitude of the pump pulse by a factor of $\sqrt{2}$. $\Omega_{1}(t)$ and $\Omega_{2}(t)$ are canceled in the Hamiltonian making $\Omega_{3}(t)$ the only relevant quantity in the scheme.
		}\label{pulses}
	\end{figure}
	The peak effective Rabi frequency of  transform-limited pulses   $\Omega_{3(0)}$ is given by  $\Omega_{3(0)} = \Omega_{p0}^2/(4\sqrt{2}\Delta)$. It is reduced when chirping is applied to the pump and Stokes pulses with the spectral rates $\alpha_p^{\prime}$ and $\alpha_s^{\prime}$ respectively. The relative amplitudes of all the Rabi frequencies involved in the dynamics are shown in Fig.\,\ref{pulses}. In this chapter and the following one, all the frequency parameters are defined in the units of the frequency $\omega_{21}$ and time parameters are defined in the units of $\omega^{-1}_{21} $.
	
	\subsection{The C-CARS chirping scheme}
	The dynamics of this processs can be written as follows: During the interaction, at time $t = t_c + \delta/(\alpha_s - \alpha_p)$, the diagonal elements equal to zero - creating a coherent superposition state having equal populations for the states $\ket{1}$ and $\ket{2} $ and therefore, a maximum coherence. This time can be determined for a fixed value of two-photon detuning, $\delta$. At two-photon resonance, the system reaches this maximum coherence at the central time $t_c$. The system can be preserved at this state of the maximum coherence by imposing the condition that $(\alpha_s - \alpha_p)$ is zero in the second half of the interaction. A smooth realization of this is possible  by choosing the temporal chirp rates of the pump and Stokes pulses to be opposite in sign before the central time and equal in sign after that, along with the condition imposed for the chirp rate of probe, $\alpha_{pr} = \alpha_s - \alpha_p$, which was used while deriving the Hamiltonian in Eq.\,\eqref{HAM_supereff}. This chirping scheme, namely C-CARS, can be summarized as: $\alpha_p = -\alpha_s$ and $\alpha_{pr} = 2\alpha_s$ for $t \leq t_c$, and $\alpha_p = \alpha_s$ and $\alpha_{pr} = 0$ for $t > t_c$. 
	
	\subsubsection{Wigner-Ville distributions}
	The Wigner-Ville distribution is one of the important methods for time-frequency analysis. For a function $f(t)$, the Wigner-Ville Distribution is given by:
	\begin{eqnarray}\label{wigner_equation}
		W_f(t,w) = \int_{-\infty}^{\infty} f\left(t+\frac{t'}{2}\right)\bar{f}\left(t-\frac{t'}{2}\right)e^{-i\omega t'} \,dt'\,.
	\end{eqnarray}
	To find the Wigner-Ville distribution of the incident pulses, substitute the pulse equations \eqref{pulse_equations} in \eqref{wigner_equation}. In rotating wave approximation, the terms with $2\omega_q$ can be ignored as they oscillate at a higher frequency compared to the terms with $\omega_q$ and average away to zero. The equation then becomes:
	\begin{eqnarray}
		\begin{aligned}
			W_{E_q}(t,\omega) &= \int_{-\infty}^{\infty} \frac{1}{4}E'^2_{q_0}e^{-(t-t_c)^2/\tau^2-t'^2/4\tau^2}\left[e^{i\omega_qt'+i\alpha_q(t-t_c)t'} + e^{-i\omega_qt'-i\alpha_q(t-t_c)t'}\right]e^{-i\omega t'}dt' \\
			&=  \frac{1}{4}E'^2_{q_0}e^{-(t-t_c)^2/\tau^2}\left[\int_{-\infty}^{\infty} e^{-t'^2/4\tau^2 + [-i(\omega-\omega_q)+i\alpha_q(t-t_c)]t'}dt' \right. \\ 
			&+ \left. \int_{-\infty}^{\infty} e^{-t'^2/4\tau^2 + [-i(\omega+\omega_q)-i\alpha_q(t-t_c)]t'}dt'\right]\,.
		\end{aligned}
	\end{eqnarray}
	Using the fact that for any Gaussian integral:
	\begin{equation}
		\int_{-\infty}^{\infty}e^{ax^2+bx}dx = \sqrt{\frac{\pi}{a}}e^{b^2/4a}\,,
	\end{equation}
	the Wigner-Ville distribution equations are given by:
	\begin{equation}
		W_{E_q}(t,\omega) = \frac{\tau\sqrt{\pi}}{2}E'^2_{q_0}e^{-(t-t_c)^2/\tau^2}\left[e^{-\tau^2[\omega-\omega_q-\alpha_q(t-t_c)]^2} + e^{-\tau^2[\omega+\omega_q+\alpha_q(t-t_c)]^2}\right]\,.
	\end{equation}
	
	\begin{figure}
		\includegraphics[scale=0.85]{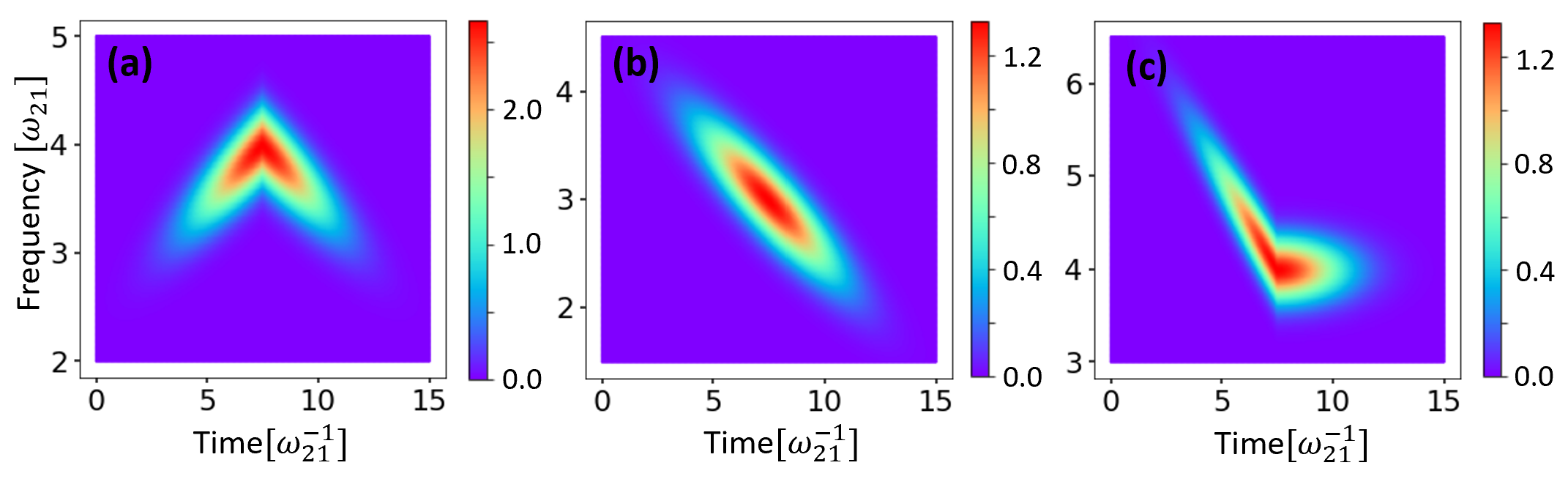}\centering
		\caption{Wigner plots of the incident pulses; pump(a), Stokes(b) and probe(c). Note that the Stokes and probe have the same amplitudes, which is different from that of the pump. The parameters used in this figure are: $\omega_p = 4.0[\omega_{21}]$, $\omega_s = 3.0[\omega_{21}]$, $\omega_{pr} = 4.0[\omega_{21}]$, $\tau = 3.0[\omega_{21}^{-1}]$, $\alpha_s = -0.2[\omega_{21}^{2}]$, and $t_c = 7.5[\omega_{21}^{-1}]$.}
		\label{wigner_plots}
	\end{figure}
	
	The positive solutions of these equations are given in Fig.\,\ref{wigner_plots}. The ``turning off'' of chirping in the second half is the essence of this scheme, resulting in a selective excitation of the molecules and suppressing any off-resonant background. If $\alpha_p$ is not reversed, the coherence is not preserved leading to population reversal between states $\ket{1}$ and $\ket{2}$.
	
	\subsubsection{Analysis of populations and coherence}
\begin{figure}
		\includegraphics[scale=0.6]{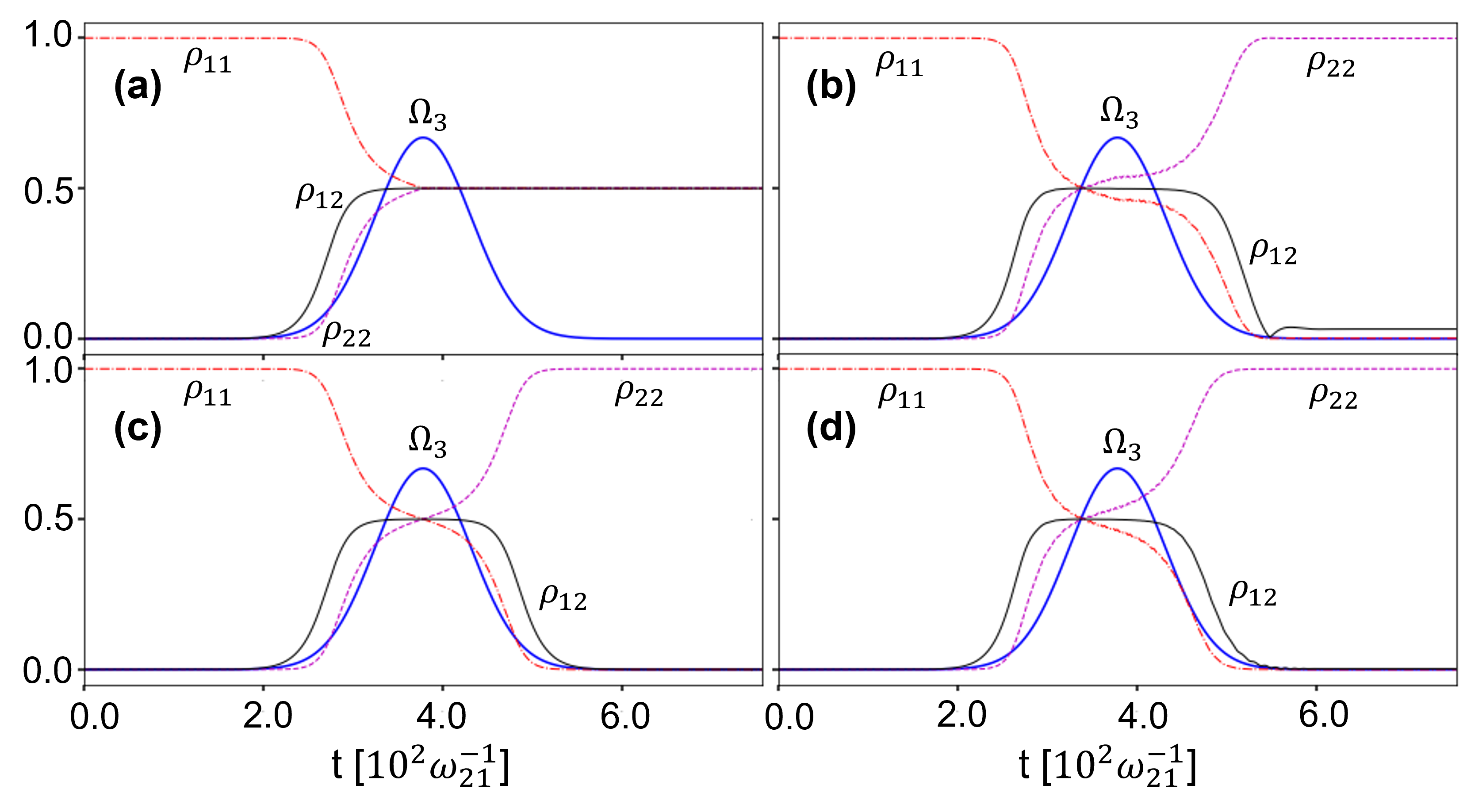}\centering
		\caption{The evolution of the populations and coherence demonstrating selective coherent excitation in C-CARS: in (a) and (b), C-CARS scheme is applied to the resonant case ($\delta =0$)  (a)  and the off-resonant case ($\delta = 0.1$) (b). Coherence is preserved at the maximum value in resonant case, while it is destroyed in the detuned case. This is in contrast with the chirping scheme where the pump and Stokes pulses are oppositely chirped, $\alpha_{p} = -\alpha_{s}$, for the whole pulse duration, shown for the resonant case ($\delta =0$) in (c)  and  off-resonant ($\delta = 0.1$) case in (d). The dynamics is similar and the coherence is zero in both these cases demonstrating the need for turning off the chirp at central time. 
			The parameters are: $\Omega_{3(0)} = 5.0[\omega_{21}]$, $\tau_0 = 10[\omega^{-1}_{21}]$, $\Delta = 1.0[\omega_{21}]$ and $\alpha_s'/\tau_0^2 = -7.5$. 
		}\label{populations}
	\end{figure}

	To demonstrate the selective excitation of molecules using C-CARS, the time evolution of populations $\rho_{11}$ and $\rho_{22} $ and coherence $\rho_{12}$ is presented in Fig.\,\ref{populations} for four different cases described below. The C-CARS control scheme is applied for the resonant ($\delta = 0$) and off-resonant ($\delta \neq 0$) cases respectively in figures \ref{populations}(a) and \ref{populations}(b). The coherence reaches maximum at a central time in the resonant case, which is preserved till the end of dynamics owing to the zero net chirp rate attained by reversing the sign of $\alpha_p$. On the contrary, the time of maximum coherence does not coincide with the central time in the non-resonant case, which results in a population transfer to the upper state and zero coherence. To emphasize the significance of reversing the sign of $\alpha_p$ in C-CARS control scheme we compare it with the scheme when the pump and Stokes are oppositely chirped for the whole pulse duration, $\alpha_p =-\alpha_s$. The  dynamics of the system in such case is plotted in figures \ref{populations}(c) and \ref{populations}(d) for $\delta = 0$ and $\delta = 0.1$ respectively.   In (c), even though the system reaches a perfect coherence at $t=t_c$, it drops to zero because population is  further adiabatically transferred to state $|2\rangle$. Coherence in (d) behaves similar to that of (b). 
	
	\begin{figure}
		\includegraphics[scale=1.0]{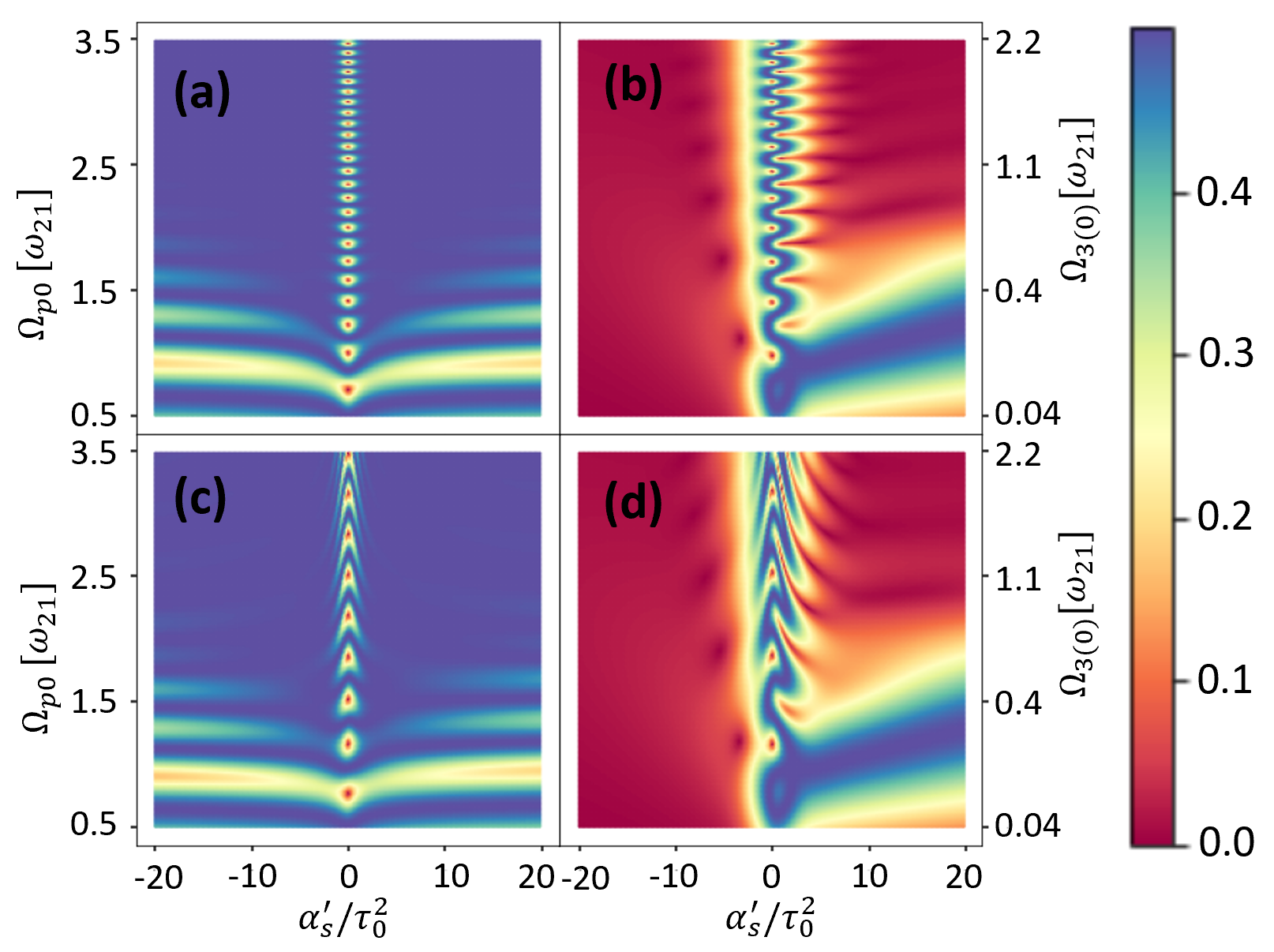}\centering
		\caption{Vibrational coherence as a function of spectral chirp and peak Rabi frequency when C-CARS chirping scheme is used: the above figures (a and b) are plotted using the super-effective two-level Hamiltonian, Eq.\,\eqref{HAM_supereff}, and below figures (c and d) are plotted using the exact four-level Hamiltonian, Eq.\,\eqref{Ham4level}. In figures (a) and (c), $\delta = 0$ and (b) and (d), $\delta = 0.1$. The similarity between the results of two Hamiltonians indicates the validity of the adiabatic approximation which is used to derive the chirping scheme. In the case of resonance, the coherence is maximum (blue) for most of the Rabi frequencies and spectral chirp rates, meaning that the chirping scheme is very robust against the changes in input parameters. In the absence of resonance, the coherence is zero (red) for most values of parameters, implying that the chirping scheme is effective in selectively exciting the system. The parameters used in this figure are: $\tau_0 = 10[\omega^{-1}_{21}] $ and $\Delta_s = \Delta_{as} = 1.0[\omega_{21}]\,.$
		}\label{coherence_maps}
	\end{figure}
	
	
	
	\subsubsection{Comparison with the exact four-level system}
	The validity of adiabatic approximation, which led to a derivation of the super-effective Hamiltonian, can be tested by comparing the results of the super-effective two-level system with the exact solution using the Liouville von Neumann equation for the four-level systems. To this end, the field interaction Hamiltonian of the four level system, after imposing the condition for chirping of the probe pulse  $\alpha_{pr} = \alpha_s - \alpha_p$, can be written as:
	
	\begin{equation}
		\mathbf{H}_{ex}(t) = \frac{\hbar}{2} \left( \begin{array}{cccc} 2\alpha_p(t-t_c)   & 0    &  \Omega_{p_0} (t) 
			&\Omega_{as_0}(t)\\
			0 &  2[\alpha_{s}(t-t_c)-\delta] &  \Omega_{s_0} (t)  & \Omega_{pr_0} (t) \\
			\Omega_{p_0} (t)  &  \Omega_{s_0} (t) &  -2\Delta_{s}  & 0\\
			\Omega_{as_0}(t) & \Omega_{pr_0} (t)  &  0  & 2[\alpha_p(t-t_c)-\Delta_{as}]  \\
		\end{array} \right).
		\label{Ham4level}
	\end{equation}
	
	
	Figure\,\ref{coherence_maps} shows the contour-plot of vibrational coherence $\rho_{12}$ at the end of dynamics as a function of the peak Rabi frequency $\Omega_{3(0)}(t) $ and dimensionless spectral chirp rate $\alpha_{s}'/\tau_0^2 $. Figures\,(a) and (b) represent the $\delta = 0$ and $\delta = 0.1$ cases, respectively, of the super-effective two level system, and (c) and (d) represent the same cases obtained by the exact solution of the four-level system using the same set of parameters. In all the figures, the one-photon detuning is $\Delta_s = \Delta_{as} = \Delta = 1.0$. For the adiabatic elimination procedure to work properly, the terms in the Hamiltonian containing detunings should be larger than the terms containing Rabi frequencies $|\Delta| \gg \Omega_3(0) $. Around the region where $\alpha_s^{'}/\tau_0^2 = 0$, $|\Delta|$ is comparable to the peak effective Rabi frequency, and the adiabatic approximation disagrees with the exact solution. As the magnitude of spectral chirp rate increases, the adiabatic approximation is accurate owing to the reduction in the peak effective Rabi frequency, because of the presence of high temporal chirp. In the resonant cases (a) and (c), the coherence is at the maximum - color blue - for the most part, indicating the robustness of C-CARS chirping scheme in preparing the system in a coherent superposition. In the off-resonant coherence, zero coherence - color red - is seen for the most part; it is in stark contrast with that of the resonant case, revealing the selective nature of coherent excitation using the C-CARS chirping scheme.  
	
	\begin{figure}
		\includegraphics[scale=1.0]{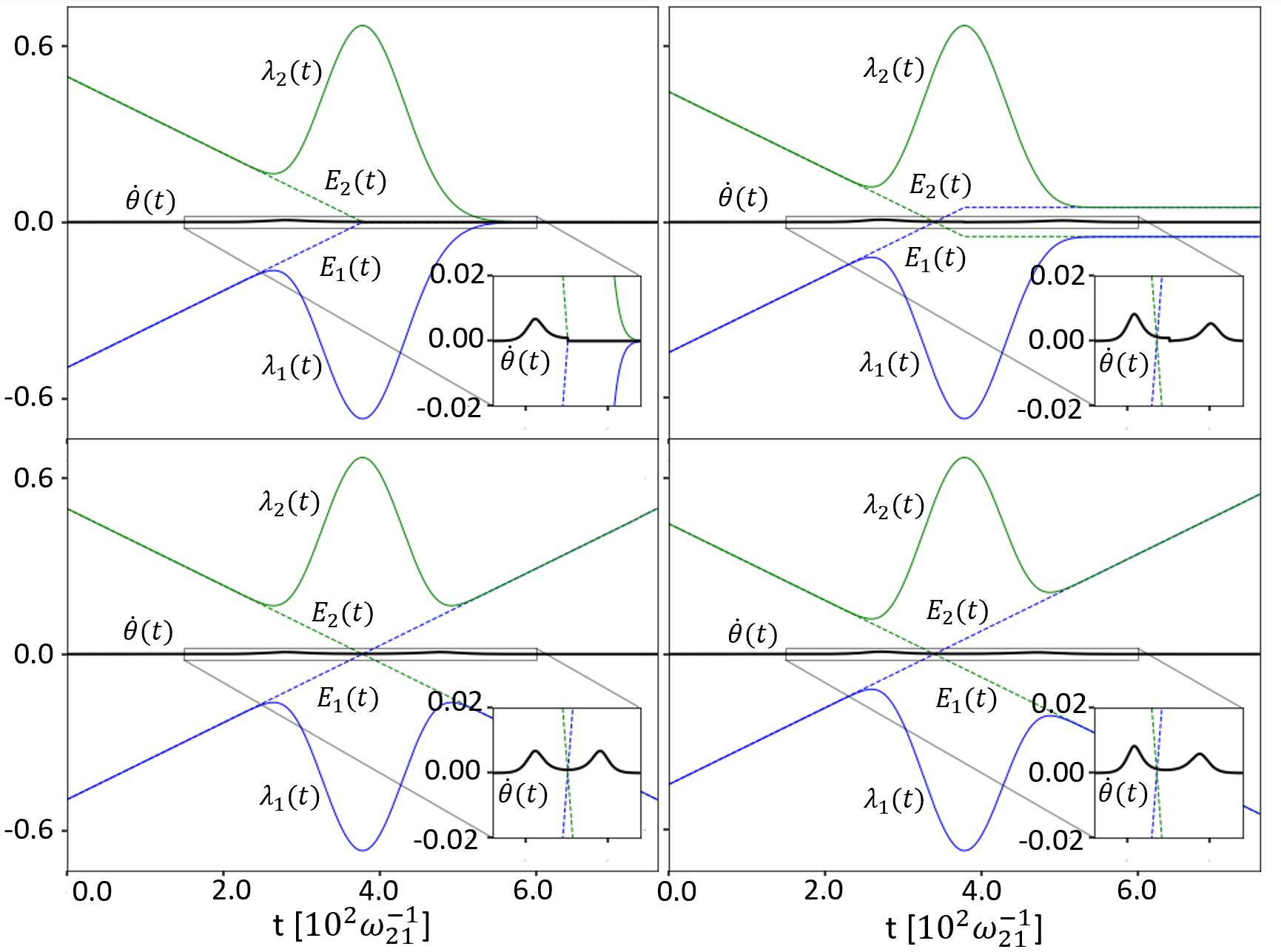}\centering
		\caption{The evolution of the bare state and the dressed state energies and the non-adiabatic parameter: $E_1(t)$ and $E_2(t)$ (dashed lines) are the bare state energies and $\lambda_1(t)$ and $\lambda_2(t)$ (solid lines) are the dressed state energies. Figures (a) and (b) are the resonant and off-resonant cases respectively when C-CARS scheme is used. Figures (c) and (d) show the resonant and off-resonant cases, when the pump and Stokes pulses are oppositely chirped for the whole pulse duration. In contrast to all the other cases, figure (a) shows the non-adiabatic parameter $\dot{\theta}(t)$, the dark solid line, remaining at zero after the central time. The parameters used are: $\Omega_{3(0)} = 5.0[\omega_{21}]$, $\tau_0 = 10[\omega^{-1}_{21}]$, $\Delta = 1.0[\omega_{21}]$ and $\alpha_s'/\tau_0^2 = -7.5$.  
		}\label{dressed_energies}
	\end{figure}
	
	\subsection{Analysis of Dressed States and Adiabatic Passage}
	When light interacts with any quantum system, the eigenstates undergo some shifts resulting in a set of quantum states that are said to be `dressed' by the light. These states are called dressed states and the initial states that were `untouched' by the light are called bare states \cite{Berman_dressed_1982}. The robustness of the C-CARS chirping scheme stems from the adiabatic nature of the interaction,  which can be demonstrated by analyzing the evolution of dressed state energies. To this end, the density matrix $\boldsymbol{\rho}(t) $ is transformed to a dressed density matrix using the transformation $\boldsymbol{\rho}_d(t) = \mathbf{T}(t)\boldsymbol{\rho}(t) \mathbf{T}^{\dagger}(t), $ where $\mathbf{T}(t) $ is an orthogonal matrix given by:
	\begin{equation}\label{T}
			\mathbf{T}(t) = 
			\begin{pmatrix}
				\cos\theta(t)	&	-\sin\theta(t)	\\
				\sin\theta(t)	&	\cos\theta(t) 	\\
			\end{pmatrix}\,.
	\end{equation}
	Since $\mathbf{T}$ is an orthogonal matrix, $\mathbf{T}\mathbf{T}^\dagger = \mathbf{T}^\dagger\mathbf{T} = 1$, and
	\begin{equation}
		\boldsymbol{\rho} = \mathbf{T}^\dagger\boldsymbol{\rho}_d \mathbf{T}\,,
	\end{equation}
	the Liouville von-Neumann equations can now be transformed to the dressed frame:
	\begin{align}
		i\hbar \dot{\boldsymbol{\rho}} = \left[\mathbf{H}_{se},\, \boldsymbol{\rho}\right]
	\end{align}
	\begin{align}
		i\hbar \dv{t}\boldsymbol{\rho} = \left[\mathbf{H}_{se},\, \mathbf{T}^{\dagger}\boldsymbol{\rho}_d \mathbf{T}\right]\label{bare_liouville}
	\end{align}
	\begin{align}
		\begin{aligned}
			i\hbar \dv{t}\left(\mathbf{T}^{\dagger}\boldsymbol{\rho}_d \mathbf{T}\right) &= i\hbar \left[\left(\mathbf{T}^{\dagger}\boldsymbol{\rho}_d\right)\dot{\mathbf{T}} + \dot{\left(\mathbf{T}^{\dagger}\boldsymbol{\rho}_d\right)} \mathbf{T}\right]\\
			&= i\hbar \left[\mathbf{T}^{\dagger}\boldsymbol{\rho}_d\dot{\mathbf{T}} + \dot{\mathbf{T}}^\dagger\boldsymbol{\rho}_d\mathbf{T} + \mathbf{T}^\dagger\dot{\boldsymbol{\rho}}_d\mathbf{T}\right]
		\end{aligned}
	\end{align}
	Rewriting the Eq.\eqref{bare_liouville},
	\begin{align}
		\begin{aligned}
			i\hbar\left[ \mathbf{T}^{\dagger}\boldsymbol{\rho}_d\dot{\mathbf{T}} + \dot{\mathbf{T}}^\dagger\boldsymbol{\rho}_d\mathbf{T} + \mathbf{T}^\dagger\dot{\boldsymbol{\rho}}_d\mathbf{T}\right] &= \mathbf{H}_{se}\left(\mathbf{T}^{\dagger}\boldsymbol{\rho}_d \mathbf{T}\right) - \left(\mathbf{T}^{\dagger}\boldsymbol{\rho}_d \mathbf{T}\right)\mathbf{H}_{se}\\
			i\hbar\left[ \mathbf{T}^\dagger\dot{\boldsymbol{\rho}}_d\mathbf{T}\right] &= \mathbf{H}_{se}\left(\mathbf{T}^{\dagger}\boldsymbol{\rho}_d \mathbf{T}\right) - \left(\mathbf{T}^{\dagger}\boldsymbol{\rho}_d \mathbf{T}\right)\mathbf{H}_{se} \\
			&- i\hbar \left[\mathbf{T}^{\dagger}\boldsymbol{\rho}_d\dot{\mathbf{T}} + \dot{\mathbf{T}}^\dagger\boldsymbol{\rho}_d\mathbf{T}\right]
		\end{aligned}
	\end{align}
	Multiplying this equation with $\mathbf{T}$ from the left side and $\mathbf{T}^\dagger$ from the right side gives:
	\begin{align}
		\begin{aligned}
			i\hbar \dot{\boldsymbol{\rho}}_d &= \mathbf{T}\mathbf{H}_{se}\mathbf{T}^{\dagger}\boldsymbol{\rho}_d - \boldsymbol{\rho}_d \mathbf{T}\mathbf{H}_{se}\mathbf{T}^\dagger - i\hbar \boldsymbol{\rho}_d\dot{\mathbf{T}}\mathbf{T}^\dagger - i\hbar \mathbf{T}\dot{\mathbf{T}}^\dagger\boldsymbol{\rho}_d
		\end{aligned}
	\end{align}
	Beacuse $\mathbf{T} $ is orthogonal, $\dot{\mathbf{T}}\mathbf{T}^\dagger = -\mathbf{T}\dot{\mathbf{T}^\dagger}$, using this in the above equation gives:
	\begin{align}
		\begin{aligned}
			i\hbar \dot{\boldsymbol{\rho}}_d &= \mathbf{T}\mathbf{H}_{se}\mathbf{T}^{\dagger}\boldsymbol{\rho}_d - \boldsymbol{\rho}_d \mathbf{T}\mathbf{H}_{se}\mathbf{T}^\dagger + i\hbar \boldsymbol{\rho}_d\mathbf{T}\dot{\mathbf{T}}^\dagger - i\hbar \mathbf{T}\dot{\mathbf{T}}^\dagger\boldsymbol{\rho}_d
		\end{aligned}
	\end{align}
	Rearranging the equation gives the density matrix equations in the dressed frame:
	\begin{align}
		\begin{aligned}
			i\hbar \dot{\boldsymbol{\rho}}_d &= \mathbf{T}\mathbf{H}_{se}\mathbf{T}^{\dagger}\boldsymbol{\rho}_d  - i\hbar \mathbf{T}\dot{\mathbf{T}}^\dagger\boldsymbol{\rho}_d -  \boldsymbol{\rho}_d \mathbf{T}\mathbf{H}_{se}\mathbf{T}^\dagger + i\hbar \boldsymbol{\rho}_d\mathbf{T}\dot{\mathbf{T}}^\dagger \\
			&= \left(\mathbf{T}\mathbf{H}_{se}\mathbf{T}^{\dagger}  - i\hbar \mathbf{T}\dot{\mathbf{T}}^\dagger\right)\boldsymbol{\rho}_d -  \boldsymbol{\rho}_d\left( \mathbf{T}\mathbf{H}_{se}\mathbf{T}^\dagger - i\hbar \mathbf{T}\dot{\mathbf{T}}^\dagger\right) \\
			&= \left[\mathbf{T}\mathbf{H}_{se}\mathbf{T}^{\dagger}  - i\hbar \mathbf{T}\dot{\mathbf{T}}^\dagger\,, \boldsymbol{\rho}_d\right]\\
			i\hbar \dot{\boldsymbol{\rho}}_d &= \left[\mathbf{H}_d\,, \boldsymbol{\rho}_d\right]
		\end{aligned}
	\end{align}
	where $\mathbf{T}(t)\mathbf{H}_{se}(t)\mathbf{T^ \dagger}(t)$ is a diagonal matrix. For adiabatic passage to occur, the dressed state Hamiltonian $\mathbf{H}_d(t)$ should give the dressed state energies separation to be greater than  $\mathbf{T}(t)\mathbf{\dot{T}^{\dagger}}(t)$ to avoid coupling between dressed states \cite{Berman_dressed_1982, Bergmann_1989, STIRAP_Shore_2015}. The dressed state Hamiltonian is found to be:
	
	\begin{equation}\label{dressed_Hamiltonian}
			\mathbf{H}_d(t) = \frac{\hbar}{2}
			\begin{pmatrix}
				-\sqrt{(-\delta+\alpha_{pr}(t-t_{c}))^2+(2\Omega_{3}(t))^2}	&	-i\dot{\theta}(t)	\\
				i\dot{\theta}(t)	&	\sqrt{(-\delta+\alpha_{pr}(t-t_{c}))^2+(2\Omega_{3}(t))^2}	\\
			\end{pmatrix}\,,
		\end{equation}
		where the non-adiabatic parameter $\dot{\theta}(t)$, which comes from the matrix $\mathbf{T}(t)\mathbf{\dot{T}^{\dagger}}(t)$, is given by:
		\begin{align}
			\dot\theta(t) &=  \frac{(-\delta+\alpha_{pr}(t-t_{c}))\dot\Omega_3(t) - 2\Omega_3(t)\alpha_{pr}}{(-\delta+\alpha_{pr}(t-t_{c}))^2 + 4\Omega_3(t)^2}\,.
		\end{align}
		
		\begin{figure}
			\includegraphics[scale=0.6]{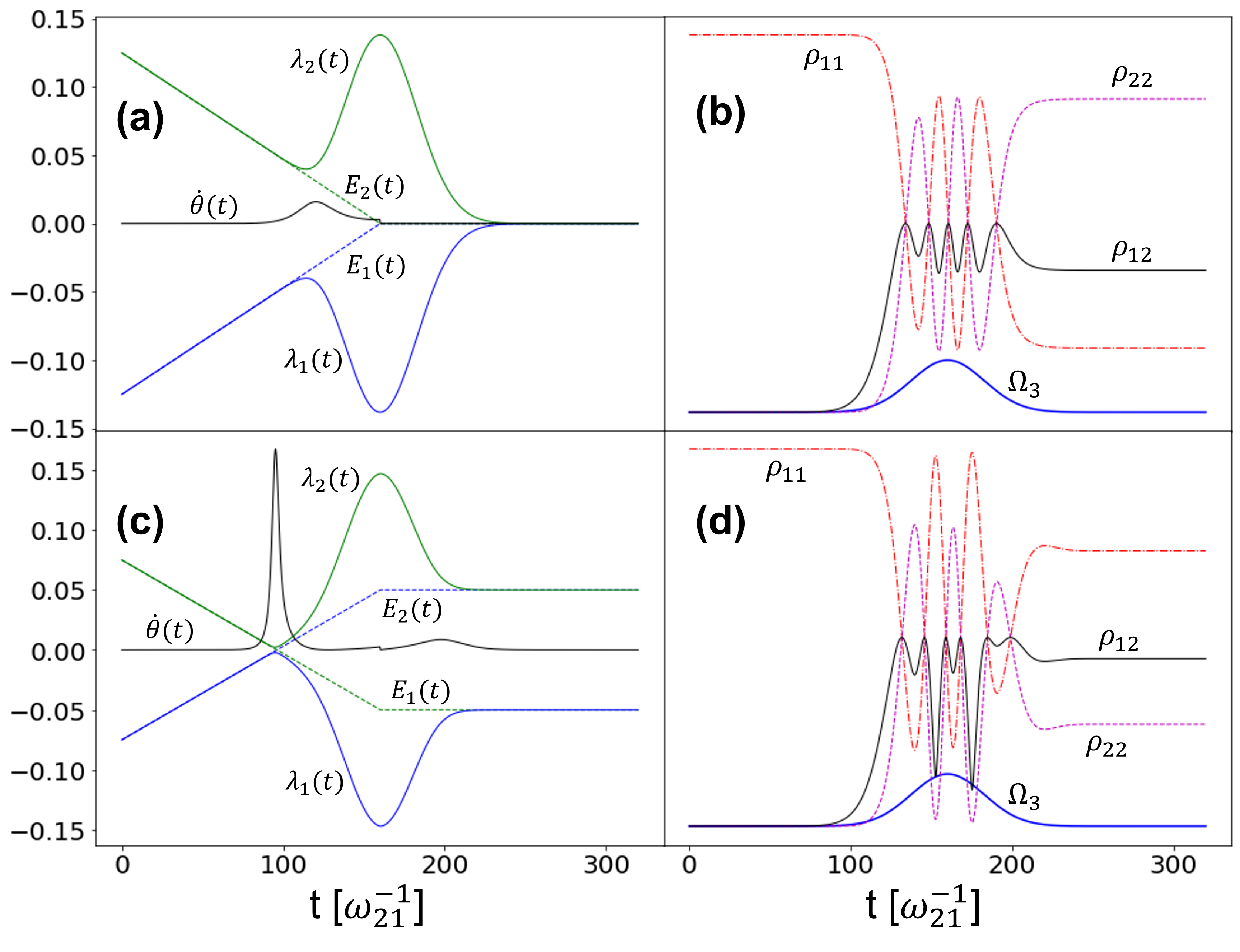}\centering
			\caption{Evolution of energies, populations and coherence for for field parameters $\Omega_{3(0)} = 0.18$, $\tau_0 = 25$ and $\alpha_s'/\tau_0^2 = -0.8$. Figure (a) show the dressed and bare state energies and (b) show the population dynamics in the case of resonance ($\delta =0$). Figures (c) and (d) show the same plots in the absence of resonance ($\delta =0.1 $).
			}\label{populations-2}
		\end{figure}
		
		Analyzing the equation for $\dot{\theta}(t)$ reveals the selective nature of adiabatic passage in the case of resonance. In the resonant case, the C-CARS chirping scheme ensures that the process be adiabatic in the second half of the pulse by keeping the  non-adiabatic coupling parameter $\dot{\theta}(t)$ at zero during this time period. But adiabaticity is not guaranteed in the off-resonant case due to the non-zero factor $\delta$ in the equation. This is demonstrated in Fig.\, \ref{dressed_energies}, where the bare state energies are given by: $E_1(t) = {H}_{se_{11}}(t)$ and  $E_2(t) = {H}_{se_{22}}(t)$  and the dressed state energies are given by: $\lambda_1(t) = {H}_{d_{11}}(t)$ and $\lambda_2(t) = {H}_{d_{22}}(t)$. The figures (a) and (b) represent resonant ($\delta =0$) and off-resonant ($\delta \neq 0$) cases when C-CARS chirping scheme is used. Clearly, the $\dot{\theta}(t)$, dark sold line, has non-zero values in the second half when the system is detuned. The perfectly adiabatic nature of interaction in Fig.\,\ref{dressed_energies}(a) corresponds to the maximum coherence in Fig.\,\ref{populations}(a) and the non-adiabatic nature in Fig.\,\ref{dressed_energies}(b) corresponds to the population inversion in Fig.\,\ref{populations}(b). The parameters used in Fig.\,\ref{dressed_energies} are the same as that used in Fig.\,\ref{populations}.  In figures \ref{dressed_energies}(c) and \ref{dressed_energies}(d), the same quantities are plotted for $\delta =0$ and $\delta \neq 0$ respectively for the scheme when the pump and Stokes are chirped oppositely for the whole pulse duration. 
		The process is perfectly adiabatic only in the second half of (a) since a smooth realization of $\dot{\theta}(t) = 0$ was made possible owing to the developed C-CARS chirping scheme. The dynamics is greatly different and the selective excitation does not happen when the effective Rabi frequency, $\Omega_3(t)$, is not strong enough - as shown in Fig.\,\ref{populations-2} where $\Omega_3(0) = 0.18$, $\alpha_s'/\tau_0^2 = -0.8$ and $\tau_0 = 25[\omega^{-1}_{21}]$. In both (a), $\delta=0$ and (b), $\delta=0.1 $ cases, the non-adiabatic coupling parameter is much greater compared to Fig.\,\ref{dressed_energies}. In the off-resonant case (d), the coherence oscillates much stronger compared to the resonant case (c) showing the absence of adiabatic passage.
		
\begin{figure}
			\includegraphics[scale=1.0]{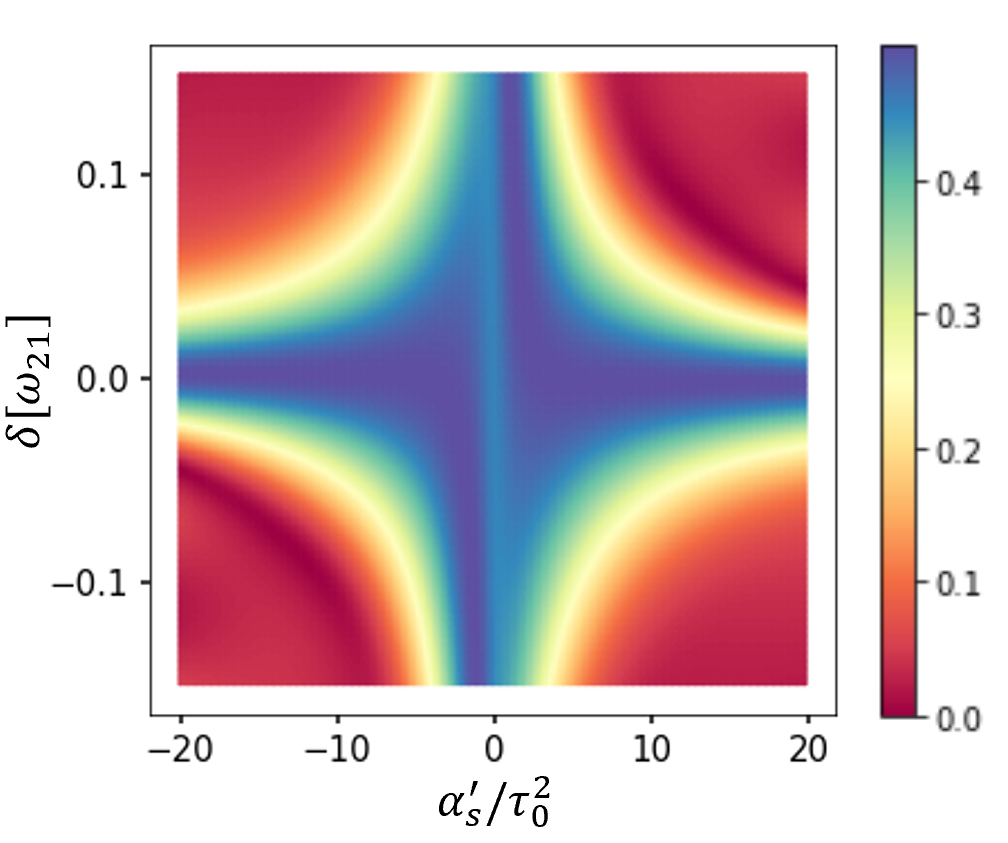}\centering
			\caption{Coherence as a function of two-photon detuning and dimensionless spectral chirp rate. As the chirp rate increases, the system becomes selective to lesser values of $\delta$. For the values of spectral chirp rate close to zero, $\alpha_s^{\prime}/\tau_0^2 \approx 0 $, a vertical line is present as the selectivity is degraded where effectively there is no chirping.
				Flipping the signs of detuning and chirp rates results in flipping of energy levels in the two-level Hamiltonian, creating a symmetry for the coherence values of ($\alpha_s^{\prime}/\tau_0^2$, $\delta$) and ($-\alpha_s^{\prime}/\tau_0^2$, $-\delta$). The parameters used in the figure are: $\Omega_{3(0)}=1.6[\omega_{21}]$ and $\tau_0 = 4.66[\omega^{-1}_{21}]$. 
			}\label{delta_vs_chirp}
		\end{figure}

		
		To investigate how the value of two-photon detuning and spectral chirp rate are related to the selectivity in C-CARS method, the end-of-pulse vibrational coherence is plotted as a function of $\delta $ and $\alpha_s'/\tau_0^2 $ in Fig.\,\ref{delta_vs_chirp}. At two-photon resonance, the coherence is maximum for all the values of chirp rate. If the detuning is large, a small chirp rate can selectively excite the system while for smaller values of detuning, the chirp rate needs to be increased in order to suppress the non-resonant background. This provides a way to control the selectivity by adjusting the values of chirp rate. The plot is symmetric across the diagonal lines as flipping the signs of both $\alpha_{s}$ and $\delta$ will not change the dynamics; it would only result in switching of the diagonal elements in Hamiltonian \eqref{HAM_supereff}. The plot is also nearly symmetric across the $\delta = 0$ line, indicating that the selectivity holds for both red and blue detunings. When the chirp rate is close to zero, the pulses are transform limited, implying that the maximum coherence and selectivity provided by the chirping scheme is absent in this region. This explains the vertical line present at the 0 of abscissa.
		
		\subsection{Section Summary}	
		Here we presented a control scheme to prepare the ground electronic-vibrational states in the four-level system of CARS in a maximally coherent superposition. We derived the Hamiltonian for a `super-effective' two-level system employing the adiabatic approximation. This two-level Hamiltonian is used to derive the conditions for adiabatic passage necessary for the implementation of a selective excitation of spectrally close vibrations. The amplitudes of the Stokes and probe pulses have to be equal and should be $\sqrt{2} $ times that of the pump pulse. The pump pulse should be chirped at the same rate as the Stokes pulse before the central time and at opposite rate after that. The probe pulse has to be chirped at a rate equal to the difference between the chirp rates of the Stokes and the pump pulses for the whole pulse duration. The solutions of Liouville-von Neumann equation show that vibrational coherence is preserved until the end of dynamics in the resonant case due to the adiabatic nature of the interaction. At two-photon resonance, vibrational coherence is maximum, 0.5, for a wide range of field parameters revealing the robustness of the method. Conversely, coherence is almost zero in the off-resonant case for most of the peak Rabi frequency values and the chirp parameters. A comparison of the coherence in the four-level and the two-level systems reveals that the adiabatic approximation is valid except when the chirp rate is almost equal to zero. A dressed-state analysis further reveals the presence of adiabatic passage in the two-photon resonance case.
		
		Liquid crystal shapers have been successfully used for chirping of femtosecond pulses. Such a shaper can be used for realization the presented C-CARS chirping scheme. This method can find important applications in sensing and imaging of molecular species because it creates a maximally coherent superposition of vibrational states in coherent anti-Stokes Raman scattering allowing the system to emit an optimized signal suitable for detection. The robustness of this method against changes in Rabi frequencies and chirp rates is helpful in experiments. The method helps suppress the background species and excite only the desired species; the resolution needed for this distinction can be controlled by the chirp parameter.

		\section{APPLICATION OF ULTRAFAST C-CARS FOR REMOTE DETECTION}
		
		
		In this section, we present the theory of generation of the anti-Stokes signal in CARS and apply the control scheme we developed in the last section to optimize the signal for remote detection of molecules. To resolve the ultrafast dynamics and optimize the output signal in CARS, we use femtosecond control pulses. We take into account the field propagation effects in a cloud of molecules. The motivation is to demonstrate the buildup of the anti-Stokes signal which may be used as a molecular signature in the backward CARS signal. The theory is based on the solution of the coupled sets of Maxwell's and the Liouville von Neumann equations and focuses on the quantum effects induced in the target molecules by the shaped laser pulse trains.  We analyze the enhancement of the backscattered anti-Stokes signal upon multiple scattering of radiation from the target molecules, which modifies propagating fields.  We examine the impact of decoherence induced by spontaneous decay and collisional dephasing. We demonstrate that decoherence due to spontaneous decay can be mitigated by applying the control pulse trains with the train period close to the decay time.
		
		The novelty of the study is in the demonstration of the buildup of coherent anti-Stokes signal as a result of controllability of vibrational coherence in the target molecules upon four chirped pulse trains propagation subject to multiple scattering events,	in utilizing the pulse train properties to mitigate decoherence and in implementing the Deep Convolution Network approach to evaluate the phase of the propagating fields, which provides with the information about the relative phase change between the pump, the Stokes, the probe and the anti-Stokes pulses. 
		
		As a case study we use the methanol vapor.  Methanol molecules have Raman active symmetric 2837 $cm^{-1}$ (85.05 THz) and asymmetric 2942 $cm^{-1}$ (88.20 THz) stretch modes. These values are within the range of molecular group vibrations  in various biochemical species, which span from 2800 to 3100 $cm^{-1}$ making the methanol a suitable choice as a surrogate molecule to allow for non-hazardous experiments in the lab. Thus, the results of methanol studies would be useful for the development of remote detection schemes 
		as well as for the environmental analyses.
		
		Various setups are available to perform CARS experiments satisfying the phase-matching conditions to separate the directional anti-Stokes signal from the incident fields. However for particles having a size comparable to the wavelength, the phase-mismatched factor is small and it was shown that the non-phase-matched CARS can provide an effective method to probe complex molecules \cite{scully-cars,Xie02}. For methanol, the ratio $4\pi \rho_0/\lambda \ll 1$, where $\rho_0\sim 10^{-10} m$ is the target molecule diameter; it relaxes the phase-matching condition and permits consideration of the collinear copropagating fields configuration.
		
		This section is organized as follows. In Section 3.1, a theoretical framework is formulated. Section 3.2 contains the numerical results for the methanol and a discussion. The section concludes with a Summary.
		
		\subsection{Theoretical framework}
		
		\subsubsection{Maxwell - Liouville von Neumann formalism}
		
		CARS is a third order nonlinear process in which three beams, the pump, the Stokes and the probe, at frequencies $\omega_p$, $\omega_s$ and $\omega_{pr}$ respectively, interact with the electronic vibrational - vibronic -  states of the target molecules to generate the anti-Stokes field at frequency $\omega_{as} = \omega_p + \omega_{pr} - \omega_s $, Fig(\ref{CARS_scheme}). In our control scheme, we use linearly chirped pulse trains which read
		\begin{equation}
			E_i(t) = \sum_{k=0}^{N-1} E_{i0} exp\{-\dfrac{(t-t_c-kT)^2}{2\tau^2}\} \cos\{\omega_{i0}(t-t_c-kT)+\alpha_i \dfrac{(t-t_c-kT)^2}{2}\}.
		\end{equation} 
		Here $T$ is the pulse train period, $t_c$ is the central time when the peak value of the Gaussian field envelope is $E_0$, $\tau$ is the chirp-dependent pulse duration, $\omega_{i0}$ is the carrier frequency, and   $\alpha_i$, $i = p, s, pr,$ is the linear chirp rate of an individual pump, Stokes and probe pulse in the respective pulse train. The values of $\alpha_i$ are chosen in accordance with the control scheme described in the last section, which means  $\alpha_s=-\alpha_p$ and $\alpha_{pr}= \alpha_s - \alpha_p$  for  $t \le t_c$; and  $\alpha_s=\alpha_p$ and $\alpha_{pr}= 0 $ for $t > t_c$ \cite{Pandya20}.
		Such chirped pulses induce the maximum coherence between vibronic states in the target molecules via adiabatic passage  provided the two-photon detuning $\delta=0$. Any slightly different vibrational mode not satisfying the two-photon resonance condition, $\delta \ne 0$, is suppressed as we explained in the last section. The selectivity of the mode excitation is determined by the condition $\tau \delta  \ge 1$. The chirped pulse duration $\tau$ relates to the transform-limited pulse duration $\tau_0$ as $\tau=\tau_0 ( 1+\alpha^{\prime 2}/\tau_0^4)^{1/2}$, and the temporal $(\alpha)$ and the spectral $(\alpha^\prime)$ chirps relate as $\alpha=\alpha^\prime \tau_0^{-4}/(1 + \alpha^{\prime 2}/\tau_0^4).$
		
		The matrix Hamiltonian written in the interaction representation and in the rotating wave approximation (RWA) was given in Eq. \eqref{HAM_4_level_int}.
		\begin{figure}
			{\includegraphics[scale=0.4]{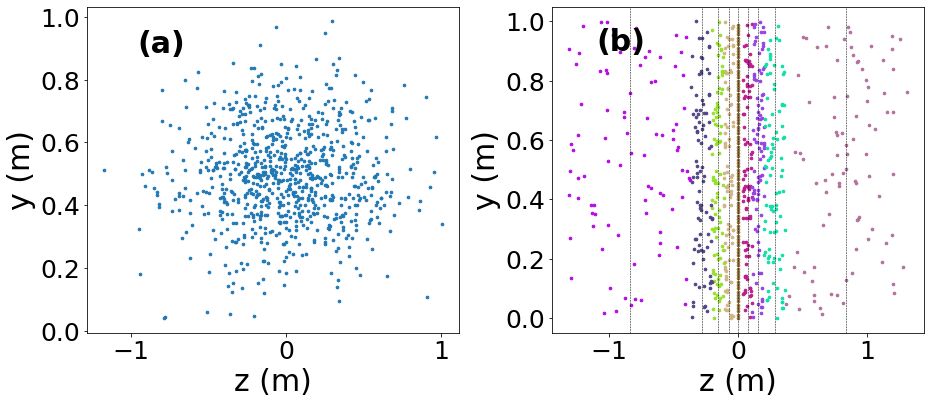}}
			\centering
			\caption{The Gaussian distribution of the target molecules in (a) based on the density of molecules and in (b) is converted into multi-layer model; molecules are given different colors to distinguish the layers. Each layer in the multi-layer model is characterized by the fractional number density $\eta$ and a distance to it's adjacent layer $(\Delta z)_\eta$. If $N_s$ is the number of the target molecules and $N$ is the number of total molecules associated with the layer, the fractional number density of that layer is defined as $\eta = N_s/N$.  The distance between the adjacent layers $(\Delta z)_\eta$ changes according to the Gaussian distribution of molecules. The incoming pulses go through a series of scattering events with the target molecules within each layer to produce a detectable backscattered CARS signal. 
			}\label{molecules}
		\end{figure}
		To account for the propagation effects in the scattering process, we combine the Liouville von Neumann equation for the states with Maxwell's equations for the fields. The displacement current is determined as $D= \epsilon_0 E + P$, where P is the expectation value of the induced dipole moment and $\epsilon_0$ is the permittivity of free space. The effects arising from magnetization are neglected giving  $B = \mu_0 H$, where $\mu_0 $ is permeability of free space. The wave equation for a field propagating in the $\hat{z}$ direction and having linear polarization in the Y plane reads:
		
		\begin{eqnarray}\label{waveeqn}
			\left(\pdv{}{z} + \frac{1}{c}\pdv{}{t} \right) \left( -\pdv{}{z} + \frac{1}{c} \pdv{}{t} \right)E =  - \mu_0 \pdv[2]{P}{t} \label{_Sv4}
		\end{eqnarray}
		
		Assuming the field is $E(z,t) = \frac{1}{2} (E_0(z,t)e^{-i[\omega t - kz - \phi(z,t)]}+c.c) $ and considering $E_0(z,t)$ and $\phi(z,t)$ as slowly varying functions of position and time, Eq.(\ref{waveeqn}) can be written as:
		\begin{equation}
			-2k \left(\pdv{ E_0(z,t)}{z}+ \frac{1}{c}  \pdv{ E_0(z,t)}{t} \right) \sin{(\omega t - k z - \phi(z,t))}= - \mu_0 \pdv[2]{}{t}P(z,t)
		\end{equation}
		
		Substituting $P(z,t) = \frac{1}{2} (P_0(z,t)e^{-i[\omega t - kz - \phi(z,t)]}+c.c )$ in the RHS, Eq.(\ref{waveeqn}) becomes:
		
		\begin{equation}\label{_slow}
			-2k (\pdv{ E_0(z,t)}{z}+ \frac{1}{c}  \pdv{ E_0(z,t)}{t})=
			\mu_0 \omega^2  Im\left[ P_0(z,t)\right].
		\end{equation}
		
		In quantum theory, the macroscopic polarization P is given by the expectation value of the electric dipole moment $\hat{\mu}$; $ \langle P(z,t) \rangle = N_sTr \{\langle\rho (z,t)\cdot\mu \rangle\} $, where $N_s$ is the molecular density of the target molecules. Applied to the four-level system of CARS, the four components of P can written as: $P_{0p}(z,t)=N_s \mu_{13}\rho_{13}(z,t)$, $P_{0s}(z,t)=N_s \mu_{23}\rho_{23}(z,t)$, $P_{0pr}(z,t)=N_s \mu_{24}\rho_{24}(z,t)$, and $P_{0as}(z,t)=N_s \mu_{14}\rho_{14}(z,t)$.
		The space components can be eliminated by the following procedure: If $\bar{t} = (t-\frac{z}{c})$, then $ \dv{t}{z} = (\dv{\bar{t}}{z} + \frac{1}{c}), $ which leads to  $\pdv{}{z} = \pdv{}{t}\pdv{t}{z} = \frac{1}{c} \pdv{}{t}$. Using the above expressions of polarizations, Eq.(\ref{_slow}) casts into:
		
		\begin{equation}
			\frac{1}{c} \pdv{E_{q}}{t} = -N_s \mu_0 \mu_{ij} \frac{E_{q} (t)}{\hbar}  \Im {\rho_{ij}} \label{_eq:3}
		\end{equation}
		where $q=p, s, pr, as$ and $i,j$ are the indexes of the states involved in the respective transitions. The following transformations are applied to the density matrix elements:
		\begin{eqnarray}
			\begin{aligned}
				\rho_{12}&=\tilde{\rho}_{12} e^{i ( \alpha_p  -\alpha_s) t^2 /2} \nonumber \\
				\rho_{13}&=\tilde{\rho}_{13} e^{i (\Delta_s t + \alpha_p t^2 /2)} \nonumber \\
				\rho_{14}&=\tilde{\rho}_{14} e^{ i \Delta_{as} t}\nonumber \\
				\rho_{23}&=\tilde{\rho}_{23} e^{i (\Delta_s t + \alpha_s t^2 /2)} \nonumber \\
				\rho_{24}&=\tilde{\rho}_{24} e^{i (\Delta_{as} t + \alpha_{pr} t^2 /2)} \nonumber \\
				\rho_{34}&=\tilde{\rho}_{34} e^{i (\Delta_{as} - \Delta_{s}) t - i\alpha_p t^2 /2}
			\end{aligned}
		\end{eqnarray}
		and the 'tilde' is removed. Then the density matrix elements $\rho_{ij}$ for the corresponding transitions are found using the Liouville von Neumann equation $i\hbar\dot{\rho} =[H,\rho]$ with the Hamiltonian from Eq.(\ref{HAM_4_level_int}). After applying the rotating wave approximation and the adiabatic elimination of the excited states assuming that $ \dot{\rho}_{13},\dot{\rho}_{14},\dot{\rho}_{23},\dot{\rho}_{24}, \dot{\rho}_{34} \approx 0, \; \rho_{34} \approx 0, \; \rho_{33},\rho_{44} \ll \rho_{11},\rho_{22} $ \quad and \quad $\dot{\rho}_{33},\dot{\rho}_{44} \approx 0 $, and using the control condition on the chirp parameters $\alpha_s -\alpha_p=\alpha_{pr}$, the density matrix elements $\rho_{13},\rho_{23},\rho_{14},\rho_{24}$ read in terms of $\rho_{11},\rho_{22}$ and $\rho_{12}$ in the field interaction representation as follows:
		
		\begin{equation}
			\begin{aligned}
				\rho_{13} &=  \dfrac{1}{2(\Delta_s + \alpha_p t)}\Omega_{p0}(t)\rho_{11} + \dfrac{1}{2(\Delta_s + \alpha_p t)}\Omega_{s0}(t)\rho_{12}\\
				\rho_{23} &=  \dfrac{1}{2(\Delta_s + \alpha_s t)}\Omega_{s0}(t)\rho_{22} + \dfrac{1}{2(\Delta_s + \alpha_s t)}\Omega_{p0}(t)\rho_{21} \\
				\rho_{14} &=  \dfrac{1}{2\Delta_{as}}\Omega_{as0}(t)\rho_{11} + \dfrac{1}{2\Delta_{as}}\Omega_{pr0}(t)\rho_{12} \\
				\rho_{24} &=  \dfrac{1}{2(\Delta_{as} + \alpha_{pr} t)}\Omega_{pr0}(t)\rho_{22} + \dfrac{1}{2(\Delta_{as} + \alpha_{pr} t)}\Omega_{as0}(t)\rho_{21}\,.
			\end{aligned}  \label{_eq:4}
		\end{equation}
		Further, substituting Eq.\eqref{_eq:4} into Eq.\eqref{_eq:3} and rewriting the equations in terms of Rabi frequencies lead to the following Maxwell's equations:
		
		\begin{equation}
			\label{MMM}
			\begin{aligned}
				\pdv{\Omega_{p0}}{t} &= c\pdv{\Omega_{p0}}{z} = -\dfrac{\eta}{2(\Delta_s+\alpha_p t)} \kappa_{13} \omega_p \Omega_{s0}(t) \Im[\rho_{12}] \\
				\pdv{\Omega_{s0}}{t} &= c\pdv{\Omega_{s0}}{z} = \dfrac{\eta}{2(\Delta_s+\alpha_s t)}
				\kappa_{23} \omega_s \Omega_{p0}(t) \Im[\rho_{12}] \\
				\pdv{\Omega_{pr0}}{t} &= c\pdv{\Omega_{pr0}}{z} = \dfrac{\eta}{2(\Delta_{as}+\alpha_{pr} t)} \kappa_{24}\omega_{pr} \Omega_{as0}(t) \Im[\rho_{12}] \\
				\pdv{\Omega_{as0}}{t} &= c\pdv{\Omega_{as0}}{z} = -\dfrac{\eta}{2(\Delta_{as})} \kappa_{14} \omega_{as} \Omega_{pr0}(t) \Im[\rho_{12}].
			\end{aligned}
		\end{equation}
		Here $\kappa_{ij} = n \mu_0 \mu_{ij}^2 c^2/ (3\hbar)$, n is the number density of molecules  given by $N_A/V_0$ under the ideal gas conditions, where $N_A$ is the Avogadro's Number, $V_0$ is the molar volume, and $\eta$ is the fractional number density which will be described in detail in the next section.
		The factor 1/3 comes from the averaging over all orientations of the molecular dipole $\left<\mu_x\mu_y\right>=\left<\mu_x\mu_z\right>=\left<\mu_y\mu_z\right>=0$ and $\left<\mu_j\right>= (1/3 ) \mu^2, j = x, y, z$ \cite{Su17}. Considering dipole moment of methanol $\mu_{ij} = 1.70D$, the constant $\kappa_{ij}$ is found to be       $3.636 \times 10^{-3} [\omega_{21}]$.
		
		The Eqs.(\ref{MMM}) coupled with the multi-layer model described below are numerically solved using the transform-limited and the control pulse trains to find the scattered anti-Stokes signal. Note, that the right side of Eqs.(\ref{MMM}), which describes the induced polarization in the target molecules, depends only on the imaginary part of coherence $\rho_{21}$ out of all density matrix elements. Thus, the maximum value of this coherence provides the optimal amplitude of the scattered signal.  
		
		
		\begin{figure}
			\includegraphics[scale=0.3]{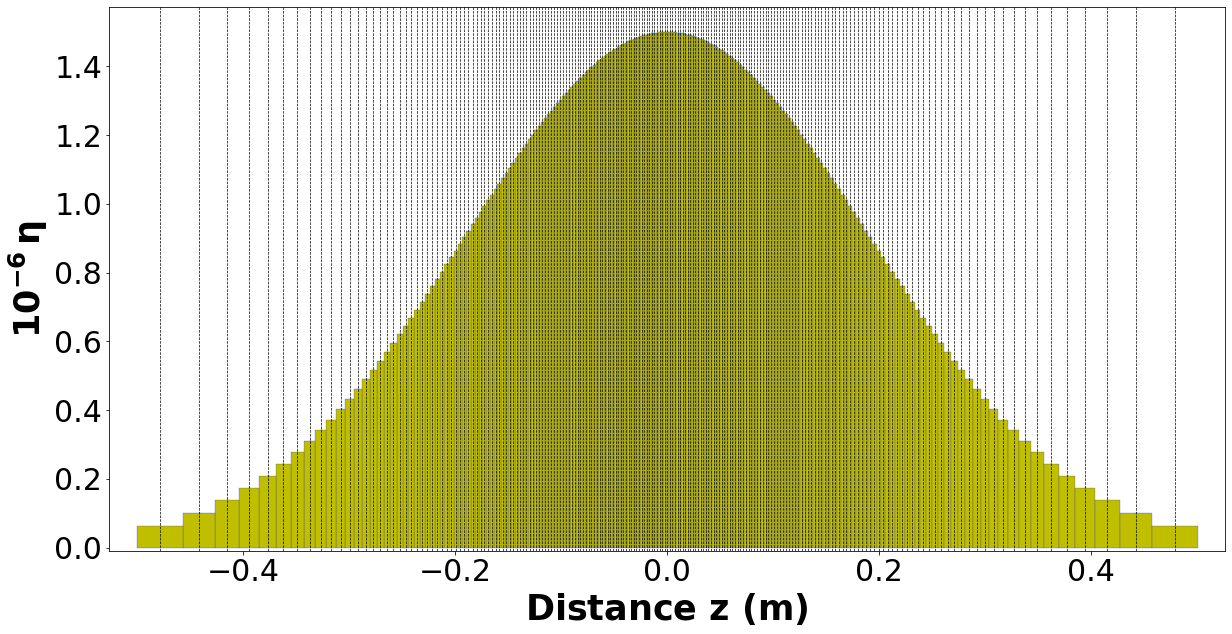}
			\centering
			\caption{An example of the multi-layer model of a molecular distribution for the width of the Gaussian distribution in Eq.(\ref{gauss}) of the target molecules $\sigma=0.19$ m. Here, each of 200 vertical lines represents the location of the scattering event and the scattering layers become more dense as the density peaks at the center.} \label{layers}
		\end{figure}
		
		To analyze the impact of decoherence due to spontaneous decay and collisional dephasing of molecules, the Liouville von Neumann equations are augmented by the relaxation terms. 
		Spontaneous decay from state $|i\rangle$ to state $|j\rangle$ is denoted by $\gamma_{ij}$, while collisional dephasing between states $|i\rangle$ and $|j\rangle$ is denoted by $\Gamma_{ij}$. Spontaneous decay impacts state populations and coherence via the diagonal and off-diagonal reduced density matrix elements respectively, while collisional dephasing assumed to be weak enough not to change state populations but to cause dipole phase interruption via off-diagonal reduced density matrix elements. Vibrational energy relaxation  \cite{Fu05,Ma08} within the ground electronic state is accounted through parameter $\gamma_{21}$. We neglect vibrational energy relaxation within the excited electronic state since the respective vibrational states $|3\rangle$ and $|4\rangle$ are negligibly populated during dynamics. Vibrational energy relaxation is an important topic in chemical physics, since it relates to fundamental reaction processes \cite{Be88,Gr04}, conformational changes \cite{Di02} or spectroscopic measurements \cite{Gi95,Ya02}, and its understanding is the first step toward controlling these phenomena.

		\begin{equation}\label{lvn_gen}
			\begin{aligned}
				\dot{\rho}_{11} &= - i/\hbar  [H,\rho]_{11} + \gamma_{21}\rho_{22} + \gamma_{31}\rho_{33} + \gamma_{41}\rho_{44} \\
				\dot{\rho}_{12} &= - i/\hbar [H,\rho]_{12}-(\gamma_{21}/2 + \Gamma_{21}) \rho_{12} \\
				\dot{\rho}_{13} &= - i/\hbar [H,\rho]_{13} - (\gamma_{31}/2 +\gamma_{32}/2+\gamma_{21}/2 +\gamma_{41}/2+\Gamma_{31})\rho_{13} \\
				\dot{\rho}_{14} &= - i/\hbar [H,\rho]_{14} - (\gamma_{41}/2 +\gamma_{42}/2+\gamma_{21}/2 +\gamma_{31}/2+ \Gamma_{41})\rho_{14} \\
				\dot{\rho}_{22} &= - i/\hbar [H,\rho]_{22} - \gamma_{21}\rho_{22} + \gamma_{32}\rho_{33} + \gamma_{42}\rho_{44} \\
				\dot{\rho}_{23} &= - i/\hbar [H,\rho]_{23} - (\gamma_{31}/2+\gamma_{32}/2+\gamma_{21}/2+\gamma_{42}/2+ \Gamma_{32})\rho_{23} \\
				\dot{\rho}_{24} &= - i/\hbar [H,\rho]_{24} - (\gamma_{41}/2+\gamma_{42}/2 +\gamma_{21}/2+\gamma_{23}/2  +\Gamma_{42})\rho_{24} \\
				\dot{\rho}_{33} &= - i/\hbar [H,\rho]_{33} - (\gamma_{31}+\gamma_{32})\rho_{33}\\
				\dot{\rho}_{34} &= - i/\hbar [H,\rho]_{34} -\Gamma_{43} \rho_{34} \\
				\dot{\rho}_{44} &= - i/\hbar [H,\rho]_{44} - (\gamma_{41}+\gamma_{42})\rho_{44}.
			\end{aligned}
		\end{equation}
		
		\subsubsection{The target molecules distribution}
		
		We consider the target molecules as a cluster of molecules with its center located a large distance away from the source and its density following the Gaussian distribution. We introduce a multi-layer model to analyze the propagation and scattering of the pump, Stokes, probe and  anti-Stokes pulses through this spatial distribution of molecules. The model mimics the distribution of a cloud of molecules in the air and allows us to solve the propagation and scattering tasks in an elegant and simple way. In this model, each layer is characterized by the fractional number density $\eta$ and a distance to it adjacent layer $(\Delta z)_\eta$. The distance between the layers changes according to the Gaussian distribution of molecules. If $N_s$ is the number of the target molecules and $N$ is the number of total molecules associated with the layer, the fractional number density of that layer is defined as $\eta = N_s/N$. Suppose all target molecules in the central layer are arranged vertically next to each other with no background molecules between them, then the area occupied by these molecules is $S$ = $\pi (d/2)^2 N_s  $ giving $N_s = 4S/\pi d^2$, where $d$ is an approximate diameter of the target molecule. 
		If  $(\Delta z)_\eta$ is the width of this layer, the total number of molecules $N$ is $(S(\Delta z)_\eta/V_0)N_A$, where $V_0$ is the molar volume and $N_A$ is the Avogadro's number. This gives
		\begin{equation}\label{dist}
			\eta = \dfrac{N_s}{N} = \dfrac{\dfrac{4S}{\pi d^2}}{(\dfrac{S(\Delta z)_\eta}{V_0})N_A} = \dfrac{4V_0}{\pi d^2 (\Delta z)_\eta N_A}. 
		\end{equation}
		We consider $N_s$ be constant within each layer. Now, we take $N=N_s$ for the central layer and calculate its width. For any subsequent layer the total number of molecules is different. 
		Given $N_s$, the increase in the layer width by $\Delta z_\eta$ increases the layers volume and, thus, decreases the target's density by the factor $(1+ \Delta z_\eta /\Delta z_0)$.
		The width of each sequential layer is calculated  using Eq.(\ref{dist}).
		Consider that the density changes as per the Gaussian distribution function having the full width at the half maximum (FWHM) $\sigma$ and its maximum value at the center $z_0$ of the cluster of molecules as
		
		\begin{equation}\label{gauss}
			\eta = \frac{N_sV_0}{SN_A\sqrt{2\pi } \sigma}e^{-(z-z_0)^2/(2\sigma^2)}. 
		\end{equation}
		
		The maximum density $\eta_{0}$ of the central layer is found by substituting $z=z_0$ in Eq.\eqref{gauss}. This value of $\eta$ is then substituted in the Eq.\eqref{dist} to find the width of the central layer $(\Delta z)_\eta = (\Delta z)_0$. Once we find the width of the central layer, the $\eta$ of the adjacent layer is found by substituting the new value of $z$, $z_0 + (\Delta z)_\eta$, in Eq.\eqref{gauss}. This process is repeated to find the entire density distribution of the cluster of molecules. The distance between scattering layers $(\Delta z)_\eta$ increases towards both ends of the distribution. So we converted the three dimensional cluster of molecules into a set of two-dimensional layers of molecules. Fig.(\ref{layers}) shows a set of layers, the distance between them and the density associated with each layer. In numerical calculations, we consider  $\sigma=0.2 m$ with its center 1 km away from the source, which together with $\eta_{0}$ determines the total number of layers to be equal to 199.

		\subsubsection{Propagation through the atmosphere}
		
		For a completeness of the picture, taking into account the effects of atmosphere as pulses propagate through the molecular distribution is needed. The propagation of femtosecond pulses through the atmosphere under various air conditions has been broadly investigated, e.g. \cite{femto1,femto2}. Various effects during the propagation including the dispersion and the nonlinear self-focusing are not within the scope of this research. We use Beer's law under the ideal conditions to account for the change in the amplitude of the pulses as they propagate through the atmosphere \cite{Beer}. Assuming there is no turbulence and the air is homogeneous, the intensity of the pulse trains attenuate exponentially due to scattering and absorption as they propagate. The intensity $I$ as a function of the distance $z$ can be written as  $I(z)=I_0 e^{-\beta_e z}$, where $\beta_e$ is the extinction coefficient that contains factors of both scattering and absorption. We use the clear air atmospheric coefficient of $0.55$ $ km^{-1}$ in numerical calculations \cite{ext-coeff} as shown later.

		\subsection{Numerical Results}
		
		Numerical analyses of the effects of the pulse shaping on the optimization of quantum coherence and mitigation of decoherence in the target molecules as well as the impact of multiple scattering from the target molecules are performed using the methanol molecule and addressing the  Raman active symmetric mode having frequency 2837 $cm^{-1}$ (85.05 THz)  \cite{Ma06}. This mode is chosen as a frequency unit $[\omega_{21}]$. The control scheme provides the selectivity of excitation of Raman active modes with the resolution up to $1/\tau$, where $\tau$ is a chirped pulse duration, which is about 2 to 3 $cm^{-1}$. Thus, the asymmetric stretch mode having frequency 2942 $cm^{-1}$ (88.20 THz) is not excited by the control scheme. The selectivity of excitation is not preserved when a broadband but transform-limited pulse trains are applied.

		First we present the results of investigation of the dependence of the population and coherence on the peak Rabi frequency of the control pulses and reveal adiabatic type of solution leading to the maximum vibrational coherence. Then we analyze the four-level system dynamics subject to the interaction with the control pulse trains in the presence of decoherence and demonstrate a sustainable value of vibrational coherence. Finally, we show the solution of the Maxwell - Liouville von Neumann equations for the control pulse trains interacting with an ensemble of methanol molecules  illustrating growth of the vibrational coherence and the anti-Stokes component of the propagating fields. Where appropriate, we compare the results with those for the transform-limited pulse trains interaction with the symmetric stretch mode in the CARS configuration. 
		
		\subsubsection{Analysis of the state populations and coherence} 
		
\begin{figure}
			\centering
			{\includegraphics[scale=0.45]{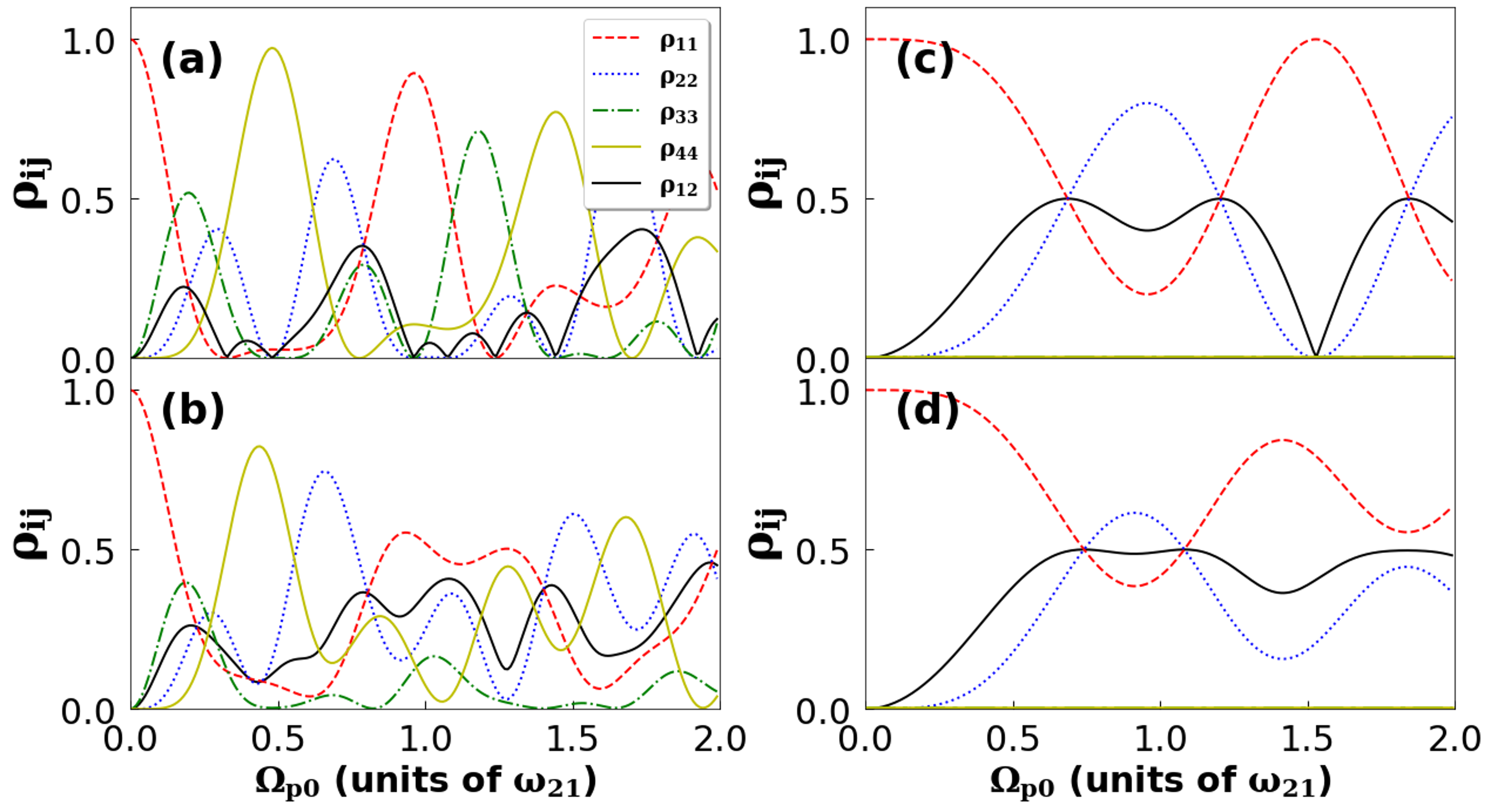}}
			\caption{The population and coherence in the four-level system as a function of the peak Rabi frequency $\Omega_p [\omega_{21}]$, which is the same for the pump, Stokes and probe pulses, $\omega_{21} = 85$ THz. Parameters used in the calculations are   $\tau_0=4.66 [\omega_{21}^{-1}], \Gamma= \gamma=0$. In  (a) the transform-limited pump, Stokes and probe pulses with zero one-photon detuning are applied, $\Delta_s=\Delta_{as}=\Delta=0$; (b) the control pump, Stokes and probe pulses with zero one-photon detuning are applied $\alpha_s^\prime/\tau_0^2=-1.0, \Delta=0$; (c) the transform-limited pulses with non-zero one-photon detuning are applied, $ \Delta=1.0 [\omega_{21}]$; (d) Control pulses with non-zero one-photon detuning are applied, $\alpha_s^\prime/\tau_0^2=-1.0, \Delta=1.0 [\omega_{21}]$. Once coherence is built by the control pulses, it never drops to zero, in contrast to the transform-limited pulses solution. The detuned control scenario is robust for applications in CARS microscopy and spectroscopy because it provides coherence resilient to fluctuations in the intensity of the Raman fields.}
			\label{Delta}
		\end{figure}

		Fig.(\ref{Delta}(a)-(d)) shows the dependence of the populations and coherence as a function of the peak Rabi frequency for the case of the transform-limited pump, Stokes and probe pulses with zero and non-zero one-photon detuning (a),(c),  and control pulses with zero and non-zero one-photon detuning (b),(d) respectively. The envelope of the Rabi frequency is the same for all three transform-limited pulses, which are also used as an initial condition for chirping in the control scheme. The values of the Rabi frequency on the abscissa are presented for the transform-limited pulse. Decoherence is not taken into account to get a clear picture of the dependence of the state population and coherence on the Rabi frequencies. Under the one-photon resonance condition shown for the transform-limited pulses in (a) and for the chirped pulses in (b), the population of the excited states is significant, which prevents from achieving half population each in the ground state $|1\rangle$ and the excited state $|2\rangle$. In the transform-limited pulse scenario in (a), coherence periodically becomes zero, which is not the case for the control pulses solution shown in (b). Such a behavior in (a) is due to the pulse area type of solution, when the probability amplitude of the states depends on the pulse area with $\pi$ value leading to the population inversion and $2 \pi$ - to the population return. In contrast, the control pulse scheme provides adiabatic type of response in the four-level system with nonzero value of coherence, which depends on the strength of the fields as shown in (b). The one-photon detuning $\Delta_s=\Delta_{as}=\Delta=1.0 [\omega_{21}]$ minimizes the transitional population of the excited states $|3\rangle$ and $|4\rangle$ for both transform-limited and the control pulse scenario shown in (c) and (d) respectively. The one-photon detuning shifts the point of first zero coherence toward higher values of the Rabi frequencies in the transform-limited case in (c). In the control case in (d), the first point of equal population giving the maximum vibrational coherence occurs at the peak Rabi frequency $\Omega_{p0}=0.75 [\omega_{21}]$ and is achieved due to two-photon adiabatic passage with a negligible involvement of the excited state manifold into population dynamics. Beyond this point, coherence value varies within the range from 0.5 to 0.35 for the peak Rabi frequency $\Omega_{p0}=1 [\omega_{21}]$ and higher. Once coherence is built, it never drops to zero, in contrast to the transform-limited pulses solution. 
		Thus, the detuned chirped pulse control scheme is more robust for the applications in CARS microscopy and spectroscopy because it provides one with a sustainable value of coherence resilient to fluctuations in the intensity of the Raman fields.
		
		\begin{figure}
			\centering
			{\includegraphics[scale=0.25]{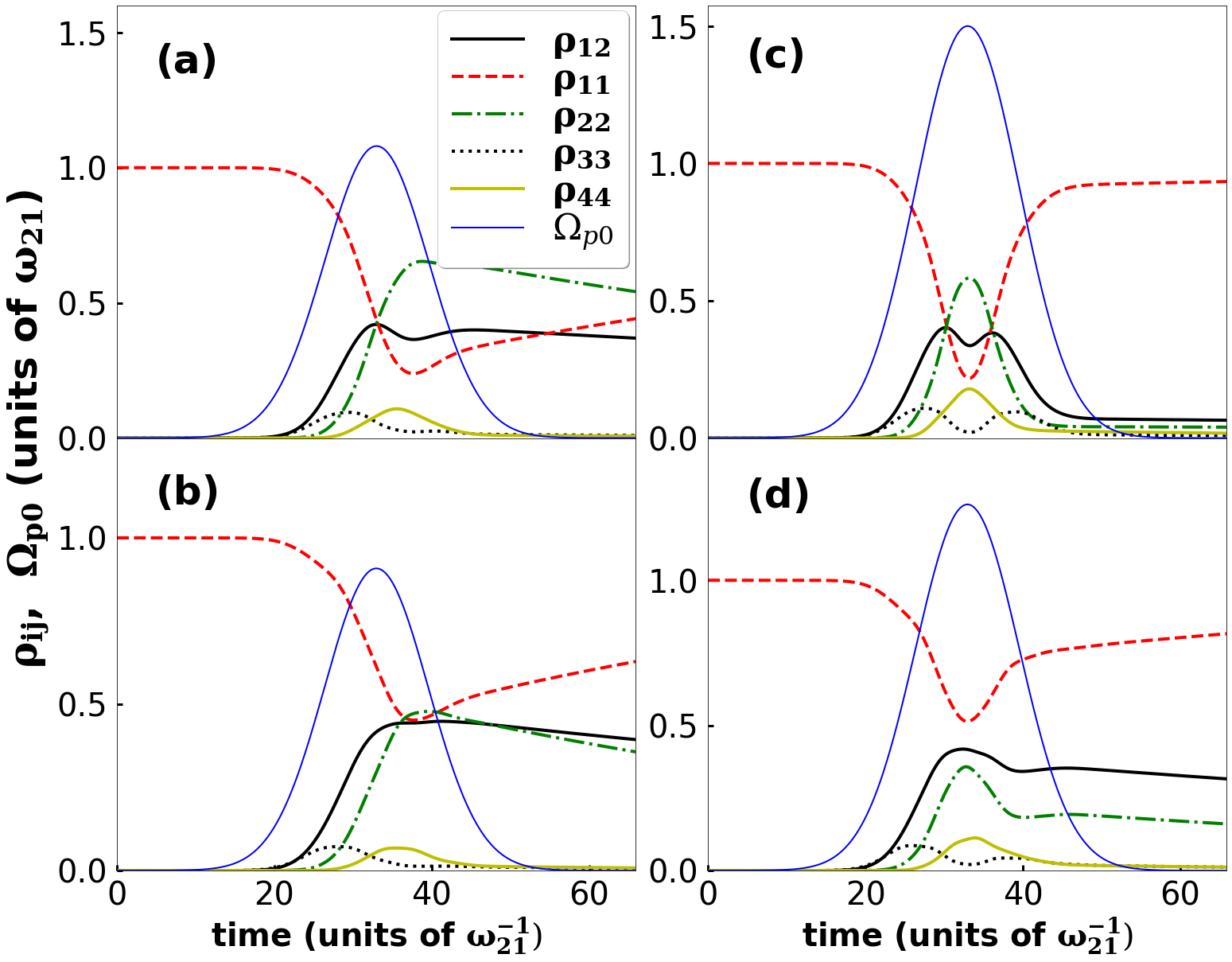}}
			\caption{Dynamics of the population of four states $\rho_{11}$ (dashed red), $\rho_{22}$ (dash-dotted green), $\rho_{33}$ (dotted black), $\rho_{44}$ (solid yellow) and coherence $\rho_{21}$ (solid black) in the four-level system interacting with the transform-limited pump, Stokes and probe pulses in  (a),(c); and  the control pulses in (b),(d), $\alpha_s^\prime/\tau_0^2=-1.0$ for the peak Rabi frequency of the pump, the Stokes and the probe pulses, (before chirping for the control scheme), $\Omega_p=1.08$ $[\omega_{21}]$ in (a),(b), and 1.5 $[\omega_{21}]$ in (c),(d). Other parameters are $\tau_0=4.66$ $ [\omega_{21}^{-1}],$ all $\gamma_{ij}=\gamma=1.176 \times 10^{-2} [\omega_{21}], \Gamma=0, \Delta=1.0 [\omega_{21}]$.}
			\label{Fig1}
		\end{figure}
		
		To demonstrate adiabatic passage generated under the condition of nonzero one-photon detuning, a time-dependent picture is presented in Fig.(\ref{Fig1}(a-d)). The time dependence of the population and coherence in the four level system interacting with the transform-limited pump, Stokes and probe pulses, (a),(c), and with the control pulses, (b),(d) shows population dynamics and coherence for two values of the peak Rabi frequency $\Omega_p=1.08$ and $1.5 [\omega_{21}]$. The value of the Rabi frequency $\Omega_{p0}=1.08 [\omega_{21}]$  is chosen according to the Fig.(\ref{Delta}(d)), which generates the second equal population between the ground state $|1\rangle$ and the excited state $|2 \rangle$ and the maximum coherence $\rho_{21}$ in the control pulses scenario. It leads to adiabatic population transfer from the ground state $|1\rangle$ to the excited state $|2\rangle$. Meanwhile, the value of the peak Rabi frequency $\Omega_{p0}=1.5 [\omega_{21}]$ is chosen because  it gives the first zero coherence for the transform-limited pulse scenario in Fig.(\ref{Delta}(c)), which is not the case for the control scheme in Fig.(\ref{Delta}(d)). Parameter $\gamma$ is non-zero in order to see how spontaneous decay impacts state dynamics for the chosen representative values of the Rabi frequency. The time dependence of the populations and a significant coherence is still observed in (d) demonstrating benefits of the control scheme. 
		
		\subsubsection{Analysis of the system dynamics subject to the interaction with the control pulse trains in the presence of decoherence}	
		
\begin{figure}
			\centering
			{\includegraphics[scale=0.23]{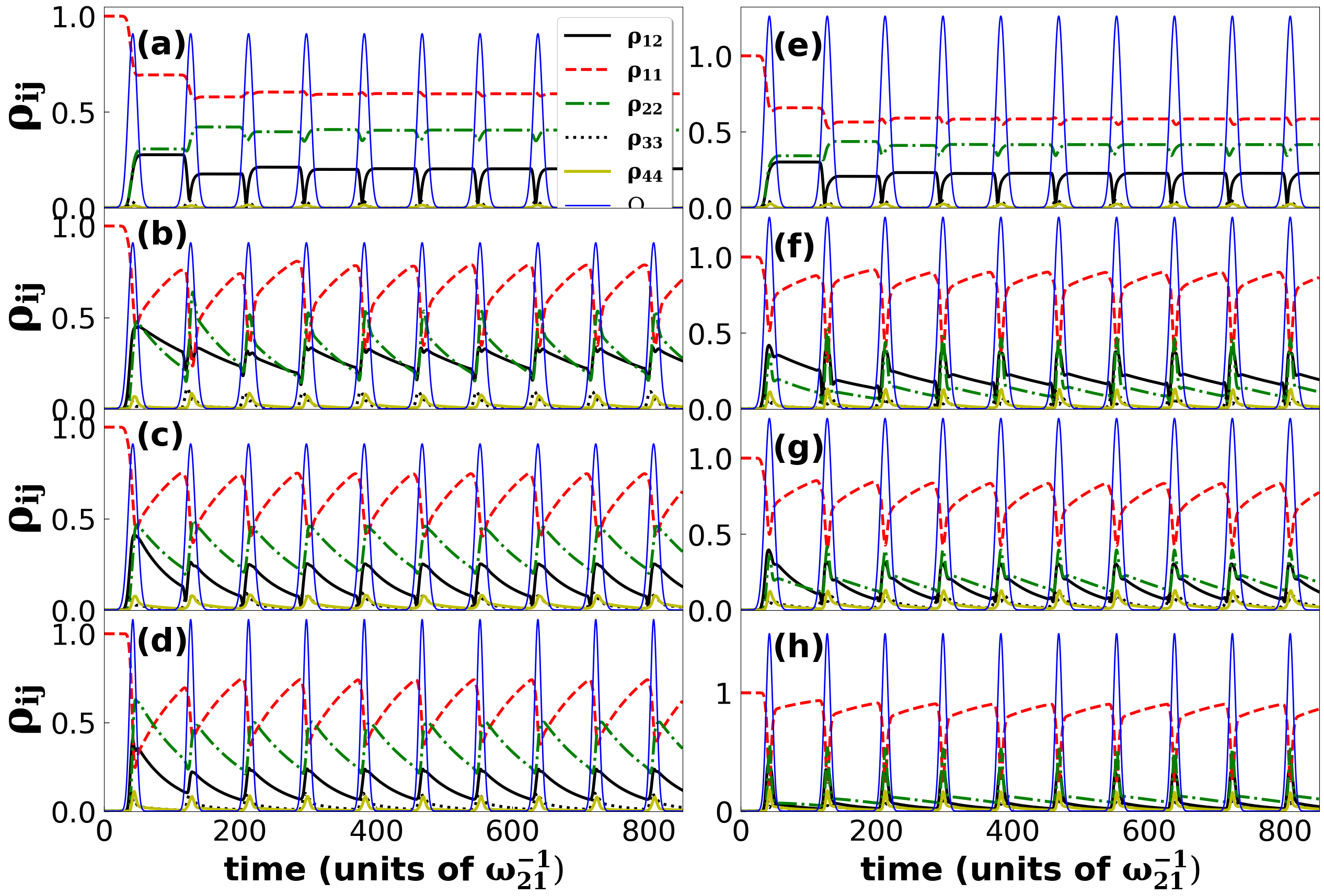}}
			\caption{Dynamics of the population of four states  $\rho_{11}$ (dashed red), $\rho_{22}$ (dash-dotted green), $\rho_{33}$ (dotted black), $\rho_{44}$ (solid yellow) and coherence $\rho_{21}$ (solid black) in the four-level system interacting with the control pulse trains having a repetition rate of 1 THz and peak Rabi frequency in (a-d) equal to $\Omega_{p0}=1.08 [\omega_{21}]$, and in (e-h) equal to $\Omega_{p0}=1.5 [\omega_{21}]$. In (a),(e)  $\gamma_{4i}=\gamma_{3i}=\Gamma_{4i}=\Gamma_{3i}=10^{14} Hz$, i=1,2, with no vibrational relaxation, $\gamma_{21}=\Gamma_{21}=0$; in (b),(f)  $\gamma_{4i}=\gamma_{3i}=\gamma_{21}=10^{12} Hz; \Gamma_{4i}=\Gamma_{3i}=\Gamma_{21}=0;$ in (c),(g) $\gamma_{4i}=\gamma_{3i}=\Gamma_{4i}=\Gamma_{3i}=\gamma_{21}=\Gamma_{21}=10^{12} Hz;$ and (d),(h) the four-level system interacting with the transform-limited pulse trains and $\gamma_{4i}=\gamma_{3i}=\Gamma_{4i}=\Gamma_{3i}=\gamma_{21}=\Gamma_{21}=10^{12} Hz$. The rest field parameters are $\tau_0=4.66 [\omega_{21}^{-1}], 
				\Delta_s=\Delta_{as}=1.0 [\omega_{21}]$ and $\alpha_s^\prime/\tau_0^2=-1.0$ for the control pulse scenario. }
			\label{decoh}
		\end{figure}

		We analyze the impact of decoherence in the four-level system through its interaction with the control pump, Stokes and probe pulse trains each consisting of ten pulses in Fig.(\ref{decoh}). The results in (a-d) are given for the peak Rabi frequency $\Omega_{p0}=1.08 [\omega_{21}]$, and the results in (e-h) for $\Omega_{p0}=1.5 [\omega_{21}]$. The value $\Omega_{p0}=1.08 [\omega_{21}]$ provides the maximum coherence (1/2) for the control pulse and high value of coherence (0.45) for the transform-limited pulse according to Fig.(\ref{Delta}c,d), and the  $\Omega_{p0}=1.5 [\omega_{21}]$ gives a contrast value of coherence for the control and the transform-limited pulse application, 0.39 and 0.07 respectively. 
		We analyze the controllability and sustenance of vibrational coherence in the four-level system subject to a fast spontaneous decay and collisions  ($\sim 10 fs$); then we investigate the impact of vibrational relaxation considering the decay on the order of $1 ps$ and demonstrate how the loss of coherence due to this process may be mitigated by periodically restoring population of the excited vibrational state $|2\rangle$ of the ground electronic state; and then we compare this result to the case when collisional dephasing is on the same order of magnitude ($\sim 1 ps$).
		
		Fast spontaneous decay and collisional dephasing rates $(10^{14} Hz)$ of the transitional excited states $|3\rangle$ and $|4\rangle$  impact population dynamics and coherence even though these states are negligibly populated, shown in Fig.(\ref{decoh}(a),(e)). Here populations and coherence $\rho_{21}$ are presented as a function of time for $\gamma_{4i}=\gamma_{3i}=\Gamma_{4i}=\Gamma_{3i}=10^{14} Hz$, i=1,2. Population of states $|1\rangle \approx 0.6$ and $|2\rangle \approx 0.4$ is stable between pulses, but, even though $|3 \rangle$ and $|4\rangle$ states are negligibly populated owing to the control scheme applied, their fast decoherence while pulse is on (chirped pulse duration is $55 fs$) impacts populations of states $|2\rangle$ and $|1\rangle$ and coherence $\rho_{21}$ periodically drops to $\sim 0.02$. Between pulses, such a fast relaxation from the excited states leads to a reduced but stable value of coherence $\rho_{21} \sim 0.2$.
		
		Fig.(\ref{decoh}(b),(f)) shows the system dynamics in the presence of the vibrational relaxation of state $|2\rangle$ described by $\gamma_{21} = 10^{12} Hz$. Spontaneous decay from the excited states is also present, $\gamma_{4i}=\gamma_{3i}=\gamma_{21}=10^{12} Hz; \Gamma_{4i}=\Gamma_{3i}=\Gamma_{21}=0.$ Figure demonstrates that coherence $\rho_{21}$ is periodically built up by the chirped pulses, and then insignificantly reduces its value between the pulses in the trains. Spontaneous decay rate $\gamma=1THz$ from the excited state $|4\rangle$ to $|3\rangle$  does not make any contribution to the population dynamics and was neglected. 
		However,  because the pulse train period is chosen to match the decay time $T = 1/\gamma_{21} = 1ps,$ (and no collisional dephasing, $\Gamma_{ij}=0$), the population of state $|2\rangle$ decreased due to spontaneous decay is periodically restored by control fields providing a sustainable value of coherence. When vibrational relaxation is much faster (e.g., $10^{14} Hz$) than the pulse repetition rate ($10^{12} Hz$),  coherence $\rho_{21}$ becomes negligibly small between pulses (not shown here). 
		Switching on collisional dephasing such that $\Gamma_{21}=\gamma_{21}=1 THz$ results in a more dramatical reduction of coherence $\rho_{21}$ as it is shown in Fig.(\ref{decoh}(c),(g)) because collisional dephasing cannot be mitigated by this mechanism being represented by off-diagonal density matrix elements. However, the resultant coherence $\rho_{21}$ does not drop to zero between pulses. This is due to the choice of the pulse repetition rate as well as the control scheme leading to a negligible population of the excited states $|3\rangle$ and $|4\rangle$ in the dynamics. In contrast, the simultaneous application of the transform-limited pump, Stokes and probe pulse trains shown in Fig.(\ref{decoh}(d),(h))  results in strong dependence of coherence on the peak Rabi frequency  in accordance with the pulse area solution. The simultaneous application of the transform-limited pulses in this calculation aims to compare with the chirped pulses scenario. (Note, that within a different control scheme, e.g., F-STIRAP \cite{Shore}, which imposes a time delay between the Stokes and the pump pulses, the transform-limited pulses generate the maximum coherence.) The results of calculations presented in Fig.(\ref{decoh}) for various values of the Rabi frequency of the control pulses and the transform-limited pulses led to a conclusion that for the control scheme there is vibrational coherence in the system for any value of the peak Rabi frequency within the adiabatic range, while for the related transform-limited pulse scenario this is not the case.
		
		\subsubsection{Impact of Beer's law on the average intensity}
		
		\begin{figure}
			\includegraphics[scale=0.4]{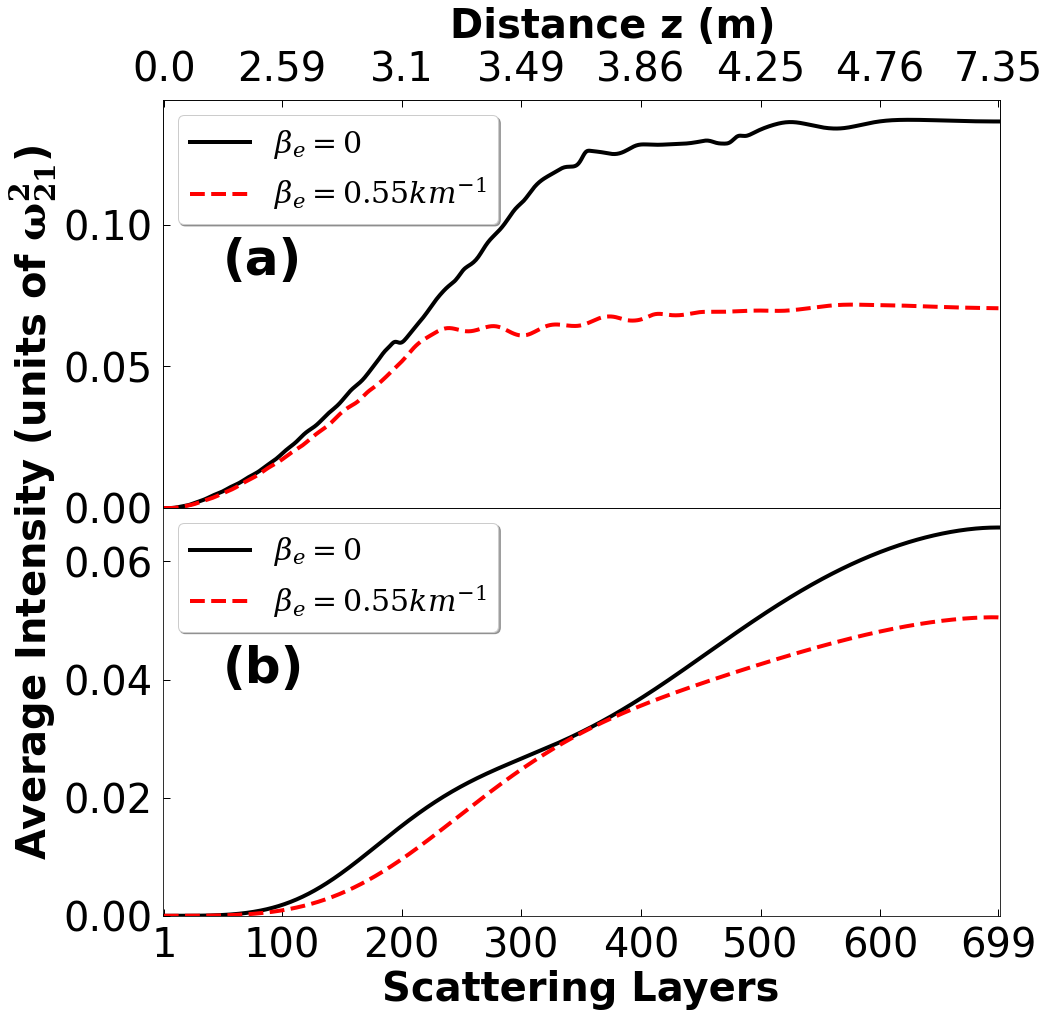}
			\caption{An average intensity of the anti-Stokes pulses as a function of the number of scattering layers. The pulses are calculated by modeling the propagation of a transform limited pulse train containing 10 pulses using Beer's law. The black solid curve represents the change in the average intensity as pulses undergo scattering through layers for the case of $\beta_e=0$ (without taking air into consideration), and red dashed curves shows the intensity for $\beta_e=0.55$  $ km^{-1}$. The one-photon detuning is $\Delta = 1 [\omega_{21}]$ in (a), and $\Delta = 10 [\omega_{21}]$ in (b). The width of the target molecules distribution is $\sigma= 1 m$. The depreciation of intensity is due to scattering and absorption in the air.} \label{anti-stokes-air}
		\end{figure}
		
		We apply Beer's law under the ideal conditions to evaluate the change in the amplitude of the anti-Stokes signal as pulses propagate through the atmosphere. I apply ten transform-limited pulses in the pulse train. Numerical analysis shows that the amplitude of the pump,  Stokes and probe pulse trains is reduced upon propagation, while the average intensity of the anti-Stokes pulse trains is amplified as shown in Fig.(\ref{anti-stokes-air}) for propagation through 699 layers for both cases, with and without impact from the air. The intensity of the anti-Stokes pulse trains in the presence of the air is depreciated due to the scattering and absorption effects.
		
		\subsubsection{Analysis of the Maxwell - Liouville von Neumann equations and demonstration of the anti-Stokes signal generation}
		
		Using Maxwell's equations Eqs.(\ref{MMM}) coupled to the Liouville von Neumann equations Eqs. (\ref{lvn_gen}) we numerically analyzed the propagation effects of the control pump, Stokes, probe and the generated anti-Stokes fields scattered from the target molecules and observed the amplification of the anti-Stokes component. A detailed description of this analysis with chirped pulses, where a deep learning technique is implemented, is included in the next section.
		
		We also analyzed propagation effects using the transform-limited pump, Stokes, and probe pulse trains having the peak Rabi frequency $\Omega_{p(s,pr)} = 85 THz=\omega_{21}$ and been largely detuned from the one-photon transitions, the detuning is $\Delta_s=\Delta_{as}=\Delta= 10 \omega_{21}= 850 THz$ for the adiabatic regime. We consider 10 pulses in the pulse train having period $T = 1 ps$. 	The increase of the peak value of the anti-Stokes Rabi frequency $\Omega_{as}(t)$  by two orders of magnitude is observed 1 meter (199 layers) away from the peak molecular density. Coherence is increasing from pulse to pulse and the population is adiabatically transferred from the ground state $|1\rangle$ to the excited state $|2\rangle$ in the four-level system during the interaction with four fields in the CARS configuration. Here adiabatic regime is achieved due to a large one-photon detuning $\Delta= 10 \omega_{21}$ and the choice of the peak Rabi frequency $\Omega_{p(s,pr)}=\omega_{21}$, which result in a negligible population of the transitional states $|3\rangle$ and $|4\rangle$.
		
		\begin{figure}
			\centering
			{\includegraphics[scale=0.25]{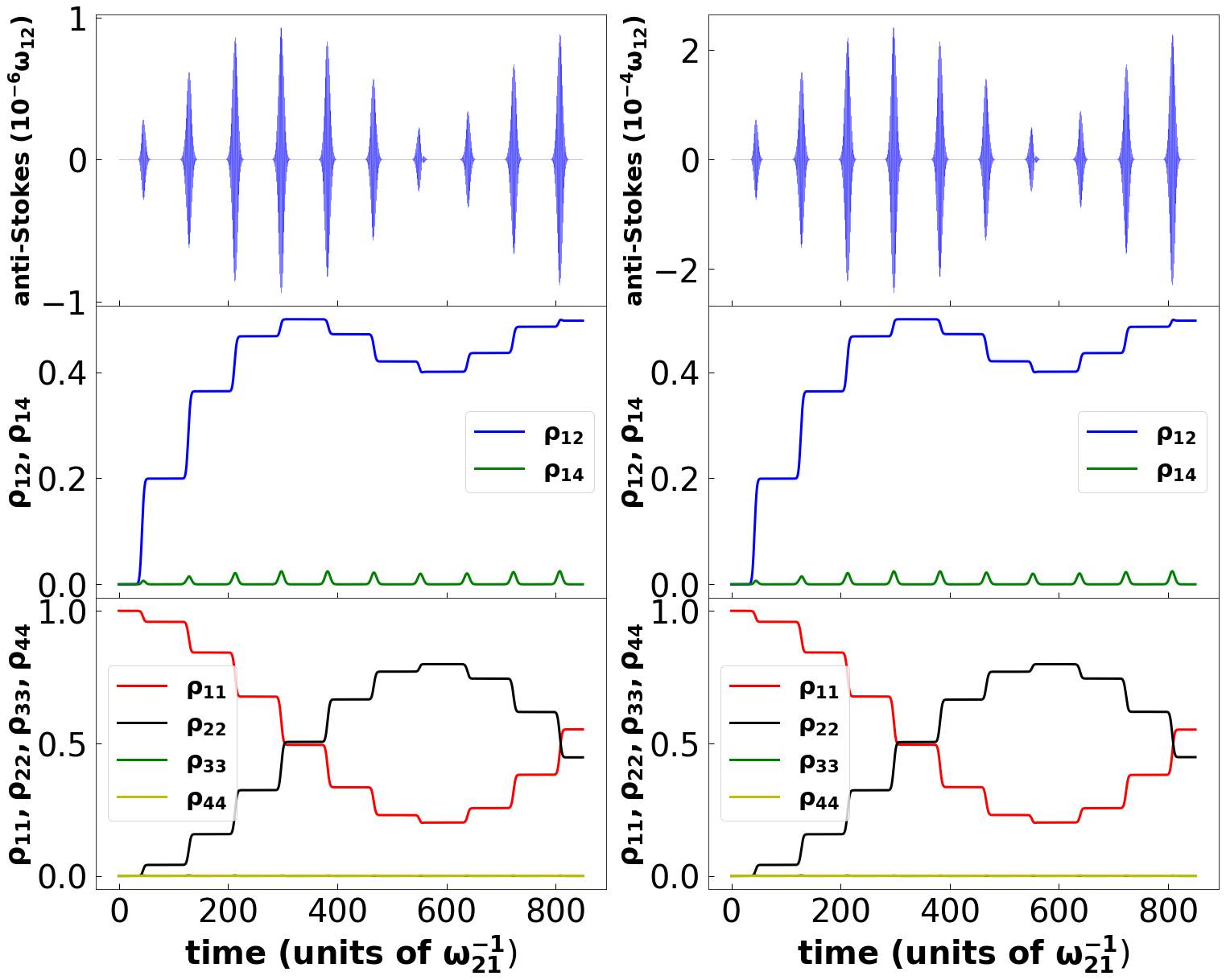}}
			\caption{Scattering dynamics using the transform-limited pump, Stokes, and probe pulse trains with the peak Rabi frequency equal to the frequency between states $|1\rangle$ and $|2\rangle$, $\Omega_{p(s,pr)_0} = \omega_{21}$, and been largely detuned from the one-photon transitions, the detuning  is $\Delta_s=\Delta_{as}=\Delta= 10 \omega_{21}= 850 THz$ for the adiabatic regime. There are 10 pulses in each pulse train. The first column shows ten anti-Stokes pulses (top), the state  coherence (middle) and populations (bottom) after the first scattering event; the second column shows the same after the 199th scattering event. Parameters $\sigma = 0.2 m$; 199 layers provide a distance of  1 m away from the peak molecular density; $\tau_0 =54.8 fs$; $T=1 ps$.  
			}\label{scattering}
		\end{figure}
		
		From the results above it follows that the implementation of the control pulse trains in the four-wave mixing of CARS is more robust for the  generation of a sustainable anti-Stokes backscattered signal compared to the use of a set of transform-limited pulses. This is due to the adiabatic regime of light-matter interaction which preserves vibrational coherence and facilitates a build-up of the anti-Stokes signal. For the case of the phase-matching conditions relaxed, given the size of the molecules is less than the wavelength of the incident fields, a collinear copropagating configuration of CARS may be created using the methanol as a surrogate target. Because the anti-Stokes radiation is generated as a result of the stimulated Raman scattering process, it is highly directional and is built up in the forward and the backward directions dominantly \cite{scully-cars,liu-detection}. Therefore, the backscattered anti-Stokes signal will reach a detector near the laser source. The following parameters of the fields may be used in an experiment: the pulse duration of order $ 100 fs$, the peak field amplitude of $E_{0p(s,pr)} \sim 1.6 \times 10^9 V/m$; the control pulse chirps obeying the relationship $\alpha_s=-\alpha_p$, and  $\alpha_{pr}= \alpha_s - \alpha_p$ for the first half of the pulse duration $t \le t_c$, and  $\alpha_s=\alpha_p$, $\alpha_{pr}= 0$ for $t > t_c$; the value of $\alpha_s=-7 THz/fs$, the pulse train period of order of spontaneous decay time and the one-photon detuning of order $\Delta \sim 1/fs$.
		
		\subsection{Section Summary}
		
		We presented a semiclassical theory of the four-wave mixing process in the coherent anti-Stokes Raman scattering implementing the control pulse trains. The theory is based on a set of Maxwell's equations for propagation of the pump, the Stokes, the probe and the anti-Stokes components of the fields coupled to the Liouville von Neumann equations with relaxation for dynamics in the target molecules. It is intended for the investigations of the remote detection of biochemical molecules. The multi-layer model is developed to account for the spatial distribution of the target molecules in the air mimicking the environmental conditions. The machine learning approach is developed to analyze the evolving phase of the pulse trains as they undergo scattering within each layer. The approach makes use of the deep Convolutional Neural Networks (discussed in the next section). The quantum control method for the incident pulse shaping is implemented, which optimizes the macroscopic induced polarization in the target molecules by maximizing vibrational coherence. The method implies chirping of the incident pulse trains, which induce adiabatic population transfer within four states in the CARS scheme leading to a sustainable, high vibrational coherence. Importantly, the transitional excited states get negligibly populated, thus minimizing the impact of spontaneous decay and associated losses of coherence from these states. Moreover, the choice of the pulse train period to match the spontaneous decay time permits for mitigation of the vibrational decay.  The enhancement of the anti-Stokes field is observed upon propagation through the ensemble of the target molecules, achieved by the control pulse trains as well as by the transform-limited pulse trains with a large detuning and a carefully chosen Rabi frequency. The coherent enhancement of  the anti-Stokes signal and mitigation of decoherence by chirped control fields form a foundation for the propagation of the anti-Stokes signal through distances on a kilometer scale.

		\section{DEEP NEURAL NETWORKS APPLICATIONS IN QUANTUM CONTROL}
		
		
		In the previous section, we developed a control scheme that helps us optimize the signal from target molecules by maximizing the vibrational coherence. In order to apply this scheme effectively and to investigate the controllability of population dynamics and vibrational coherence in the target molecules by propagating electromagnetic fields, we need to know the key fields' parameters evolution after each scattering event. This allows us to accurately calculate the quantum coherence and the induced polarization at the sequential steps of numerical calculation. 
		In the case of using the chirped pulse control scheme within the multi-layer model of molecule distribution, the Maxwell - Liouville von Neumann equations alter the initial, pre-determined phase of the incident pulses impacting the response of the target molecules. Thus, extracting the analytical phase from the numerical solutions of Eqs.(\ref{MMM}) and verifying that the pre-determined chirping scheme is applied to each scattering event becomes an extremely important task for evaluating the response from the quantum system. To accomplish this goal, we need to develop a mechanism to extract the chirp parameters from the scattered pulses. To this end, we created a generic machine learning model that classifies a given pulse into one of the three categories based the kind of phase it has and do the regression analysis to reveal the phase of pulse. In this section, we first present this deep learning technique and then apply it to the formalism we developed in the last section to simulate the output signal using chirped pulses. 
		
		\subsection{Deep learning and applications}
		
		Although the idea of artificial intelligence has decades of history, it has picked up momentum recently with the development of machine learning and deep learning techniques along with the advancement in computational power. Deep learning is rapidly transforming almost all the industries. It helps reduce human intervention and scale up the speed in solving complex problems. 
		
		Deep learning, in general terms, can be thought of as training a computer to solve a certain type of problem by feeding enormous data. At its core, it is the optimization of numerous parameter values of a mathematical function to fit the training data. The idea of applying deep learning into quantum control techniques is novel. A deep neural network consists of several layers each having certain number of layers.
		
		We developed a deep neural network for classifying different kinds of pulses from the numerical data, based on their chirping and extracting the chirp parameters from these classified pulses using a machine learning technique \cite{MachLearn1,MachLearn2}. This approach of extracting the information about the phase of the pulses from the numerical grid and obtaining an accurate value of the chirp parameters is principally novel and may have a wide range of applications in the quantum control and spectroscopy.
		
		There are different kinds of neural networks, each being used for specific purposes. The machine learning model that we created is the deep Convolutional Neural Network (CNN). A CNN is generally used for analyzing visual imagery. As the plots of pulses with different chirps can be analyzed visually, the CNN one of the best choices in this case as well.
		
		\subsection{Classification and regression of chirped pulses using Convolutional Neural networks (CNN)}
		
		\begin{figure}
			\centering
			{\includegraphics[scale=0.8]{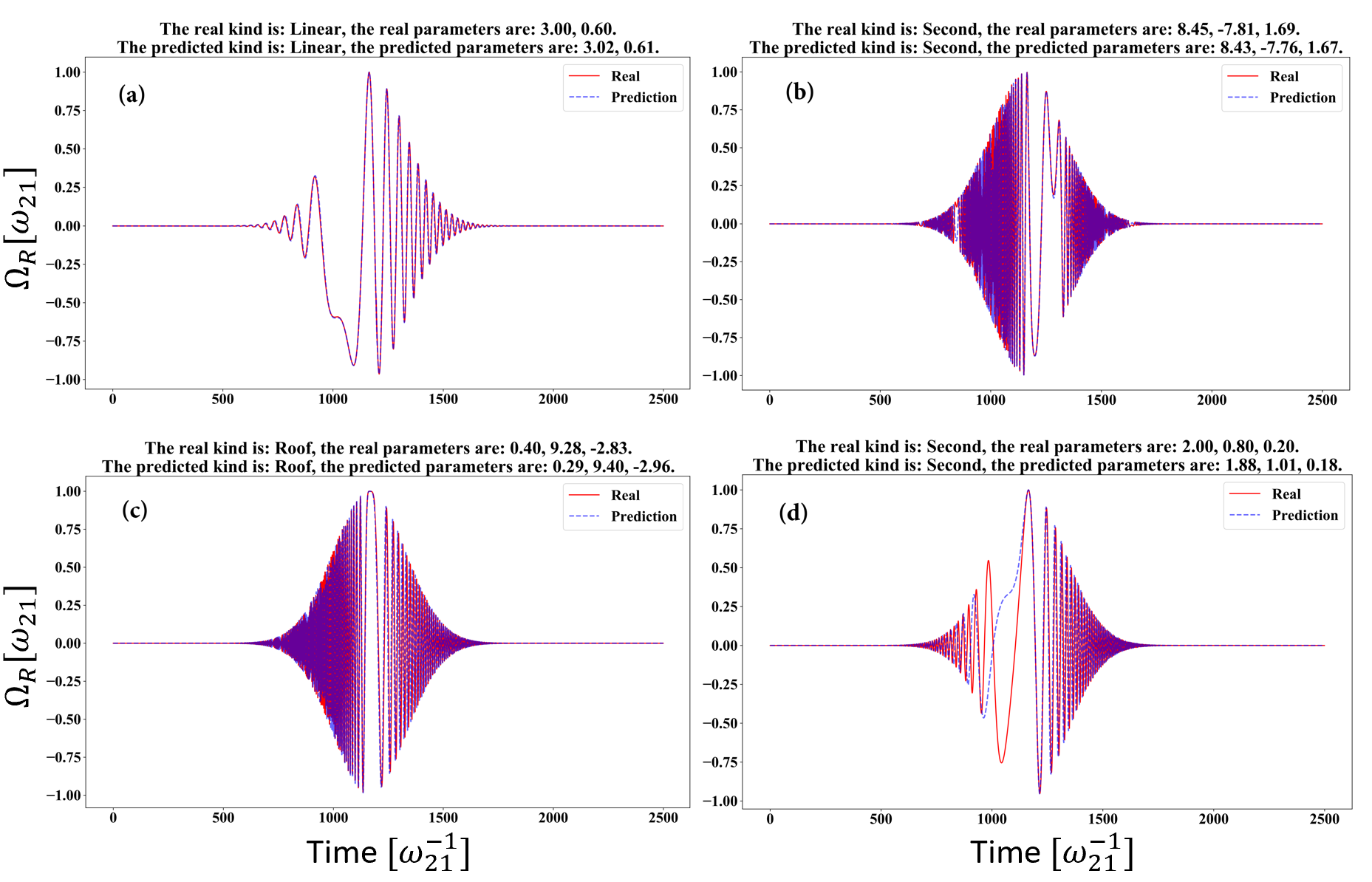}}
			\caption{ Different shapes of the phase of the field obtained numerically (solid line) and using the deep convolution neural network model (dashed line) with different types of the phase of the input pulse: (a) Linear chirp, $\phi(t) = a_1 t + a_2 t^2$; (b) Quadratic dependence of the phase on time having $a_2 < 0$ in $\phi(t) = a_1 t + a_2 t^2 + a_3 t^3$; (c) 'Roof' chirp having positive chirp rate for the first and negative chirp rate for the second part of the pulse \cite{Pandya20}, $\phi(t)= a_1 t + \tilde{a}_2 t^2$ for $t \le 0$, and $\phi(t) = a_1 t  + \tilde{\tilde{a}}_2 t^2$ for $t > 0$; (d) Quadratic dependence of the phase on time having $a_2 > 0$ in $\phi(t) = a_1 t + a_2 t^2 + a_3 t^3.$ The values of parameters are printed in the titles of the pictures. Note that there is no discrepancy in determination of the kind of the phase, only parameters have rare errors.
			} \label{CNN}
		\end{figure}
		
		A CNN is built to classify a given pulse into one of three kinds: linear, quadratic and the chirp shape according to our control scheme $\alpha_s=-\alpha_p$ and $\alpha_{pr}= \alpha_s - \alpha_p$  for  $t \le t_c$; and  $\alpha_s=\alpha_p$ and $\alpha_{pr}= 0 $ for $t > t_c$. Another CNN is built to do the regression work, it calculates the parameters of the fields and shares a similar structure as the classification neural network. The structure of CNN used will be discussed later in the section.

		Of principle importance for studying the phase of the numerical pulses is the availability of training data. Massive training data is a necessary requirement for deep learning training to concur a 
		problem \cite{overfitting}. Since it is difficult to collect thousands of actual data from the experiments, we created a program that generated the scattered laser pulses randomly based on an arbitrary laser pulse model
		\begin{equation}\label{pulse_equ}
			E(t)=E_0 e^{- \frac{t^2}{2\tau^2}} \cos[\omega_L t +M(t)].
		\end{equation}
		Here $\tau$ is a single pulse duration, $E_0$ is the peak value of the field having the Gaussian envelope, and $\omega_L t +M(t)$ is the phase of the field having the modulation $M(t)$, which is the key to quantum control. A different parity of the phase modulation 
		leads to different control scenarios \cite{goswami, gliu}. 
		Here we present $M(t)$ as an expansion in the Taylor series
		\begin{equation}
			M(t) = a_0+a_1t^1+a_2t^2+a_3t^3+...
		\end{equation}
		Since in most cases the higher orders have a very limited contribution, 
		we created data for three kinds of the phase using terms up to the third power in time: 'The Linear', which is determined by two parameters: the carrier frequency ($a_1$) and the linear chirp ($a_2$), then the field phase reads $\phi(t) = a_1 t + a_2 t^2$; 
		'The Second', which is determined by three parameters: the carrier frequency ($a_1$), the linear chirp ($a_2$), and the second order chirp ($a_3$), then the phase reads $\phi(t) = a_1 t + a_2 t^2 + a_3 t^3;$ and 'The Roof', which is comprised of two parts, before central time and after, and is determined by three parameters: the carrier frequency ($a_1$), the linear chirp ($\tilde{a}_2$) for the first half of the pulse and the linear chirp ($\overset{\approx}{a}_2$) for the second half of the pulse, then the  constructed phase of the field reads $\phi(t)= a_1 t + \tilde{a}_2 t^2$ for $t \le 0$, and $\phi(t) = a_1 t  + \overset{\approx}{a}_2 t^2$ for $t > 0$. 
		
		We simulated the pulses with these three kinds of phases using characteristic values of the field parameters and generated training data in quantity of  
		$5 \times 10^4$ for each kind by varying the carrier frequency and the chirp rate. 
		During the training process, we applied the Adam Optimizer algorithm with the learning rate of 0.1, and the regularization of 0.02 \cite{Adam}. The loss function of the classification model is the cross entropy, but the mean squared error for the regression model. The early stop technique was also used to control the overfitting \cite{caruana2001overfitting}.
		The details of the construction of the neural networks for both the classification and the regression models are presented in the next section.
		
		After training the classification and the regression models, they are combined to be used as directed. The classification block classifies the random pulse and sends it to the corresponding regression block to solve for the analytical parameters of one of three kinds of the phase. The classification reaches the accuracy of 97.93\%, and the overall root mean square error of the regression is smaller than 0.1, providing the deep learning model's results accurate enough. Both the classification and regression models are evaluated via a separate test data set, which contains $3 \times 10^3$ samples. To demonstrate high accuracy of the analytical fit to the numerical data of the phase of the field we show several prototypical phases in Fig.(\ref{CNN}).
		
		\subsection{The structure of the CNN used}
		Both the classification and the regression neural networks share the same core structure. Since the numerical pulses, which we generated as the training data, have 2500 time steps, all models have the input shape of $2500\times1$. There are three blocks of the mini-convolutional neural network in the models. The first block contains three 1D convolutional layers with the kernel size of 3. The second block has two layers of the 1D convolutional network with kernel size of 5. The third block has a single 1D convolutional layer of kernel of 7. All the convolutional layers are activated by the Rectified Linear Units Function \cite{Hi10} and the Group Normalization \cite{he}. There is a maximum pooling layer of pool size 4 after each block. There is a linear layer of size 1024 after the output of the convolutional blocks is flattened. 
		
		\begin{figure}\centering
			\includegraphics[width=13cm]{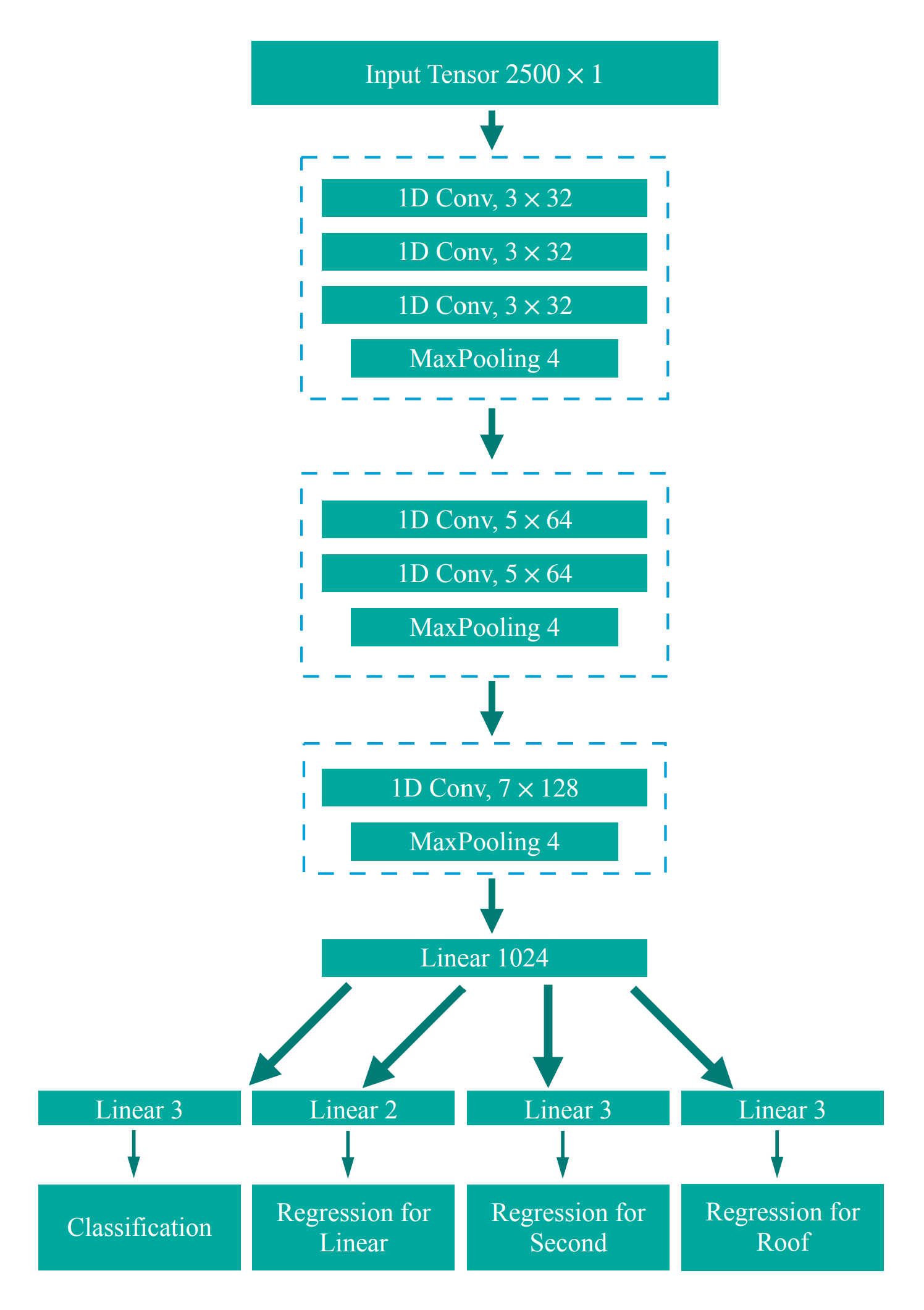}
			\caption{The structure of the Deep Neural Network. The same structure is shared by the phase type classifier and the three phase value regression models, except for the last output layer. Three convolutional blocks are u
				sed sequentially to extract the highly non-linear information from the input time dependent tensor. The linear layer is used after flattening the output from the last convolutional block.}
			\label{LL}
			
		\end{figure}

		The structure of the neural network, shown in Fig.(\ref{LL}), is determined by the validation results, together with the other hyperparameters, such as the learning rate, the choice of optimizer and regularization. We adjust the kernel size, the number of blocks and the number of layers in each block to have the optimal validation result. The 1D convolution layers are used because they are suitable for extracting the information within the sub-region of the whole input tensor. It is a match to our aim, which is to extract the instantaneous value of the analytical parameter from the numerical sequential, time-dependent data. Besides, we use several 1D convolution layers as a block to extract the high dimension information from the input tensor. Three kernels of size of 3 cover the same area of the input tensor as a single kernel of size of 7, but the former catches the higher dimension information than the later one. We didn't set all blocks to three layers of kernel size of 3 because we would like to control the overfitting problem.
		
		\subsection{Results}
		\begin{figure}
			\centering
			{\includegraphics[scale=0.4]{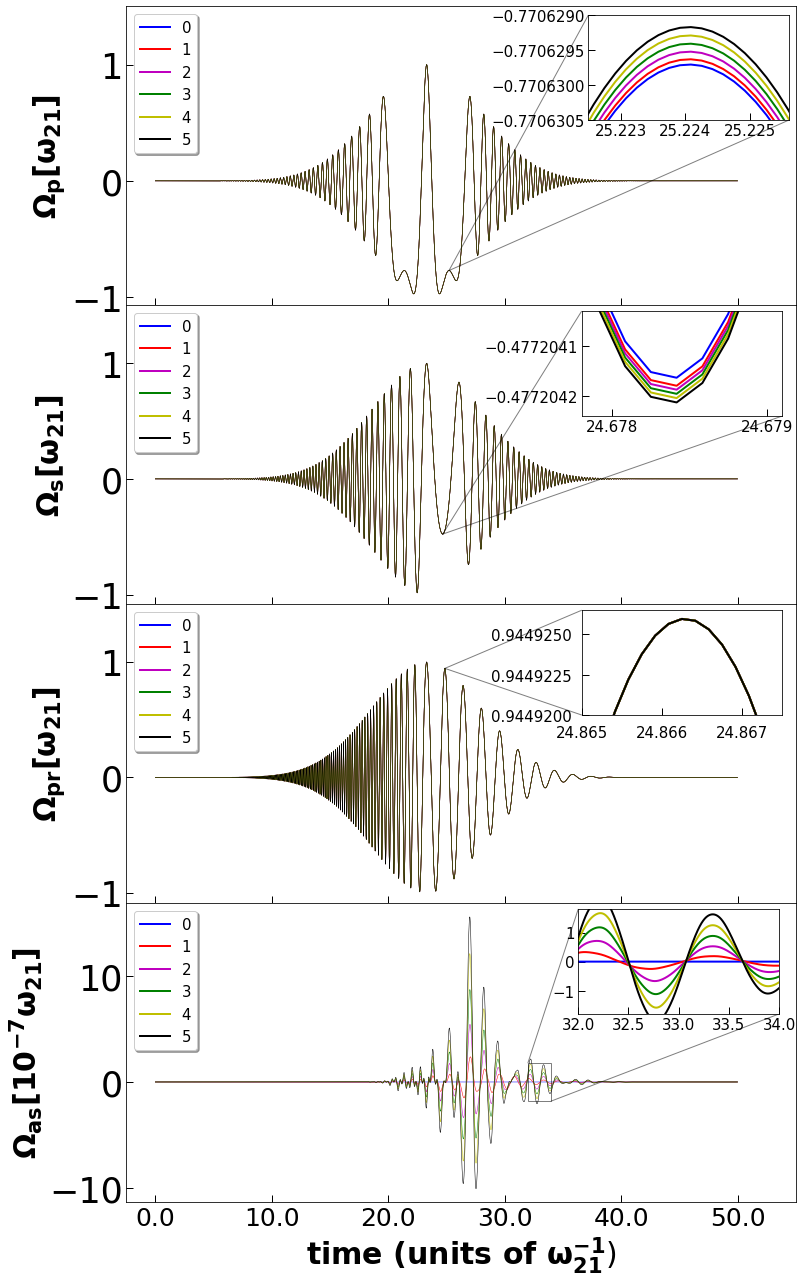}}
			\caption{The pump, the Stokes, the probe and the built-up anti-Stokes chirped pulses after each of five consecutive scattering events. 0,1,2,3,4,5 represent incoming, 1st, 2nd, 3rd, 4th and 5th scattering event respectively. The incident pulses are chirped in accordance with the control scheme. 
				The parameters of the fields are $\Omega_{p(s,pr)} = 85 THz$ $(E_{p(s,pr)_0} \sim 1.6 \times 10^9 V/m)$, $\tau_0=54.8 fs$, $\alpha_s=-7 THz / fs$, and $\Delta_s=\Delta_{as}=\Delta=850 THz$. The anti-Stokes field is built up gradually and constitutes $\sim 10^{-6}$ of the amplitude of the incident field.}
			\label{MaxwellChirp}
		\end{figure}

		The machine learning approach was implemented to reveal the modulation of the phase of four field components after each scattering. Fig.(\ref{MaxwellChirp}) shows the control pump, Stokes, probe and the build-up anti-Stokes pulses after each of five consecutive scattering events for the parameters of the fields $\Omega_{p(s,pr)} = 85 THz$ $(E_{p(s,pr)_0} \sim 1.6 \times 10^9 V/m)$, $\tau_0=54.8 fs$, $\alpha_s=-7 THz / fs$, and $\Delta_s=\Delta_{as}=\Delta=850 THz$. The neural networks explained in the previous section were optimized to work for these parameters. The classifier neural network predicted the pulses as the third kind described above and the regression neural network provided the chirping parameters. After 5 scattering events, the change in the initial chirp rate $\alpha_s$ is less than 0.001\% indicating that the control scheme would work for large number of layers. The anti-Stokes component is built up having the peak Rabi frequency about $10^{-6} \Omega_p$ after the fifth iteration.
		
		
		Machine learning is a powerful technique to solve problems in almost all branches of science. With the availability of immense amount of data and increased computational efficiency, the machine learning is transforming academic research and every major industry. We showed how deep neural networks can be used to analyze and understand chirped pulses. The analysis helped us to verify that control pulses can be used for optimization of the signal in detection of molecular systems without losing the phase values. As signal optimization is the essence of any sensing and detection methods, the technique we developed here could find variety of applications in quantum control methods.
		
		\section{CREATION OF MAXIMALLY COHERENT STATES USING FRACTIONAL STIMULATED RAMAN ADIABATIC PASSAGE}
		
		
		Stimulated Raman Adiabatic Passage (STIRAP), which was first reported in 1990 \cite{STIRAP_Original_1990}, is a process that allows population transfer in a quantum system efficiently to an initially unpopulated state via an intermediate state which is not populated in the process. As the intermediate state is not populated and the process is adiabatic, this method is very robust and immune to the spontaneous decay. STIRAP is a two-photon process where the Stokes pulse is applied first followed by the pump pulse, which is often referred to as a ``counter-intuitive" ordering of pulses. A considerable overlap between the two pulses is necessary for the adiabatic process and efficient transfer of population. Since its discovery, STIRAP has been exploited for tremendous applications  and several reviews have been published \cite{Vitanov_2017, ShoreTut, Bergmann_2019, STIRAP_Shore_2015}. A variation of the STIRAP process, namely fractional STIRAP (F-STIRAP), was introduced by Vitanov {\em et al.} in \cite{F_STIRAP}, where they showed that a coherent superposition of the initial and final states can be prepared by keeping the amplitude of the Stokes pulse non-zero for a longer time and making both the amplitudes vanish simultaneously. In \cite{Scully_F-STIRAP}, Sautenkov {\em et al.} used F-STIRAP to create a maximally coherent superposition in Rb vapor to enhance signal generation. This technique was based on the idea of delayed CARS, where the anti-Stokes signal is generated by a probe field applied at a later time once the superposition of states is created by the pump and Stokes pulses. This is different from the process of ordinary CARS where the anti-Stokes pulse is generated due to the four-wave mixing process involving the pump, Stokes and probe pulses which are applied simultaneously.
		
		In the previous sections, we developed and applied a quantum control theory to maximize the vibrational coherence and optimize the signal in CARS, for the purpose of remote detection. The process of F-STIRAP explained above can be used for similar applications as it creates a maximally coherent superposition selectively via adiabatic processes. One variation of STIRAP is introduced in \cite{Band_1994}, where the pump and Stoke pulses were chirped to selectively populate one of the two states in a nearly degenerate system. It was shown that by changing the sign of the chirp rate, the population can be driven to a pre-determined state in the four-level system. This was a major improvement to the existing methods based of STIRAP. Just like the chirping of pulses in STIRAP results in selective population of states, the chirping of pulses in F-STIRAP can be used to selectively create coherent superpositions in a nearly degenerate system. This is the motivation for this section, because the already developed semiclassical theory may be applied for remote detection in the framework of F-STIRAP. We first explain the process of STIRAP and the effect of non-zero two-photon detuning. We investigate how chirping of pulses in STIRAP can be beneficial for populating the desired energy level in a nearly degenerate four-level system. Then the process of F-STIRAP is described, along with an explanation as to how it can be used to create arbitrary coherent superposition states. Finally, we lay the ground work for a chirped-fractional-STIRAP scheme which can be used to improve the existing methods to achieve selective coherent superposition in a nearly degenerate system.
		
		\subsection{The Stimulated Raman Adiabtic Passage (STIRAP)}
		
		\begin{figure}
			\centering
			\includegraphics[scale=0.6]{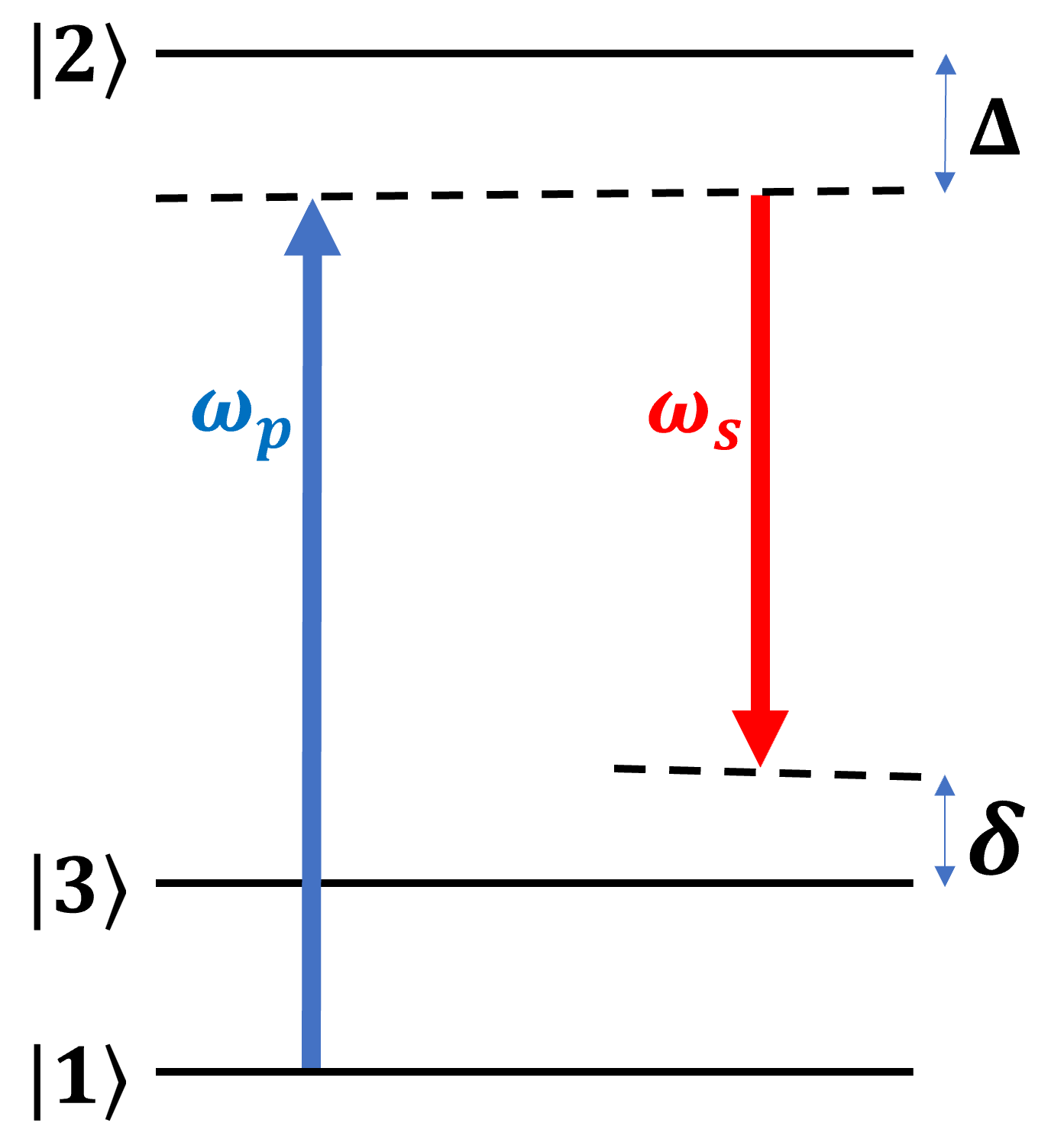}
			\caption{The three-level system for STIRAP. The population is transferred from the state $\ket{1}$ to $\ket{3}$ without populating the intermediate state $\ket{2}$. The STIRAP process is applicable to a ladder system as well, in which case the state $\ket{3}$ lies above $\ket{2}$. The detunings in both cases are given by: $\Delta =  \omega_{21} - \omega_p$ and $\delta = \omega_{31} - (\omega_p - \omega_s) $.} \label{5_3-level}
		\end{figure}
		
		The schematic diagram of the three-level system for STIRAP is shown in Fig. \ref{5_3-level}. The system is driven by two pulses: pump and Stokes, having frequencies $\omega_p$ and $\omega_s$ respectively. The population in state $\ket{1}$ is tranferred completely to the state $\ket{3}$ via the intermediate state $\ket{2}$. An important characteristics of this process is that the intermediate state $\ket{2}$ does not receive any population and it makes the transfer of population immune to any spontaneous decay. Another peculiarity of this method is the ordering of pulses: the Stokes pulse, which couples the initially unpopulated states $\ket{2}$ and $\ket{3}$ begins earlier than the pump pulse which couples states $\ket{1}$ and $\ket{2}$. A considerable overlap between the two pulses is necessary to provide a smooth adiabatic transfer as the mixing angle should vary very slowly. This will be further explained in the next section.
		
		The basic STIRAP Hamiltonian in Schr\"odinger representation is can be written as:
		
		\begin{equation}\label{5_STIRAP_HAM_3-level}
			\small{
				\mathbf{H}(t) = \hbar
				\begin{pmatrix}
					\omega_1	&	\tfrac{\mu_{21}}{\hbar}E_{p}(t)	&	0\\
					\tfrac{\mu_{21}}{\hbar}E_{p}^*(t)	&	\omega_2 	&	\tfrac{\mu_{23}}{\hbar}E_{s}^*(t)\\
					0	&	\tfrac{\mu_{23}}{\hbar}E_{s}(t)	&	\omega_3 \\
			\end{pmatrix} }
		\end{equation}
		where the pump and Stokes fields, as a general case, are considered to be chirped with chirp rates $\alpha$ and $\beta$. The equations of pulses, having Gaussian envelopes with time duration $\tau$  are given by:
		\begin{align*}
			E_{p}(t) &= \tfrac{1}{2}E_{p0} e^{-\tfrac{(t-t_p)^2}{\tau^2}} e^{i\omega_p(t-t_p) + i\frac{1}{2}\alpha(t-t_p)^2} + c.c. \\
			E_{s}(t) &= \tfrac{1}{2}E_{s0} e^{-\tfrac{(t-t_s)^2}{\tau^2}} e^{i\omega_p(t-t_s) + i\frac{1}{2}\beta(t-t_s)^2} + c.c.
		\end{align*}
		\begin{figure}
			\centering
			\includegraphics[scale=0.6]{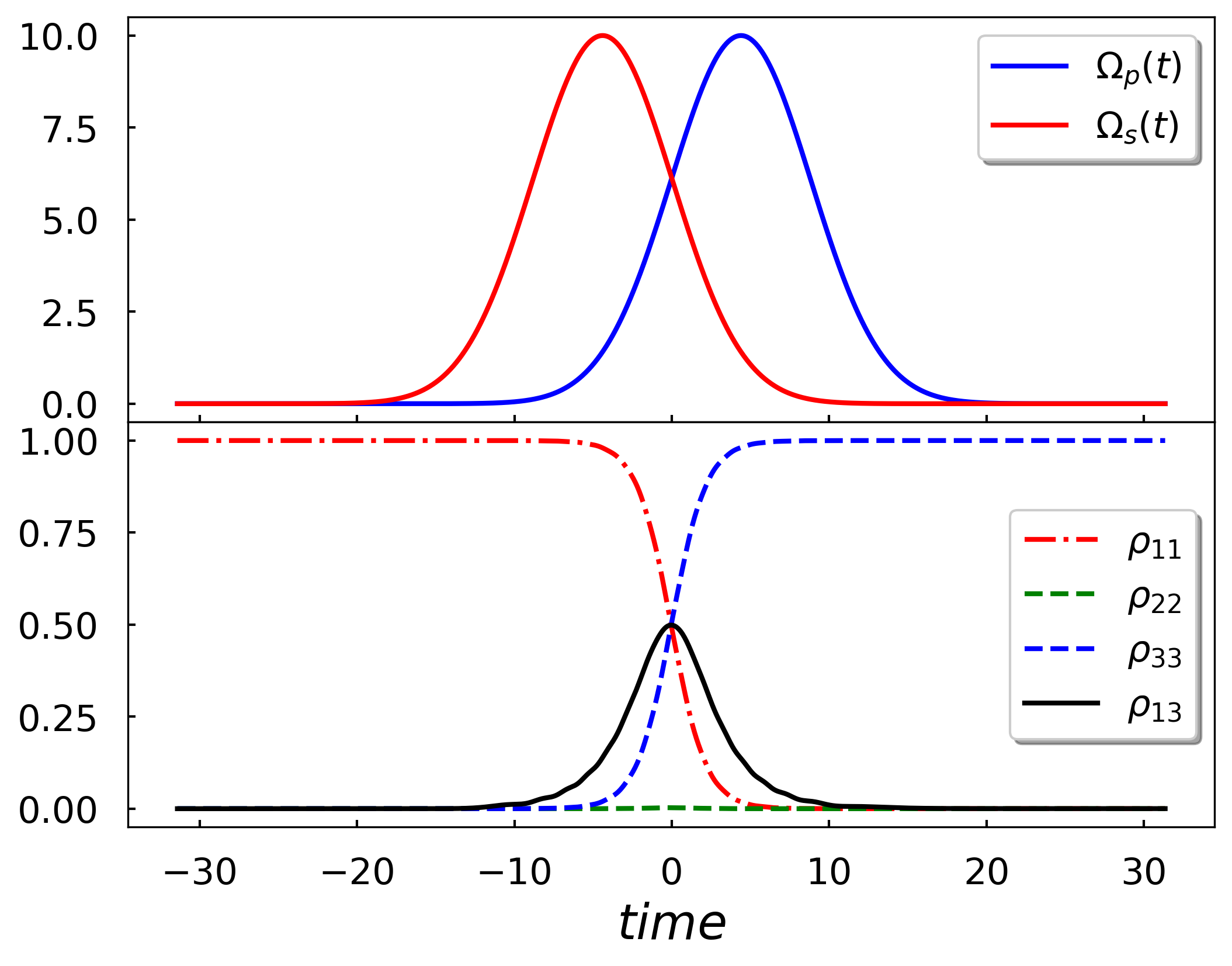}
			\caption{The STIRAP process. The pump pulse is followed by the Stokes pulse, allowing an adiabatic passage of population from state $\ket{1}$ to $\ket{3}$, without populating state $\ket{2}$.} \label{5_STIRAP_scheme}
		\end{figure}\label{STIRAP_process}\.
		where $t_p$ and $t_s$ are the central times of pump and Stokes respectively satisfying $t_s < t_p$.
		
		To transform the above Hamiltonian into field-interaction representation, consider the Schr\"odinger equation:
		\begin{equation}\label{shrodinger}
			i\hbar \mathbf{\dot{a}}(t) = \mathbf{H(t)a(t)}
		\end{equation}
		and apply the transformations:
		\begin{equation}\label{5_trans_chirped}
			\begin{aligned}
				a_1 &= \tilde{a}_1 e^{i\omega_p(t-t_p) + i\frac{1}{2}\alpha(t-t_p)^2} \\
				a_2 &= \tilde{a}_2  \\
				a_3 &= \tilde{a}_3 e^{i\omega_s(t-t_s) + i\frac{1}{2}\beta(t-t_s)^2}
			\end{aligned}
		\end{equation}
		and shift the diagonal elements to receive:
		\begin{equation}\label{5_STIRAP_Ham_chirped}
			\small{
				H = \hbar
				\begin{pmatrix}
					\alpha(t-t_p)	&	\tfrac{1}{2}\Omega_{p0}(t)	&	0	\\
					\tfrac{1}{2}\Omega_{p0}^*	&	\Delta 	&	\tfrac{1}{2}\Omega_{s0}^*(t)\\
					0	&	\tfrac{1}{2}\Omega_{s0}(t)	&	\delta +\beta(t-t_s)	\\
			\end{pmatrix}}
		\end{equation}
		where $\Delta$ and $\delta$ are the one-photon and two-photon detunings defined by: $\Delta = \omega_{21} - \omega_p$ and $\delta = \omega_{31} - (\omega_p-\omega_s )$ respectively.
		
		Apparently, taking $\alpha = \beta = 0$ gives the conventional STIRAP Hamiltonian without chirp:
		\begin{equation}\label{5_STIRAP_Ham_nochirp}
			\small{
				H = \hbar
				\begin{pmatrix}
					0	&	\tfrac{1}{2}\Omega_{p0}(t)	&	0	\\
					\tfrac{1}{2}\Omega_{p0}^*	&	\Delta 	&	\tfrac{1}{2}\Omega_{s0}^*(t)\\
					0	&	\tfrac{1}{2}\Omega_{s0}(t)	&	\delta	\\
			\end{pmatrix}}\,.
		\end{equation}
		
		
		To investigate the conditions for adiabatic passage in STIRAP, it is useful to diagonalize   the Hamiltonian in \eqref{5_STIRAP_HAM_3-level} by using an orthogonal matrix. Assume that the probability amplitudes of bare states $\mathbf{c}(t)$ evolve according to the Schr\"odinger equation $i\hbar \mathbf{\dot{c}}(t)= \mathbf{H}(t)\mathbf{c}(t)$. The dressed (adiabatic) states with probability amplitudes $\mathbf{a}(t)$ can be defined by the equation $\mathbf{c}(t) = \mathbf{T}(t)\mathbf{a}(t)$ where $\mathbf{T}(t)$ is an orthogonal matrix given by:
		\begin{equation}\label{HAM1}
				\textbf{T}(t) = 
				\begin{pmatrix}
					\sin\theta(t) \sin\phi(t)	&	\cos\theta(t)	&	\sin\theta(t) cos\phi(t)\\
					\cos\phi(t)	&	0 	&	-\sin\phi(t)\\
					\cos\theta(t) \sin\phi(t)	&	-\sin\theta(t)	&	\cos\theta(t) \cos\phi(t)\\
				\end{pmatrix}
		\end{equation}
		where the mixing angles are defined as: 
		\begin{equation}
			\tan\theta(t) = \frac{\Omega_{p0}(t)}{\Omega_{s0}(t)}
		\end{equation}
		and 
		\begin{equation}
			\tan2\phi(t) = \frac{\sqrt{\Omega_{p0}^2(t) + \Omega_{s0}^2(t)}}{\Delta(t)} = \frac{\Omega_{rms}(t)}{\Delta(t)}\,.
		\end{equation}
		The dressed state amplitudes $\mathbf{a}(t)$ follow the Schr\"odinger equations $i\hbar \mathbf{\dot{a}}(t)= \mathbf{H_a}(t)\mathbf{a}(t)$, where 
		\begin{equation*}
			\mathbf{H_a}(t) = \mathbf{T^ \dagger}(t)\mathbf{H}(t)\mathbf{T}(t) - i \hbar \mathbf{T}^{\dagger}(t)\mathbf{\dot{T}}(t) 
		\end{equation*}
		which gives:
		\begin{equation*}
			\begin{aligned}
				i\hbar \dot{\mathbf{a}}(t) &= [\mathbf{T^ \dagger}(t)\mathbf{H}(t)\mathbf{T}(t) - i \hbar \mathbf{T}^{\dagger}(t)\mathbf{\dot{T}}(t)]\mathbf{a}(t)\\
				i\hbar \dot{\mathbf{a}}(t) &= \mathbf {H_d}(t)\mathbf{a}(t) + i \mathbf {\dot{\Theta}}(t) \mathbf{a}(t)\,.
			\end{aligned}
		\end{equation*}
		Here, $\mathbf{H_d}(t)$ is a diagonal matrix and $\dot{\mathbf{\Theta}}(t)$ is a matrix with only non-diagonal elements. For adiabatic passage to occur, the matrix $\mathbf{H_a}(t)$ should be very close to $\mathbf{H_d}(t)$, meaning values in $\dot{\mathbf{\Theta}}$ should be very small compared to the difference in diagonal values $\mathbf{H_{a}}_{11}(t)$ and $\mathbf{H_{a}}_{11}(t)$. At two-photon resonance, $\delta = 0$, The matrix $\mathbf{H_a}(t)$ looks like:
		\begin{equation}\label{H_a0}
				\mathbf{H_a}(t) = \hbar
				\begin{pmatrix}
					\tfrac{1}{2}\Omega_{rms} \cot \phi	&	i \dot{\theta}\sin\phi	&	i \dot{\phi}\\
					-i \dot{\theta}\sin\phi	&	0 	&	-i \dot{\theta}\cos\phi\\
					-i \dot{\phi}	&	i \dot{\theta}\cos\phi	& - \tfrac{1}{2}\Omega_{rms} \tan \phi\\
				\end{pmatrix}
		\end{equation}
		The diagonal elements of this Hamiltonian are the dressed (adiabatic) energies, which can be written as:
		\begin{align}
			\begin{aligned}
				\lambda_+(t) &= \tfrac{1}{2}\Omega_{rms}(t) \cot \phi(t) = \tfrac{1}{2} \left(\Delta + \sqrt{\Delta^2 + \Omega_{rms}^2(t)} \right) \\
				\lambda_0(t) &= 0 \\
				\lambda_-(t) &= -\tfrac{1}{2}\Omega_{rms}(t) \tan \phi(t) = \tfrac{1}{2} \left(\Delta - \sqrt{\Delta^2 + \Omega_{rms}^2(t)} \right)\,.
			\end{aligned}
		\end{align}\label{STIRAP_eigenvelues}
		At one-photon resonance, $\Delta = 0$, $\tan 2\phi = \infty$, $\phi = \pi/4$. In this limit, the adiabaticity condition becomes: $|\Omega_{rms}(t)| \gg |\dot{\theta}(t)| $. To satisfy the adiabatic condition in STIRAP, the mixing angle $\theta = \tan^{-1}(\Omega_{p0}(t)/\Omega_{s0}(t))$, should be varying slowly. For this, it is necessary that the overlap between the pump and Stokes pulses is not too large or not too small. The eigenstates (adiabatic states of dressed states) corresponding to these eigenvalues are:
		\begin{align}
			\begin{aligned}
				\Phi_+(t) &= \psi_1 \sin\theta(t)\sin\phi(t) + \psi_2 \cos\phi(t) + \psi_3 \sin\theta(t)\sin\phi(t)\\
				\Phi_0(t) &= \psi_1\cos\theta(t)-\psi_3\sin\theta(t) \\
				\Phi_-(t) &= \psi_1 \sin\theta(t)\cos\phi(t) - \psi_2 \sin\phi(t) + \psi_3 \cos\theta(t)\cos\phi(t)
			\end{aligned}
		\end{align}\label{STIRAP_eigenstates}
		where $\psi_1$, $\psi_2$ and $\psi_3$ are the eigenstates of bare quantum system. The eigenstate corresponding to the dressed energy zero, $\Phi_0(t)$ is called dark state. In the beginning, when $\Omega_{p0}(t) = 0$ while $\Omega_{s0}(t) > 0$, the mixing angle $\theta(t) = 0$ and the dark state $\Phi_0(t)=\psi_1$. In the end, when $\Omega_{s0}(t) = 0$ while $\Omega_{p0}(t) > 0$, the mixing angle $\theta(t) = \pi/2$ and the dark state $\Phi_0(t)=-\psi_3$. So the dark state has now gone from $\psi_1$ to $\psi_3$ without acquiring any component of $\psi_2$. In order for the dark state to not acquire any component of the excited state, the condition for adiabaticity should be satisfied.
		
		\begin{figure}
			\centering
			{\includegraphics[scale=0.5]{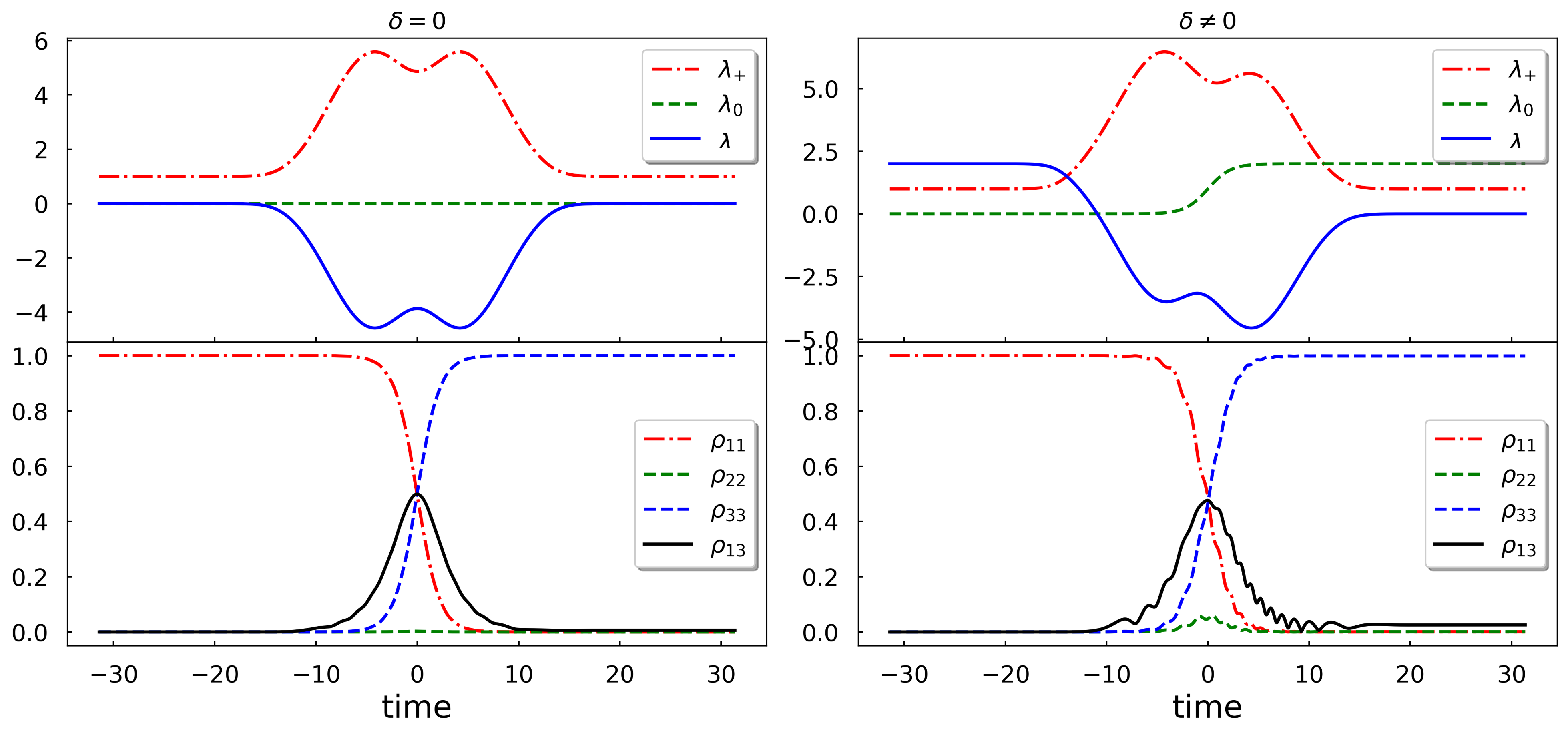}}
			\caption{STIRAP in the presence of two-photon resonance (left) and in the absence of two-photon resonance (right). In the left figure, $\Delta\neq 0$ and $\delta=0$. The system remains aligned with the adiabatic state $\Phi_0$ throughout the process. The population is fully transferred to $\psi_3$. In the right figure $\Delta \neq 0$ and $\delta \neq 0$. This process is not completely adiabatic. The system is aligned with $\Phi_-$ in the beginning and is aligned with $\Phi_+$ in the end. Even though the population is transferred completely to $\psi_{3}$, the process is completely not adiabatic as there are two crossings with non-adaiabtic coupling between the dressed states. 
			} \label{5_detuned_STIRAP}
		\end{figure}
		We have now derived the conditions for adiabaticity when both $\Delta=0$ and $\delta=0$. Finding the adiabaticity conditions for non-zero detunings is not so trivial. In the general case, when $\Delta \neq 0$ and $\delta \neq 0$, the matrix $\mathbf{H_a}(t)$ may be written as the sum of two matrices. $\mathbf{H_a}(t) = \mathbf{H_{a0}}(t) + \mathbf{H_{a1}}(t)$, where $\mathbf{H_{a0}}(t)$ is the Hamiltonian when $\delta = 0$, which is given by the Eq \ref{H_a0} when $\Delta=0$ as well, and $\mathbf{H_{a1}}(t)$ is the additional term due to the absence of two-photon resonance, which is given by: 
		\begin{equation}\label{H_a1}
				\mathbf{H}_{a1}(t) = \frac{1}{2}\hbar \delta
				\begin{pmatrix}
					\cos 2\theta \sin^2 \phi	&	- \sin 2\theta \sin \phi	&	\tfrac{1}{2}\cos 2\theta \sin 2\phi\\
					- \sin 2\theta \sin \phi	&	-\cos 2\theta 	&	- \sin 2\theta \cos \phi\\
					\tfrac{1}{2}\cos 2\theta \sin 2\phi	&	- \sin 2\theta \cos \phi	& \cos 2\theta \cos^2 \phi\\
				\end{pmatrix}\,.
		\end{equation}
		The two-photon detuning shifts all energies of adiabatic states in proportion to the $\delta$ and two-photon resonance is a necessary condition for adiabatic passage in STIRAP. For the specific case when $\Delta = 0$, $\phi = \pi/4$, and the above Hamiltonian becomes
		\begin{equation}\label{H_a1}
			\begin{aligned}
					\mathbf{H}_{a1}(t) &= \frac{1}{2}\hbar \delta
					\begin{pmatrix}
						\frac{1}{2}\cos 2\theta	&	-\frac{1}{\sqrt2} \sin 2\theta	&	\tfrac{1}{2}\cos 2\theta\\
						-\frac{1}{\sqrt2} \sin 2\theta	&	-\cos 2\theta 	&	-\frac{1}{\sqrt2} \sin 2\theta\\
						\tfrac{1}{2}\cos 2\theta	&	-\frac{1}{\sqrt2} \sin 2\theta	& \frac{1}{2}\cos 2\theta\\
					\end{pmatrix}\\
				&= \frac{1}{4}\hbar \delta \cos2\theta
				\begin{pmatrix}
					1	&	-\sqrt2 \tan 2\theta	&	1\\
					-\sqrt2 \tan 2\theta	&	-2	&	-\sqrt2 \tan 2\theta\\
					1	&	-\sqrt2 \tan 2\theta	& 1\\
				\end{pmatrix}\,.
			\end{aligned}
		\end{equation}
		The evolution of dressed state energies and populations in the case of both two-photon resonance and non-zero two-photon detuning are given in the Fig. \ref{5_detuned_STIRAP}. In the left figure, the $\Delta \neq 0$ and $\delta=0$. As shown, adiabatic passage possible in case of two-photon resonance even if the system is not in one-photon resonance, $\Delta \neq 0$. The system is aligned with the dark state $\Phi_0$ and the population is fully transferred the state $\psi_3$. There is no crossing of energy levels. But, when the $\delta$ is non-zero, the dressed state energy levels cross each other and the adiabaticity is lost. The system is aligned with the dressed state $\Phi_-$ in the beginning and becomes aligned with $\Phi_+$ by the end. Even though the population is completely transferred to $\psi_3$, the process is not fully adiabatic. So, in summary, two-photon resonance is necessary for the adiabatic passage in the process of STIRAP. 
		
		\subsection{Chirped-STIRAP: selective population of two nearly-degenerate states}
		\begin{figure}
			\centering
			\includegraphics[scale=0.7]{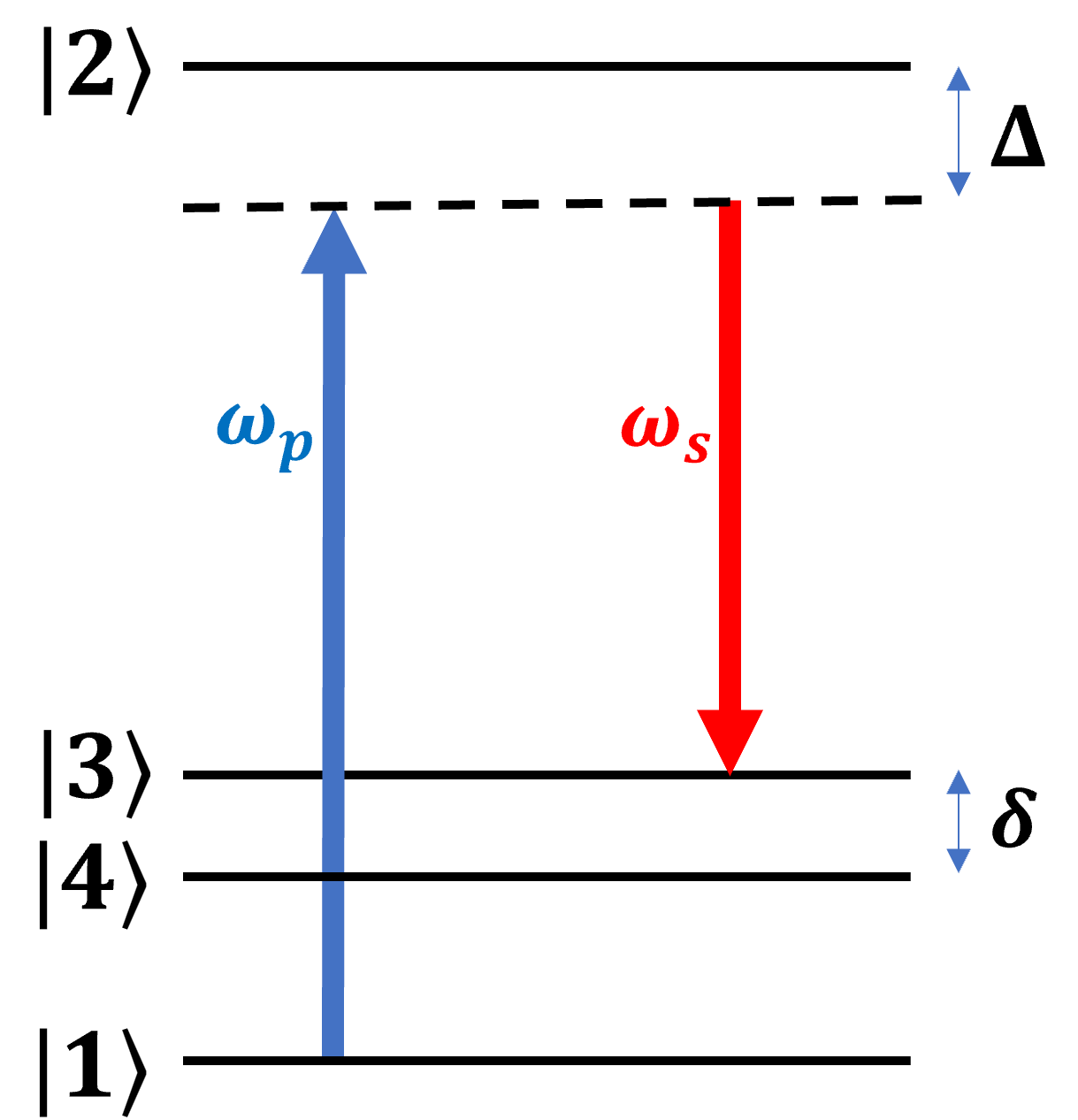}
			\caption{The four-level system for chirped-STIRAP. The states $\ket{3}$ and $\ket{4}$ are nearly degenerate with Stokes pulse in resonance with state $\ket{3}$ and detuned to $\ket{4}$. $\Delta = \omega_{21} - \omega_p$, and $\delta = \omega_{41} - (\omega_p - \omega_s) $.} \label{5_4-level}
		\end{figure}
		In the previous section, we analyzed the ordinary STIRAP process and explained the origin of adiabatic passage. In this section, we will consider a four-level system with two nearly degenerate states and use chirped pulse in STIRAP to populate one of these states. This method was first introduced in \cite{Band_1994}. The four-level system we consider for this analysis is shown in Fig. \ref{5_4-level}. The states $\ket{3}$ and $\ket{4}$ are nearly degenerate with Stokes pulse in resonance with state $\ket{3}$. The Hamiltonian for this four-level system in field-interaction representation when chirped pulses are used can be obtained from \eqref{5_STIRAP_HAM_3-level} and is given by:
		\begin{equation}\label{5_STIAP_HAM_4-level}
			\small{
				H = \hbar
				\begin{pmatrix}
					\alpha(t-t_p)	&	\tfrac{1}{2}\Omega_{p0}(t)	&	0	&	0\\
					\tfrac{1}{2}\Omega_{p0}^*	&	\Delta 	&	\tfrac{1}{2}\Omega_{s0}^*(t)	&	\tfrac{1}{2}\Omega_{s0}^*(t)\\
					0	&	\tfrac{1}{2}\Omega_{s0}(t)	& \beta(t-t_s)	&	0\\
					0	&	\tfrac{1}{2}\Omega_{s0}(t)	&	0	&	\delta+\beta(t-t_s)\\
			\end{pmatrix}}
		\end{equation}
		
		\begin{figure}
			\centering
			{\includegraphics[scale=0.6]{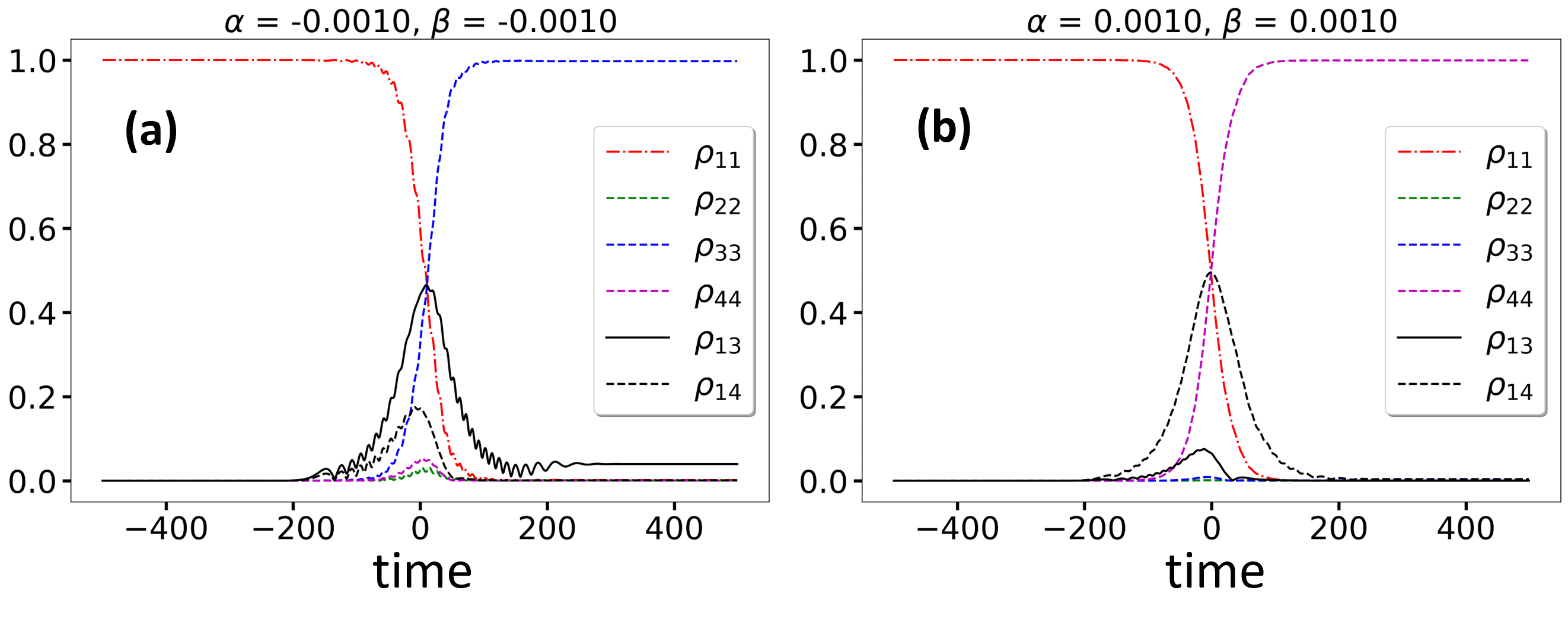}}
			\caption{Population dynamics in STIRAP when both pump and Stokes pulses are negatively chirped (a) and positively chirped (b). The population is driven exclusively to state $\ket{3}$ when the chirp rate is negative and to $\ket{4}$ when chirp rate is positive}\label{5_4-level_STIRAP}
		\end{figure}
		The evolution of populations plotted using the above Hamiltonian is shown in Fig. \ref{5_4-level_STIRAP}. In Fig. \ref{5_4-level_STIRAP}(a), both the pump and Stokes pulses are chirped with negative chirp rate $\alpha =\beta = -0.001$. In this case, state $\ket{3}$ which is in resonance with the Stokes pulse, is populated. This behavior is much like the ordinary STIRAP. But if the sign of both the chirp rates are flipped, the detuned state $\ket{4}$ is populated, as shown in Fig. \ref{5_4-level_STIRAP}(b). This means that the flow of population can be controlled and be directed to the desired energy level by chirping the pulses in STIRAP. Similar to the ordinary STIRAP, the intermediate level is not populated in the process of chirped STIRAP as well. 
		\begin{figure}
			\centering
			{\includegraphics[scale=0.6]{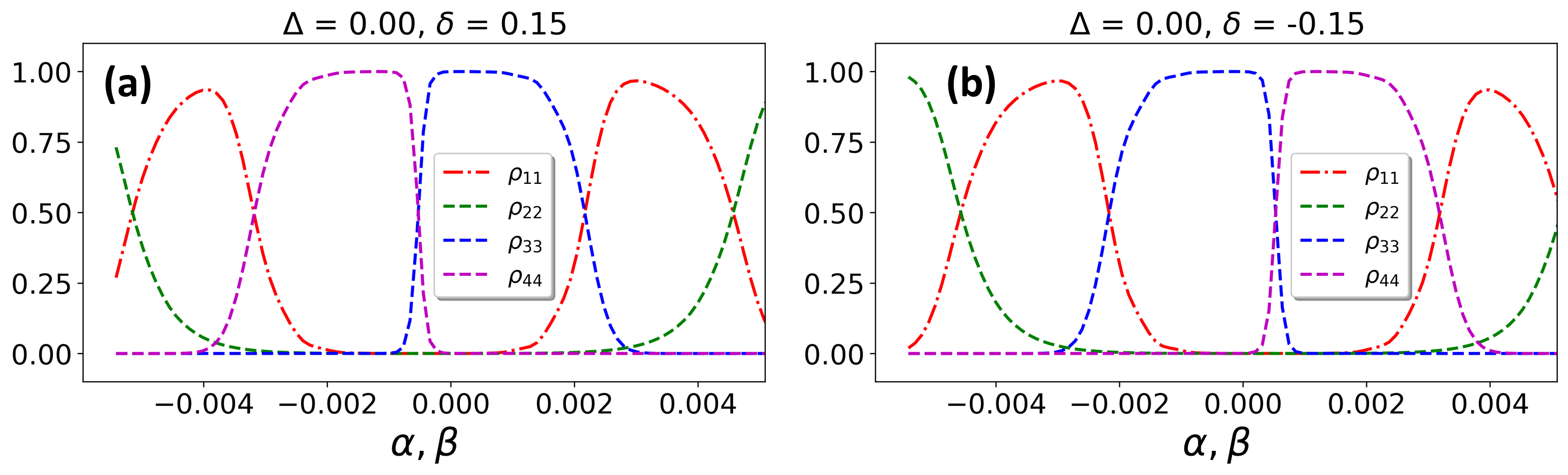}}
			\caption{Populations vs chirp rate of pump and Stokes when the two-photon detuning is positive (a) and negative (b). The chirp rates of pump and Stokes are taken to be equal, $\alpha = \beta$, in both the cases. }\label{5_4-level_delta}
		\end{figure}\label{5_4-level_chirp_rho}
		A broader analysis of this process is shown in Fig. \ref{5_4-level_chirp_rho} where the populations are plotted against the chirp rates of pump and Stokes pulses for positive, (a) and negative values of detuning (b). The chirp rates $\alpha$ and $\beta$ are equal in all the calculations here. In the Fig. \ref{5_4-level_chirp_rho}(a), $\delta > 0$ implying that $\ket{4}$ is above $\ket{3}$. In this case, a positive chirp causes the population to flow to the resonant final state $\ket{3} $ while negative chirp drives population to the detuned state $\ket{4} $. Fig. \ref{5_4-level_chirp_rho}(b) shows the opposite behavior as the detuning is now negative, meaning $\ket{4}$ is below $\ket{3}$. In short, we have shown that chirping the pulses in STIRAP is a powerful way to control the flow of population to a desired state adiabatically in a nearly degenerate four-level system.
		
		\subsection{The Fractional-STIRAP}
\begin{figure}
			\centering
			\includegraphics[scale=0.7]{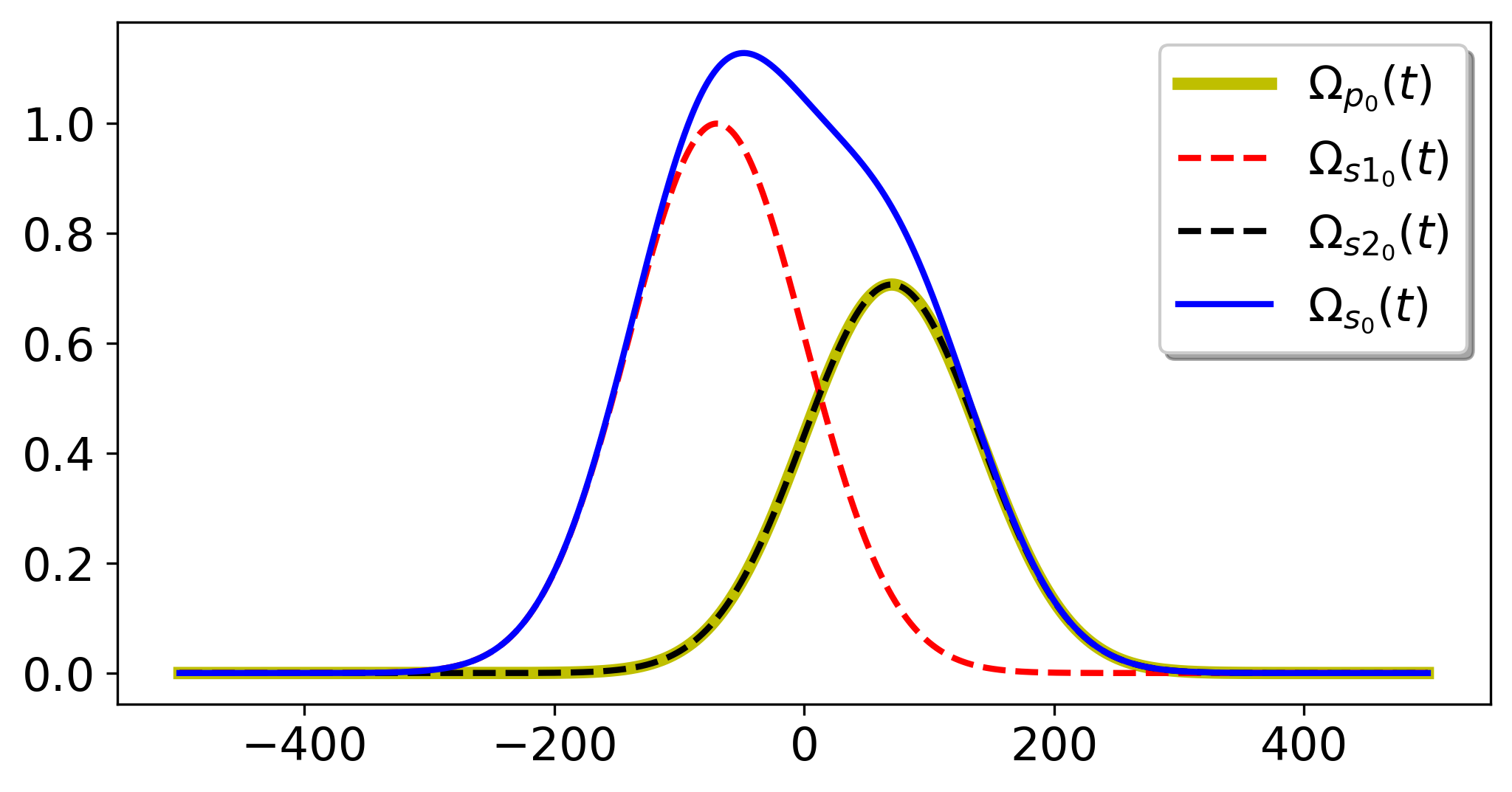}
			\caption{Fractional STIRAP using the superposition of two Stokes pulses, $\Omega_{s_0}(t)= \Omega_{s1_0}(t)+\Omega_{s2_0}(t)$, as given in Eq. \eqref{5_F-STIRAP_envelopes}. Here the constant mixing angle $A=\pi/4$. Note that the pump pulse $\Omega_{p_0}(t)$ overlaps exactly with the second Stokes pulse $\Omega_{s2_0}(t)$.} \label{5_Stokes-pulses}
		\end{figure}

		We have now seen that STIRAP is an effective and robust way to transfer population to a particular quantum state. In this section, we will see that instead of transferring the population completely to the final state, coherent superposition between the initial and final states can be created by slightly modifying the STIRAP technique. The idea is based on manipulating the amplitude of Stokes pulse so that the mixing angle $\theta(t)$ is a constant by the end of the process. Similar to STIRAP, the Stokes pulse begins earlier than the pump, but unlike STIRAP, both the pulses vanish simultaneously. This provides a coherent superposition instead of a complete population transfer while the process remain adiabatic. To derive the evolution of amplitudes in this process, take the dark state of STIRAP process given in \eqref{STIRAP_eigenstates}:
		\begin{equation}
			\Phi_0(t) = \psi_1\cos\theta(t) - \psi_3\sin\theta(t)
		\end{equation}
		At $t=-\infty$, the system is in $\psi_1$, and at $t=\infty$, the system has moved to the $\psi_3$. We need to manipulate the mixing angle $\theta(t)$ in such a way that at $t=\infty$, the system is in a coherent superposition of $\psi_1$ and $\psi_3$. Let us assume: $\theta(\infty) = A$, where $A$ is a constant. This gives:
		\begin{equation}
			\Phi(t=-\infty) = \psi_1, \ \ \ \ \
			\Phi(t=\infty) = \psi_1\cos A - \psi_3\sin A
		\end{equation}
		which means the mixing angle:
		\begin{equation}
			\theta(t=-\infty) = 0, \ \ \ \ \
			\theta(t=\infty) = \tan A\,.
		\end{equation}
		To achieve a mixing angle that satisfies this condition, two Stokes pulses can be applied; the first one at time $t=-t_p$ and second one at $t=t_p$, where $t_p$ is the central time of pump. The envelope equations of pump and Stokes satisfying this condition can be written as:
		\begin{equation}\label{5_F-STIRAP_envelopes}
			\begin{aligned}
				\Omega_{p_0}(t) &= \Omega_{0} \sin A e^{-\tfrac{(t-t_p)^2}{\tau^2}} \\
				\Omega_{s_0}(t) &= \Omega_{0} e^{-\tfrac{(t+t_p)^2}{\tau^2}} + \Omega_{0} \cos A e^{-\tfrac{(t-t_p)^2}{\tau^2}} \,. \\
			\end{aligned}
		\end{equation}
		Note that if the constant mixing angle $A=\pi/2$, the second term of the Stokes equation is zero and we are back to STIRAP, where the Stokes has a central time of $-t_p$ and pump has a central time of $t_p$. This provides a complete population transfer. On the other hand, when $A=\pi/4$, $\cos A = \sin A = 1/{\sqrt2}$ and the system is transformed to a maximally coherent superposition. In this case, the dark state $\Phi_0(t) = \tfrac{1}{\sqrt{2}}(\psi_1
		-\psi_3)$. 
		The envelopes of two Stokes pulses and their superposition according to the Eq. \eqref{5_F-STIRAP_envelopes} when $A=\pi/4$  are shown in Fig. (\ref{5_Stokes-pulses}). In this figure, $\Omega_{s1_0}(t)$ and $\Omega_{s2_0}(t)$ are the first and second Gaussian envelopes that make up the new Stokes field $\Omega_{s_0}(t)$ given in Eq. \eqref{5_F-STIRAP_envelopes}. Note that the second Stokes pulse $\Omega_{s2_0}(t)$ completely overlaps with the pump pulse $\Omega_{p_0}(t)$ because they both have the same central time. This makes sure that the tail of the resultant Stokes field overlaps with the that of the pump field, in order to achieve $\tan\theta(t) = \Omega_{p_0}(t)/\Omega_{s_0}(t) = 1$ as time $t \rightarrow \infty$.
		
		In order to understand the population dynamics in fractional-STIRAP, it is useful to deal with the field-interaction Hamiltonian in this case.
		By intuition, it can be seen that the STIRAP Hamiltonian in Eq (\ref{5_STIRAP_Ham_nochirp}) can be used for F-STIRAP as well, replacing the Stokes pulse with the new Stokes field in Eq. \eqref{5_F-STIRAP_envelopes}.
		\begin{figure}
			\centering
			\includegraphics[scale=0.9]{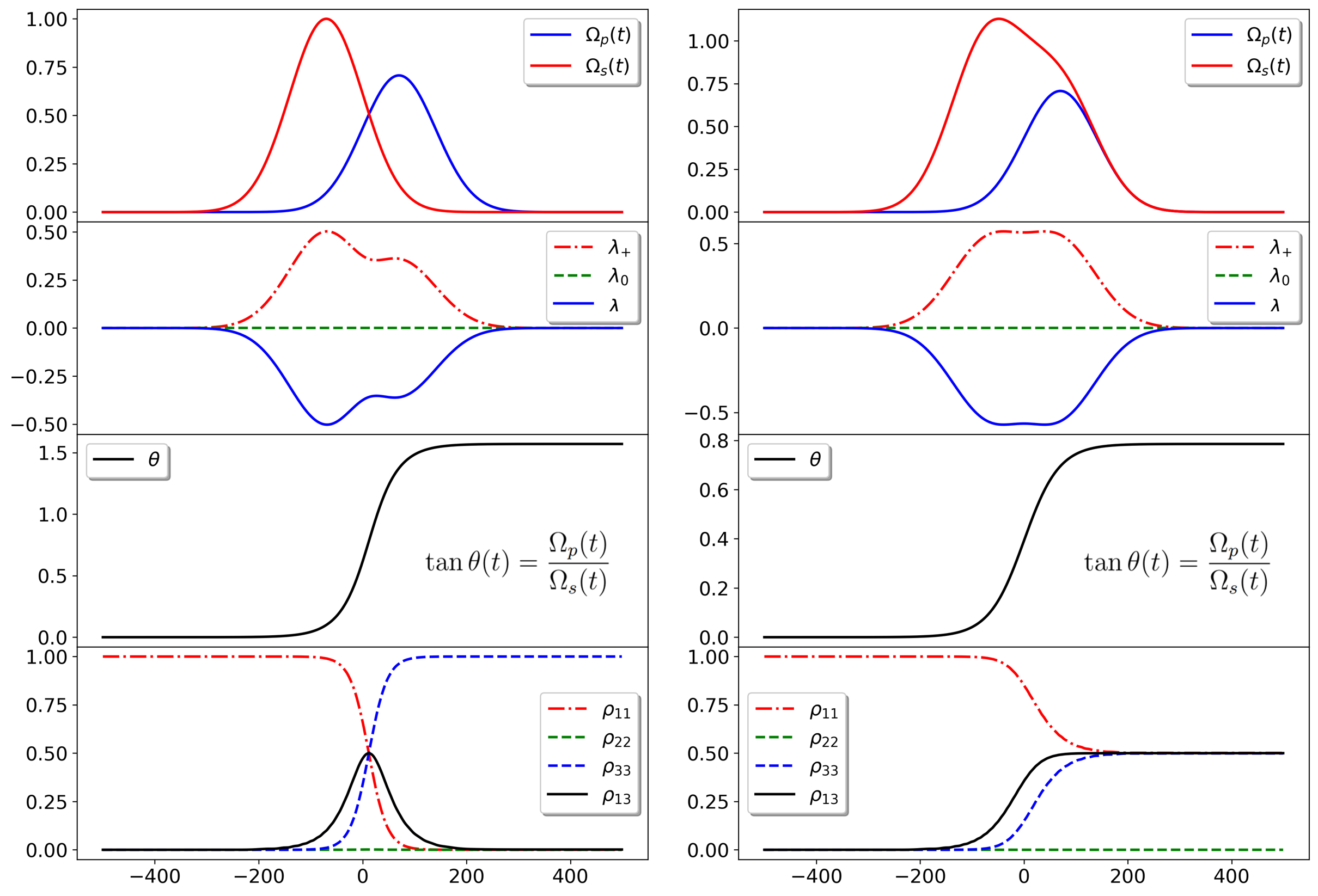}
			\caption{Comparison of the STIRAP (left) and fractional-STIRAP (right) processes. In both cases, the evolution of dressed state energies are similar and there are no crossing of energy levels. The mixing angles vary slowly in both cases, implying that the process is adiabatic. As $t \rightarrow \infty$, the mixing angle goes to $\pi/2$ and $\pi/4$ in STIRAP and F-STIRAP respectively. The coherence $\rho_{13}$ is zero in STIRAP while it is maximum, 0.5, in F-STIRAP.} \label{5_F-STIRAP-comparison}
		\end{figure}
		In the previous sections, we analyzed the dressed (adiabatic) states in STIRAP and the conditions for achieving adiabatic passage. As in STIRAP, two-photon resonance is necessary in order to have adiabatic passage in F-STIRAP as well. Apart from that, we saw that when $\Delta=0$, the variation in mixing angle should be very slow compared to the rms Rabi frequency, $|\Omega_{rms}(t)| \gg |\dot{\theta}(t)|$. The evolution of pulses, dressed states, mixing angle and populations are shown in Fig. \ref{5_F-STIRAP-comparison}. The dressed energies $\lambda_+$, $\lambda_0$ and $\lambda_-$ evolve similarly in both the processes and there is no crossing of energy levels. The mixing angle $\theta(t)$ evolve slowly in both cases indicating that $\dot{\theta}(t)$ is very small compared to the difference between the dressed states $\lambda_+$ and $\lambda_-$. Note that the final value of $\theta(t)$ is $\pi/2 (\approx 1.5)$ in STIRAP while it is $\pi/4 (\approx 0.8)$ in F-STIRAP. 
		
\begin{figure}
			\centering
			\includegraphics[scale=0.5]{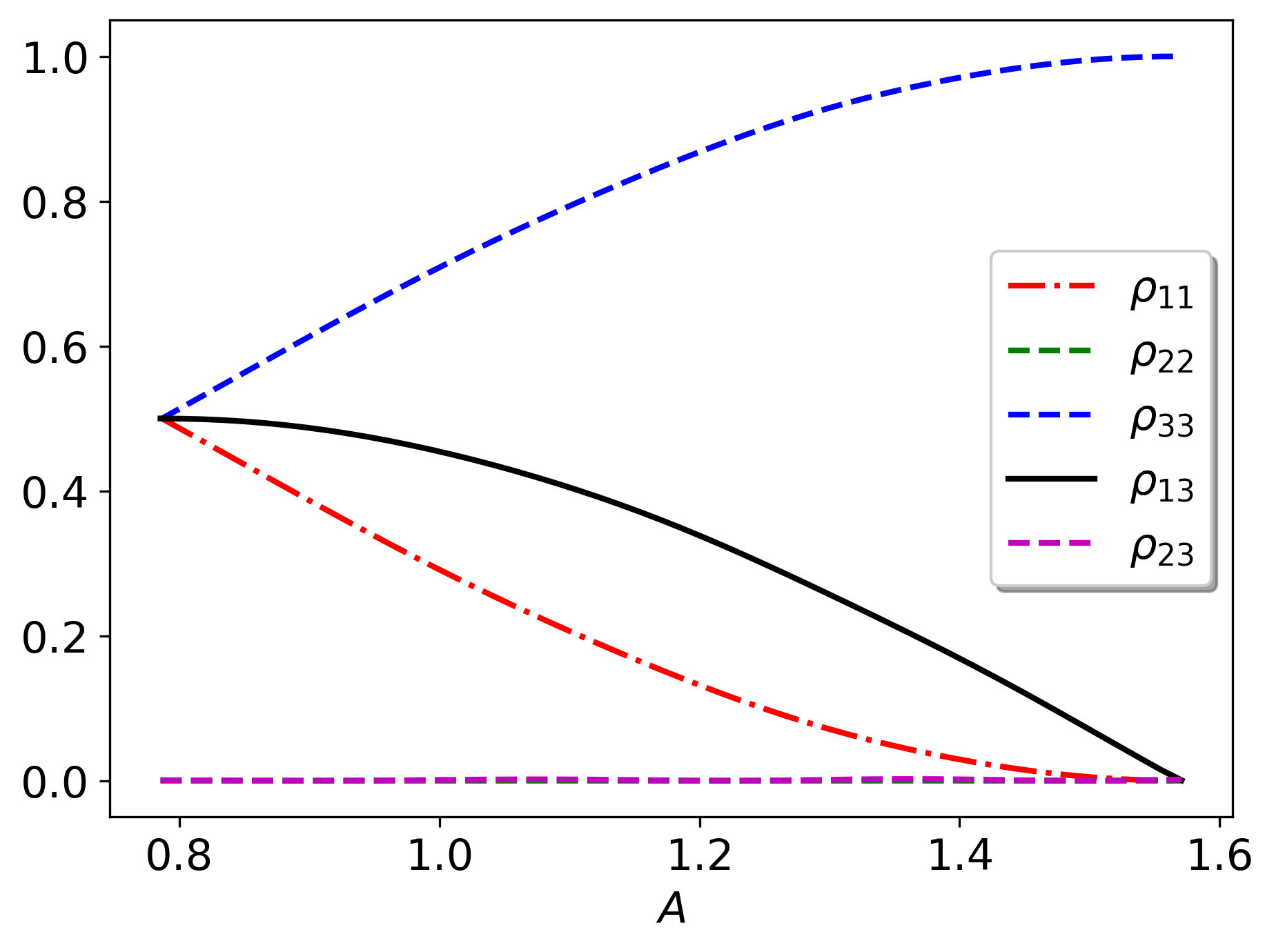}
			\caption{The plot of constant mixing angle $A$ vs populations and coherence. If $A$=$\pi/4 \approx 0.8$, the coherence is maximum and the populations initial and final states are equal. If $A=\pi/2 \approx 1.6$, the population is completely transferred to the final state. This is equivalent to the ordinary STIRAP.} \label{5_Mixing_angle}
		\end{figure}

		Fractional STIRAP can be understood as a generalized form of STIRAP. By varying the constant mixing angle, it is possible to create any arbitrary coherent superposition of the initial and final states. The final populations and coherence are plotted the constant mixing angle in Fig. \ref{5_Mixing_angle}. For $ A=\pi/4 \approx 0.8$, the coherence is maximum and for $A=\pi/2 \approx 1.6$, the coherence is zero and the final state population is 1. Any arbitrary coherence between the initial and final states can be produced by carefully choosing the angle $A$.
		
		\subsection{Appilcation of F-STIRAP for Remote Detection}
		\begin{figure}
			\centering
			\includegraphics[scale=0.8]{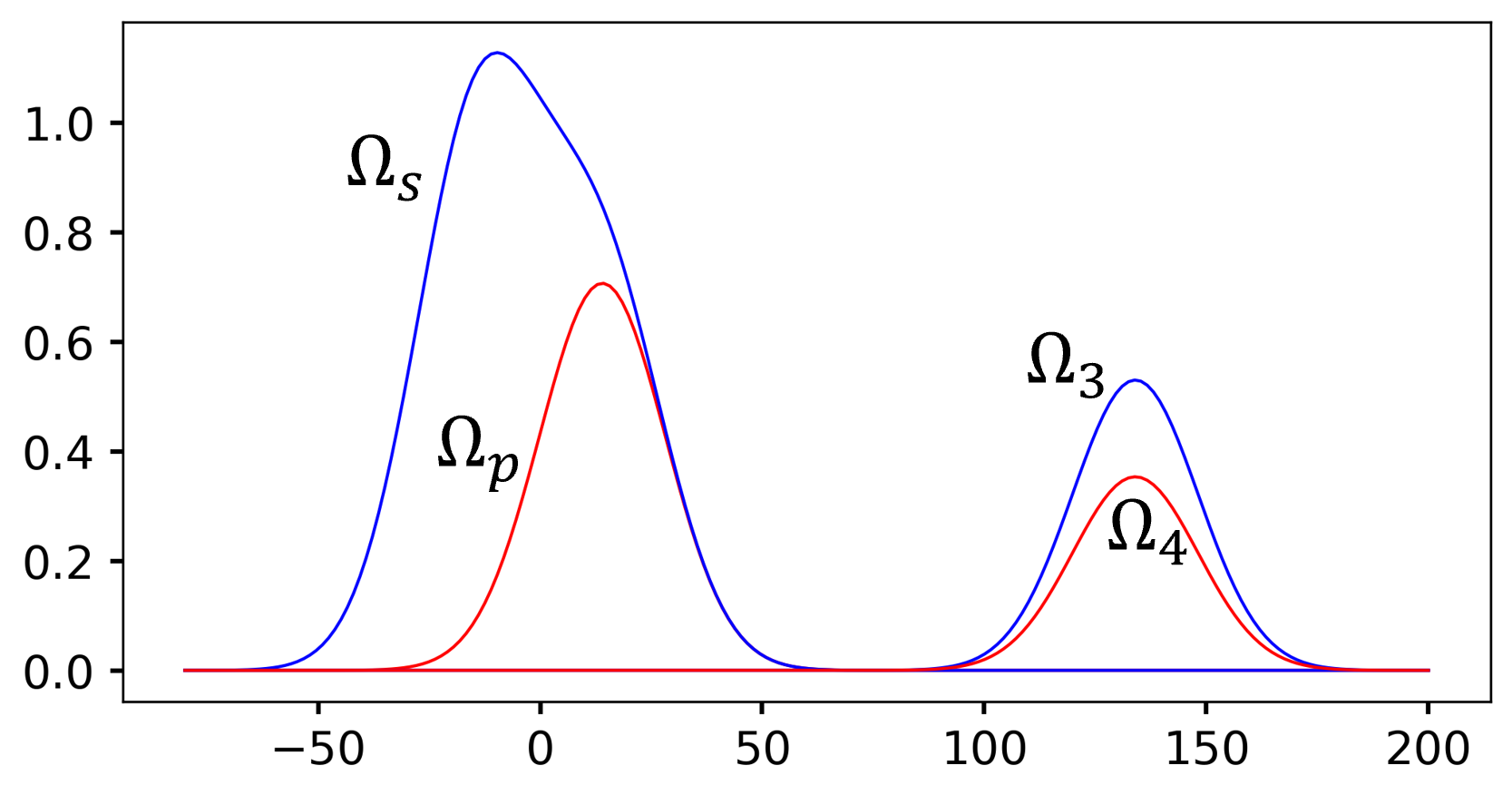}
			\caption{The schematic of using Fractional STIRAP technique to optimize the signal for remote detection. First, fields $\Omega_p$ and $\Omega_s$ distribute the populations equally and maximize coherence between the sublevels. Field $\Omega_3$ is then applied after some time, which generated field $\Omega_4$ by coherent scattering with the system.} \label{5_F-STIRAP_CARS}
		\end{figure}
		
		We showed that fractional-STIRAP is a robust and efficient technique to create maximally coherent superposition states. In CARS, a coherent superposition state is generated through driving by the pump and Stokes fields and the probe pulse interacts with this superposition to generate the anti-Stokes signal. An extension of fractional-STIRAP to the technique of CARS can be done by applying a pump and Stokes first to create the coherence, followed by a probe pulse at a later time. The schematic of this method is given in Fig. \ref{5_F-STIRAP_CARS}. This scheme of creating maximum coherence can be combined with the semiclassical theory we developed in section 3 to create control protocols that optimize the signal used sensing and detection. We saw that chirping of pulses in STIRAP is beneficial as it helps us to control the population flow to a desired state in a four-level system. In the same way, chirping of pulses in fractional-STIRAP can be used as a way to control the formation of coherent superposition between a desired pair of states in a four-level system. Combining the technique of chirping pulses in fractional-STIRAP with the semiclassical theory for remote detection is expected to make considerable improvements to the existing method for imaging, sensing and detection.

		\section{SUMMARY}
		
		In this chapter, we took a semiclassical approach to deal with light-matter interactions and developed several methods to prepare  quantum systems in a predetermined state. The primary focus was to improve the existing methods of detection and sensing by controlling the incident field parameters in order to optimize the output signal. We learned that adiabatic passage regime of interaction gives a robust way of preparing maximally coherent superpositions of quantum states in a multilevel system. We presented ways of improving the techniques of CARS and STIRAP by chirping the incident laser fields.
		
		In the introductory section, the general theory of light-matter interaction and Raman spectroscopy was discussed. In the second section, a theory of quantum control method where the amplitudes and phases of all the incident pulses in Coherent Anti-Stokes Raman Spectroscopy are carefully manipulated to satisfy the conditions for adiabatic passage was developed. First, a large one-photon detuning was assumed so that the two excited states can be eliminated and the four-level system can be simplified to a ``super-effective'' two-level system. This reveals the dynamics of energy levels and a control scheme can be developed in order to maximize the vibrational coherence. The amplitudes of probe and Stokes pulses should be equal and should be less than the amplitude of pump by a factor of $\sqrt{2}$. The chirp rates of the pump and Stokes needs to be opposite in sign before the central time and should be equal after that. The probe should be chirped at a rate of equal to the difference between pump and Stokes pulses all the time. This chirping scheme, which we called C-CARS or Chirped-CARS, is a robust method to create a maximally coherent superposition of the system via adiabatic passage while suppressing the non-resonant background, overcoming one of the major limitations of CARS spectroscopy. We also show that the selectivity of this scheme can be increased by controlling the chirp parameter in the chirping scheme.
		
		In the third section, we developed a semiclassical theory which makes use of the C-CARS scheme and presented a realistic model of the detection method by taking methanol vapor as a surrogate system. The aim was to simulate the optimized output signal from a cloud of molecules using both pulses and pulse trains incident on the system. A detailed analysis was done to show the advantages of using control pulses and pulse trains, and to understand the effects of decoherence and propagation through atmosphere. A layer model of molecular distribution was created where each layer is characterized by the fractional density of the target molecules. A set of coupled Maxwell-Liouville von Neumann equations was derived and numerically solved to find the output from each layer. The results from each layer were applied to the subsequent layers to generate the final output signal. When transform limited pulses are propagated through a molecule distribution of 199 layers, equivalent to 0.5 meters from the center of the cloud, there is an amplification of 2 orders of magnitude by the final scattering compared to the first scattering. To ensure that the control pulses do not lose their phase values during the propagation through multiple layers, a machine learning model was created to extract the chirp parameters from numerical outputs. An exclusive look at this machine learning technique, which is based on deep Convolutional Neural Networks (CNN), was given in section 4. Two CNNs, one to classify the pulses based on their phase values and another one to extract these values were created. Primary results show that, the control scheme is efficient for hundred of layers as the average change in chirp rate after each scattering is less than 0.001\%.
		
		In the final section, we discussed the theory of STIRAP process and the conditions for adiabatic passage in STIRAP. We showed that by controlling the sign of chirp rate of the incident pulses, the population can be exclusively flown to a predetermined quantum state in a nearly degenerate system. Later, we explained how a variation of STIRAP, namely Fractional STIRAP, can be used to create a maximally coherent superposition of two quantum states in a multilevel system. Later, we laid the groundwork for how the Fractional STIRAP technique can be used to optimize the signal in detection and sensing methods.
		
		\section*{Acknowledgment}
		
		The authors gratefully acknowledge support from the Office of Naval Research under awards N00014-20-1-2086 and N00014-22-1-2374. S.M. acknowledges the Helmholtz Institute Mainz Visitor Program and J. Ch. the support from Johannes Gutenberg University of Mainz.
		
		{}
		
		\pagebreak
		
		\addcontentsline{toc}{section}{Appendices}
		\section*{Appendix A} 
		{\bf Derivation of ``super-effective" two-level Hamiltonian using adiabatic elimination in density matrix frame}
		\vskip 0.3in
		The Schr\"{o}dinger Hamiltonian for the 4-level system in Fig. 1 is given by:
		\begin{equation}\label{HAM}
			\small{
				\textbf{H} = \hbar
				\begin{pmatrix}
					\omega_1	&	0	&	\mu_{31}E_p(t)	&	\mu_{41}E_{as}(t)\\
					0	&	\omega_2	&	\mu_{32}E_s(t)	&	\mu_{42}E_{pr}(t)	\\
					\mu_{31}^*E_p^*(t)	&	\mu_{32}^*E_s^*(t)	&	\omega_3	&	0\\
					\mu_{41}^*E_{as}^*(t)	&	\mu_{42}^*E_{pr}^*(t)	&	0	&	\omega_4\\
			\end{pmatrix}}
		\end{equation}
		where the pulses $E_q(t)$ are given by:
		\begin{equation}
			E_q(t) = \frac{1}{2}E_{q_0}(t)\left[e^{i\omega_q(t-t_c)+i\frac{\alpha_q}{2}(t-t_c)^2} + e^{-i\omega_q(t-t_c)-i\frac{\alpha_q}{2}(t-t_c)^2} \right]
		\end{equation}
		The Liouville-von Neumann equations are:
		\begin{align}\label{Liouville}
			\begin{aligned}
				i\hbar \dot{\boldsymbol{\rho}} &= \left[\mathbf{H},\, \boldsymbol{\rho}\right] \\
				&= \mathbf{H}\boldsymbol{\rho} - \boldsymbol{\rho}\mathbf{H}\,.
			\end{aligned}
		\end{align}
		Apply the following transformations:
		\begin{align} 
			\begin{aligned}
				\rho_{11} =& \tilde{\rho}_{11} \\ 
				\rho_{12} =& \tilde{\rho}_{12}e^{-i(\omega_1-\omega_2)(t-t_c)}\\ \rho_{13} =& \tilde{\rho}_{13}e^{i\omega_p(t-t_c)}\\ \rho_{14} =& \tilde{\rho}_{14}e^{i\omega_{as}(t-t_c)}\\ \rho_{22} =& \tilde{\rho}_{22}\\
				\rho_{23} =& \tilde{\rho}_{23}e^{-i(\omega_2-\omega_1-\omega_p)(t-t_c)}\\
				\rho_{24} =& \tilde{\rho}_{24}e^{-i(\omega_2-\omega_{1}-\omega_{as})(t-t_c)} \\
				\rho_{33} =& \tilde{\rho}_{33} \\
				\rho_{34} =& \tilde{\rho}_{34}e^{-i(\omega_p-\omega_{as})(t-t_c)} \\
				\rho_{44} =& \tilde{\rho}_{44}
			\end{aligned}
		\end{align}
		After defining the rabi frequencies, $\Omega_{q_0}(t) = \mu_{ij}E_{q_0}(t)/\hbar$, applying rotating wave approximations, and removing the \textit{tilde}, the equations can be written as:
		
		\begin{align*}
			\begin{aligned}
				i\dot{\rho}_{11} =& \tfrac{1}{2}\Omega_{p_0}(t)e^{\frac{i}{2}\alpha_{p}(t-t_c)^2}\rho_{31} + \tfrac{1}{2}\Omega_{as_0}(t)\rho_{41} - \tfrac{1}{2}\Omega_{p_0}^*(t)e^{-\frac{i}{2}\alpha_{p}(t-t_c)^2}\rho_{13} - \tfrac{1}{2}\Omega_{as_0}^*(t)\rho_{41} \, \\
				i\dot{\rho}_{12} =& \tfrac{1}{2}\Omega_{p0}(t)e^{\frac{i}{2}\alpha_{p}(t-t_c)^2}\rho_{32} + \tfrac{1}{2}\Omega_{as0}(t)\rho_{42} - \tfrac{1}{2}\Omega_{s_0}^*(t)e^{i(\omega_p-\omega_s+\omega_1-\omega_2)(t-t_c)-\frac{i}{2}\alpha_{s}(t-t_c)^2}\rho_{13} \\ 
				&-\tfrac{1}{2}\Omega_{pr_0}^*(t)e^{i(\omega_{as}-\omega_{pr}+\omega_1-\omega_2)(t-t_c)-\frac{i}{2}\alpha_{pr}(t-t_c)^2}\rho_{14}\\
				i\dot{\rho}_{13} =& (\omega_1 + \omega_p - \omega_3)\rho_{13} + \tfrac{1}{2}\Omega_{p_0}(t)e^{\frac{i}{2}\alpha_p(t-t_c)^2}\rho_{33} + \tfrac{1}{2}\Omega_{as_0}(t)\rho_{43} - \tfrac{1}{2}\Omega_{p_0}(t)e^{\frac{i}{2}\alpha_p(t-t_c)^2}\rho_{11}\\
				&-\tfrac{1}{2}\Omega_{s_0}^*(t)e^{i(\omega_2-\omega_1-\omega_p+\omega_s)(t-t_c)+\frac{i}{2}\alpha_{s}(t-t_c)^2}\rho_{12}\\
				i\dot{\rho}_{14} =& (\omega_1 + \omega_4 - \omega_{as})\rho_{14} + \tfrac{1}{2}\Omega_{p_0}(t)e^{\frac{i}{2}\alpha_p(t-t_c)^2}\rho_{34} + \tfrac{1}{2}\Omega_{as_0}(t)\rho_{44} - \tfrac{1}{2}\Omega_{as_0}(t)\rho_{11}\\
				&-\tfrac{1}{2}\Omega_{pr_0}^*(t)e^{i(\omega_2-\omega_1-\omega_{pr}-\omega_{as})(t-t_c)+\frac{i}{2}\alpha_{pr}(t-t_c)^2}\rho_{12}
			\end{aligned}
		\end{align*}
		
		\begin{align*}
			\begin{aligned}
				i\dot{\rho_{22}} =& \tfrac{1}{2}\Omega_{s_0}(t)e^{i(\omega_2-\omega_1-\omega_{p}+\omega_{s})(t-t_c)+\frac{i}{2}\alpha_{s}(t-t_c)^2}\rho_{32} \\ &+\tfrac{1}{2}\Omega_{pr_0}(t)e^{i(\omega_2-\omega_1-\omega_{as}+\omega_{pr})(t-t_c)+\frac{i}{2}\alpha_{pr}(t-t_c)^2}\rho_{42} \\ &-\tfrac{1}{2}\Omega_{s_0}^*(t)e^{-i(\omega_2-\omega_1-\omega_{p}+\omega_{s})(t-t_c)-\frac{i}{2}\alpha_{s}(t-t_c)^2}\rho_{23} \\ &-\tfrac{1}{2}\Omega_{pr_0}^*(t)e^{-i(\omega_2-\omega_1-\omega_{as}+\omega_{pr})(t-t_c)+\frac{i}{2}\alpha_{pr}(t-t_c)^2}\rho_{24}\\
				i\dot{\rho_{23}} =& -(\omega_3-\omega_1-\omega_p)\rho_{23}+ \tfrac{1}{2}\Omega_{s_0}(t)e^{i(\omega_2-\omega_1-\omega_{p}+\omega_{s})(t-t_c)+\frac{i}{2}\alpha_{s}(t-t_c)^2}\rho_{33}\\
				&+\tfrac{1}{2}\Omega_{pr_0}(t)e^{i(\omega_2-\omega_1-\omega_{as}+\omega_{pr})(t-t_c)+\frac{i}{2}\alpha_{pr}(t-t_c)^2}\rho_{43}\\
				&-\tfrac{1}{2}\Omega_{p_0}(t)e^{\frac{i}{2}\alpha_{p}(t-t_c)^2}\rho_{21}-\tfrac{1}{2}\Omega_{s_0}(t)e^{i(\omega_2-\omega_1-\omega_{p}+\omega_{s})(t-t_c)+\frac{i}{2}\alpha_{s}(t-t_c)^2}\rho_{22}\\
				i\dot{\rho_{24}} =& (-\omega_4+\omega_{1}+\omega_{as})\rho_{24} + \tfrac{1}{2}\Omega_{s_0}(t)e^{i(\omega_2-\omega_1-\omega_{p}+\omega_{s})(t-t_c)+\frac{i}{2}\alpha_{s}(t-t_c)^2}\rho_{34} \\ &+\tfrac{1}{2}\Omega_{pr_0}(t)e^{i(\omega_2-\omega_1-\omega_{as}+\omega_{pr})(t-t_c)+\frac{i}{2}\alpha_{pr}(t-t_c)^2}\rho_{44}\\
				&-\tfrac{1}{2}\Omega_{as_0}(t)\rho_{21}-\tfrac{1}{2}\Omega_{pr_0}(t)e^{i(\omega_2-\omega_1-\omega_{as}+\omega_{pr})(t-t_c)+\frac{i}{2}\alpha_{pr}(t-t_c)^2}\rho_{22}\\
				i\dot{\rho_{33}} =& \tfrac{1}{2}\Omega_{p_0}^*(t)e^{-\frac{i}{2}\alpha_{p}(t-t_c)^2}\rho_{13} + \tfrac{1}{2}\Omega_{s_0}^*(t)e^{-i(\omega_2-\omega_1-\omega_{p}+\omega_{s})(t-t_c)-\frac{i}{2}\alpha_{s}(t-t_c)^2}\rho_{23}\\
				&-\tfrac{1}{2}\Omega_{p_0}(t)e^{\frac{i}{2}\alpha_{p}(t-t_c)^2}\rho_{31}-\tfrac{1}{2}\Omega_{s_0}(t)e^{i(\omega_2-\omega_1-\omega_{p}+\omega_{s})(t-t_c)+\frac{i}{2}\alpha_{s}(t-t_c)^2}\rho_{32}\\
				i\dot{\rho_{34}} =& (\omega_3-\omega_4-\omega_p+\omega_{as})\rho_{34} + \tfrac{1}{2}\Omega_{p_0}^*(t)e^{-\frac{i}{2}\alpha_{p}(t-t_c)^2}\rho_{14}\\
				&+\tfrac{1}{2}\Omega_{s_0}^*(t)e^{-i(\omega_2-\omega_1-\omega_{p}+\omega_{s})(t-t_c)-\frac{i}{2}\alpha_{s}(t-t_c)^2}\rho_{24}\\
				&-\tfrac{1}{2}\Omega_{as_0}(t)\rho_{31}-\tfrac{1}{2}\Omega_{pr_0}(t)e^{i(\omega_2-\omega_1-\omega_{as}+\omega_{pr})(t-t_c)+\frac{i}{2}\alpha_{pr}(t-t_c)^2}\rho_{32}\\
				i\dot{\rho_{44}} =& \tfrac{1}{2}\Omega_{as_0}^*(t)\rho_{14} + \tfrac{1}{2}\Omega_{pr_0}(t)e^{-i(\omega_2-\omega_1-\omega_{as}+\omega_{pr})(t-t_c)-\frac{i}{2}\alpha_{s}(t-t_c)^2}\rho_{24}\\
				&-\tfrac{1}{2}\Omega_{as_0}(t)\rho_{41}-\tfrac{1}{2}\Omega_{pr_0}(t)e^{i(\omega_2-\omega_1-\omega_{as}+\omega_{pr})(t-t_c)+\frac{i}{2}\alpha_{pr}(t-t_c)^2}\rho_{42}
			\end{aligned}
		\end{align*}
		Define the detunings:
		\begin{align}
			\begin{aligned}
				\Delta_s &= \omega_p - (\omega_3 - \omega_1)\\
				\Delta_{as} &= \omega_{as} - (\omega_4 - \omega_1)\\
				\delta_A &= \omega_p - \omega_s - (\omega_2 - \omega_1)\\
				\delta_B &= \omega_{as} - \omega_{pr} - (\omega_2 - \omega_1)\,,
			\end{aligned}
		\end{align}
		and rewrite the equations:
		
		\begin{align} 
			\begin{aligned}
				i\dot{\rho}_{11} =& \tfrac{1}{2}\Omega_{p0}(t)e^{\frac{i}{2}\alpha_{p}(t-t_c)^2}\rho_{31} + \tfrac{1}{2}\Omega_{as0}(t)\rho_{41} \\ &-\tfrac{1}{2}\Omega_{p0}^*(t)e^{-\frac{i}{2}\alpha_{p}(t-t_c)^2}\rho_{13} + \tfrac{1}{2}\Omega_{as0}^*(t)\rho_{14} \,, \\
				i\dot{\rho}_{12} =& \tfrac{1}{2}\Omega_{p0}(t)e^{\frac{i}{2}\alpha_{p}(t-t_c)^2}\rho_{32} + \tfrac{1}{2}\Omega_{as0}(t)\rho_{42} -  \tfrac{1}{2}\Omega^*_{s0}(t)e^{i\delta_A(t-t_c)-\frac{i}{2}\alpha_s(t-t_c)^2}\rho_{13} \\ &-\tfrac{1}{2}\Omega^*_{pr0}(t)e^{i\delta_B(t-t_c)-\frac{i}{2}\alpha_{pr}(t-t_c)^2}\rho_{14} \,,\\
				i\dot{\rho}_{13} =&   \Delta_{s}\rho_{13} + \tfrac{1}{2}\Omega_{p0}(t)e^{\frac{i}{2}\alpha_{p}(t-t_c)^2}\rho_{33} + \tfrac{1}{2}\Omega_{as0}(t)\rho_{43} -  \tfrac{1}{2}\Omega_{p0}(t)e^{\frac{i}{2}\alpha_{p}(t-t_c)^2}\rho_{11} \\ &-\tfrac{1}{2}\Omega_{s0}(t)e^{-i\delta_B(t-t_c)+\frac{i}{2}\alpha_{pr}(t-t_c)^2}\rho_{12} \,,\\
				i\dot{\rho}_{14} =&   \Delta_{as}\rho_{14} + \tfrac{1}{2}\Omega_{p0}(t)e^{\frac{i}{2}\alpha_{p}(t-t_c)^2}\rho_{34} + \tfrac{1}{2}\Omega_{as0}(t)\rho_{44} -  \tfrac{1}{2}\Omega_{a0}(t)\rho_{11} \\ &-\tfrac{1}{2}\Omega_{pr0}(t)e^{-i\delta_B(t-t_c)+\frac{i}{2}\alpha_{pr}(t-t_c)^2}\rho_{12} \,,\\
				i\dot{\rho}_{22} =&    \tfrac{1}{2}\Omega_{s_0}(t)e^{-i\delta_A(t-t_c)+\frac{i}{2}\alpha_s(t-t_c)^2}\rho_{32} + \tfrac{1}{2}\Omega_{pr_0}(t)e^{-i\delta_B(t-t_c)+\frac{i}{2}\alpha_{pr}(t-t_c)^2}\rho_{42}\\
				&-\tfrac{1}{2}\Omega_{s_0}^*(t)e^{i\delta_A(t-t_c)-\frac{i}{2}\alpha_s(t-t_c)^2}\rho_{23} - \tfrac{1}{2}\Omega_{pr0}^*(t)e^{i\delta_B(t-t_c)-\frac{i}{2}\alpha_{pr}(t-t_c)^2}\rho_{24} \,,\\
				i\dot{\rho}_{23} =& \Delta_{s}\rho_{23} +  \tfrac{1}{2}\Omega_{s0}(t)e^{-i\delta_A(t-t_c)+\frac{i}{2}\alpha_s(t-t_c)^2}\rho_{33} + \tfrac{1}{2}\Omega_{pr0}(t)e^{-i\delta_B(t-t_c)+\frac{i}{2}\alpha_{pr}(t-t_c)^2}\rho_{43} \\ &-\tfrac{1}{2}\Omega_{p0}(t)e^{\frac{i}{2}\alpha_{p}(t-t_c)^2}\rho_{21} - \tfrac{1}{2}\Omega_{s0}(t)e^{-i\delta_A(t-t_c)-\frac{i}{2}\alpha_s(t-t_c)^2}\rho_{22} \,,\\
				i\dot{\rho}_{24} =& \Delta_{as}\rho_{24} +  \tfrac{1}{2}\Omega_{s0}(t)e^{-i\delta_A(t-t_c)+\frac{i}{2}\alpha_s(t-t_c)^2}\rho_{34} + \tfrac{1}{2}\Omega_{pr0}(t)e^{-i\delta_B(t-t_c)+\frac{i}{2}\alpha_{pr}(t-t_c)^2}\rho_{44} \\ &-\tfrac{1}{2}\Omega_{as0}(t)\rho_{21} - \tfrac{1}{2}\Omega_{pr0}(t)e^{-i\delta_B(t-t_c)+\frac{i}{2}\alpha_{pr}(t-t_c)^2}\rho_{22} \,,\\
				i\dot{\rho}_{33} =&  \tfrac{1}{2}\Omega_{p_0}^*(t)e^{-\frac{i}{2}\alpha_{p}(t-t_c)^2}\rho_{13} + \tfrac{1}{2}\Omega_{s_0}^*(t)e^{i\delta_A(t-t_c)-\frac{i}{2}\alpha_s(t-t_c)^2}\rho_{23}\\
				&-\tfrac{1}{2}\Omega_{p_0}(t)e^{\frac{i}{2}\alpha_{p}(t-t_c)^2}\rho_{31} - \tfrac{1}{2}\Omega_{s_0}(t)e^{-i\delta_A(t-t_c)+\frac{i}{2}\alpha_s(t-t_c)^2}\rho_{32}\,,\\
				i\dot{\rho}_{34} =&  (\Delta_{as}-\Delta_{s})\rho_{34} + \tfrac{1}{2}\Omega^*_{p0}(t)e^{-\frac{i}{2}\alpha_{p}(t-t_c)^2}\rho_{14} + \tfrac{1}{2}\Omega^*_{s0}(t)e^{i\delta_A(t-t_c)-\frac{i}{2}\alpha_s(t-t_c)^2}\rho_{24} \\ &-\tfrac{1}{2}\Omega_{as0}(t)\rho_{31} - \tfrac{1}{2}\Omega_{pr0}(t)e^{-i\delta_B(t-t_c)+\frac{i}{2}\alpha_{pr}(t-t_c)^2}\rho_{32} \,.\\
				i\dot{\rho}_{44} =& \tfrac{1}{2}\Omega^*_{as0}(t)\rho_{14} + \tfrac{1}{2}\Omega^*_{pr0}(t)e^{i\delta_B(t-t_c)-\frac{i}{2}\alpha_{pr}(t-t_c)^2}\rho_{24} \\
				&-\tfrac{1}{2}\Omega_{as0}(t)\rho_{41} - \tfrac{1}{2}\Omega_{pr0}(t)e^{-i\delta_B(t-t_c)+\frac{i}{2}\alpha_{pr}(t-t_c)^2}\rho_{42} \,,\\
			\end{aligned}
		\end{align}
		After applying the conditions for adiabatic elimination:
		\begin{align}
			\begin{aligned}
				\dot{\rho_{33}} &= 0\\
				\dot{\rho_{44}} &= 0\\
				\dot{\rho_{34}} &= 0\\
			\end{aligned}
		\end{align}
		in the equations for $\dot{\rho_{33}}$, $\dot{\rho_{44}} $ and $\dot{\rho_{34}}$, they can be written as:
		\begin{align}
			\begin{aligned}
				\rho_{13} &= \frac{\Omega_{p_0}(t)}{2\Delta_s}e^{\frac{i}{2}\alpha_{p}(t-t_c)^2}\rho_{11} + \frac{\Omega_{s_0}(t)}{2\Delta_s}e^{-i\delta_A(t-t_c)+\frac{i}{2}\alpha_{s}(t-t_c)^2}\rho_{12}\\
				\rho_{14} &= \frac{\Omega_{as_0}(t)}{2\Delta_{as}}\rho_{11} + \frac{\Omega_{pr_0}(t)}{2\Delta_{as}}e^{-i\delta_B(t-t_c)+\frac{i}{2}\alpha_{pr}(t-t_c)^2}\rho_{12}\\
				\rho_{23} &= \frac{\Omega_{p_0}(t)}{2\Delta_s}e^{\frac{i}{2}\alpha_{p}(t-t_c)^2}\rho_{21} + \frac{\Omega_{s_0}(t)}{2\Delta_s}e^{-i\delta_A(t-t_c)+\frac{i}{2}\alpha_{s}(t-t_c)^2}\rho_{22}\\
				\rho_{14} &= \frac{\Omega_{as_0}(t)}{2\Delta_{as}}\rho_{21} + \frac{\Omega_{pr_0}(t)}{2\Delta_{as}}e^{-i\delta_B(t-t_c)+\frac{i}{2}\alpha_{pr}(t-t_c)^2}\rho_{22}
			\end{aligned}
		\end{align}
		Substituting these in the equations for $\dot{\rho_{11}}$, $\dot{\rho_{22}}$ and $\dot{\rho_{12}}$ will simplify the system of equations into the below set of 3 equations which only includes the states $\ket{1}$ and $\ket{2}$:
		
		\begin{align}
			\begin{aligned}
				i\dot{\rho}_{11} =& \left(\frac{\Omega_{p0}(t)\Omega^*_{s0}(t)}{4\Delta_s}e^{i\delta_A(t-t_c)-\frac{i}{2}(\alpha_s - \alpha_{p})(t-t_c)^2}+\frac{\Omega^*_{pr0}(t)\Omega_{as0}(t)}{4\Delta_{as}}e^{i\delta_B(t-t_c)-\frac{i}{2}\alpha_{pr}(t-t_c)^2}\right)\rho_{21} \\ &-\left(\frac{\Omega^*_{p0}(t)\Omega_{s0}(t)}{4\Delta_s}e^{-i\delta_A(t-t_c)+\frac{i}{2}(\alpha_s - \alpha_{p})(t-t_c)^2} \right. \\ 
				&+\left. \frac{\Omega_{pr0}(t)\Omega^*_{as0}(t)}{4\Delta_{as}}e^{-i\delta_B(t-t_c)+\frac{i}{2}\alpha_{pr}(t-t_c)^2}\right)\rho_{12} \\
				i\dot{\rho}_{12} =& \left(\frac{\Omega_{p0}(t)\Omega^*_{p0}(t)}{4\Delta_s}\rho_{12}+\frac{\Omega_{p0}(t)\Omega^*_{s0}(t)}{4\Delta_{s}}e^{i\delta_A(t-t_c)-\frac{i}{2}(\alpha_s - \alpha_{p})(t-t_c)^2}\rho_{22}\right) \\ &+\left(\frac{\Omega_{as0}(t)\Omega^*_{as0}(t)}{4\Delta_{as}}\rho_{12}+\frac{\Omega^*_{pr0}(t)\Omega_{as0}(t)}{4\Delta_{as}}e^{i\delta_B(t-t_c)-\frac{i}{2}\alpha_{pr}(t-t_c)^2}\rho_{22}\right) \\ &-\left(\frac{\Omega^*_{s0}(t)\Omega_{p0}(t)}{4\Delta_{s}}e^{i\delta_A(t-t_c)-\frac{i}{2}(\alpha_{s}-\alpha_{p})(t-t_c)^2}\rho_{11}+\frac{\Omega_{s0}(t)\Omega^*_{s0}(t)}{4\Delta_{s}}\rho_{12}\right) \\ &-\left(\frac{\Omega^*_{pr0}(t)\Omega_{as0}(t)}{4\Delta_{as}}e^{i\delta_B(t-t_c)-\frac{i}{2}\alpha_{pr}(t-t_c)^2}\rho_{11}+\frac{\Omega_{pr0}(t)\Omega^*_{pr0}(t)}{4\Delta_{as}}\rho_{12}\right) \\
				i\dot{\rho}_{22} =& \left(\frac{\Omega_{p0}(t)\Omega^*_{p0}(t)}{4\Delta_s}e^{-i\delta_A(t-t_c)+\frac{i}{2}(\alpha_s - \alpha_{p})(t-t_c)^2}\right. \\
				&\left. +\frac{\Omega_{pr0}(t)\Omega^*_{as0}(t)}{4\Delta_{as}}e^{-i\delta_B(t-t_c)+\frac{i}{2}\alpha_{pr}(t-t_c)^2}\right)\rho_{12} \\ &-\left(\frac{\Omega^*_{s0}(t)\Omega_{p0}(t)}{4\Delta_s}e^{i\delta_A(t-t_c)-\frac{i}{2}(\alpha_s - \alpha_{p})(t-t_c)^2} \right. \\ 
				&+\left. \frac{\Omega^*_{pr0}(t)\Omega_{as0}(t)}{4\Delta_{as}}e^{i\delta_B(t-t_c)-\frac{i}{2}\alpha_{pr}(t-t_c)^2}\right)\rho_{21} \\
			\end{aligned}
		\end{align}
		In CARS, the two-photon detunings are generally equal, $\delta_A = \delta_B = \delta$. In the C-CARS scheme, the chirp rate of probe pulse is given by: $\alpha_{pr}=\alpha_{s}-\alpha_{p}$. Applying these conditions, the the above equations can be simplified to: 
		
		\begin{align}
			\begin{aligned}
				i\dot{\rho}_{11} =& \left(\frac{\Omega_{p0}(t)\Omega^*_{s0}(t)}{4\Delta_s}+\frac{\Omega^*_{pr0}(t)\Omega_{as0}(t)}{4\Delta_{as}}\right)e^{i\delta(t-t_c)-\frac{i}{2}(\alpha_s - \alpha_{p})(t-t_c)^2}\rho_{21} \\ &-\left(\frac{\Omega^*_{p0}(t)\Omega_{s0}(t)}{4\Delta_s}+\frac{\Omega_{pr0}(t)\Omega^*_{as0}(t)}{4\Delta_{as}}\right)e^{-i\delta(t-t_c)+\frac{i}{2}(\alpha_s - \alpha_{p})(t-t_c)^2}\rho_{12} \\
				i\dot{\rho}_{12} =& \left(\frac{|\Omega_{p0}(t)|^{2}}{4\Delta_s}+\frac{|\Omega_{as0}(t)|^{2}}{4\Delta_{as}}\right)\rho_{12} \\ &+\left(\frac{\Omega_{p0}(t)\Omega^*_{s0}(t)}{4\Delta_s}+\frac{\Omega^*_{pr0}(t)\Omega_{as0}(t)}{4\Delta_{as}}\right)e^{i\delta(t-t_c)-\frac{i}{2}(\alpha_s - \alpha_{p})(t-t_c)^2}\rho_{22} \\ &-\left(\frac{|\Omega_{s0}(t)|^{2}}{4\Delta_s}+\frac{|\Omega_{pr0}(t)|^{2}}{4\Delta_{as}}\right)\rho_{12} \\ &-\left(\frac{\Omega_{p0}(t)\Omega^*_{s0}(t)}{4\Delta_s}+\frac{\Omega^*_{pr0}(t)\Omega_{as0}(t)}{4\Delta_{as}}\right)e^{i\delta(t-t_c)-\frac{i}{2}(\alpha_s - \alpha_{p})(t-t_c)^2}\rho_{11} \\
				i\dot{\rho}_{22} =& \left(\frac{\Omega^*_{p0}(t)\Omega_{s0}(t)}{4\Delta_s}+\frac{\Omega_{pr0}(t)\Omega^*_{as0}(t)}{4\Delta_{as}}\right)e^{-i\delta(t-t_c)+\frac{i}{2}(\alpha_s - \alpha_{p})(t-t_c)^2}\rho_{12} \\
				&-\left(\frac{\Omega_{p0}(t)\Omega^*_{s0}(t)}{4\Delta_s}+\frac{\Omega^*_{pr0}(t)\Omega_{as0}(t)}{4\Delta_{as}}\right)e^{i\delta(t-t_c)-\frac{i}{2}(\alpha_s - \alpha_{p})(t-t_c)^2}\rho_{21} \\
			\end{aligned}
		\end{align}
		
		
		Now, define the new Rabi frequencies:
		\begin{equation}
			\Omega_{1}(t)=\frac{|\Omega_{p0}(t)|^{2}}{4\Delta_s}+\frac{|\Omega_{as0}(t)|^{2}}{4\Delta_{as}}, \ \ \ \ \
			\Omega_{2}(t)=\frac{|\Omega_{s0}(t)|^{2}}{4\Delta_s}+\frac{|\Omega_{pr0}(t)|^{2}}{4\Delta_{as}}\,,
		\end{equation}	
		and 
		\begin{equation}
			\Omega_{3}(t)=\frac{\Omega_{p0}(t)\Omega^*_{s0}(t)}{4\Delta_s}+\frac{\Omega^*_{pr0}(t)\Omega_{as0}(t)}{4\Delta_{as}}\,.
		\end{equation}
		to receive:
		\begin{align}
			\begin{aligned}
				i\dot{\rho}_{11} =& \Omega_3(t)e^{i\delta(t-t_c)-\frac{i}{2}(\alpha_s - \alpha_{p})(t-t_c)^2}\rho_{21} - \Omega^*_3(t)e^{-i\delta(t-t_c)+\frac{i}{2}(\alpha_s - \alpha_{p})(t-t_c)^2}\rho_{12} \, \\
				i\dot{\rho}_{12} =& \Omega_1(t)\rho_{12} + \Omega_3(t)e^{i\delta(t-t_c)-\frac{i}{2}(\alpha_s - \alpha_{p})(t-t_c)^2}\rho_{22} \\ &-\Omega_3(t)e^{i\delta(t-t_c)-\frac{i}{2}(\alpha_s -\alpha_{p})(t-t_c)^2}\rho_{11}-\Omega_{2}(t)\rho_{12}\\
				i\dot{\rho}_{22} =& \Omega^*_3(t)e^{-i\delta(t-t_c)+\frac{i}{2}(\alpha_s - \alpha_{p})(t-t_c)^2}\rho_{12} - \Omega_3(t)e^{i\delta(t-t_c)-\frac{i}{2}(\alpha_s - \alpha_{p})(t-t_c)^2}\rho_{21} \, \\
			\end{aligned}
		\end{align}
		
		The Hamiltonian can now be written, in the interaction representation as:
		\begin{equation}\label{supereff_HAM_int}
			\small{
				\textbf{H} = \hbar
				\begin{pmatrix}
					\Omega_1(t)	&	\Omega_3(t)e^{i\delta(t-t_c)-\frac{i}{2}(\alpha_s - \alpha_{p})(t-t_c)^2} \\
					\Omega^*_3(t)e^{-i\delta(t-t_c)+\frac{i}{2}(\alpha_s - \alpha_{p})(t-t_c)^2}	&	\Omega_2(t)	\\
			\end{pmatrix}}
		\end{equation}
		To derive the Hamiltonian in the field-interaction representation, the density matrix elements are transformed as below:
		\begin{align}
			\begin{aligned}
				\rho_{11} =& \tilde{\rho}_{11} \\ 
				\rho_{12} =& \tilde{\rho}_{12}e^{i\delta(t-t_c)-\frac{i}{2}(\alpha_s - \alpha_{p})(t-t_c)^2}\\ 
				\rho_{22} =& \tilde{\rho}_{22}
			\end{aligned}
		\end{align}
		Applying these transformations in the Liouville von-Neumann equations \eqref{Liouville} will give:
		\begin{align}
			\begin{aligned}
				i\dot{\rho_{11}} =& \Omega_3(t)\rho_{21} - \Omega^*_3(t)\rho_{12} \\
				i\dot{\rho_{12}} =& \Omega_1(t)\rho_{12} + \Omega_3(t)\rho_{22} - \Omega_3(t)\rho_{11}-\left[\Omega_2(t) - \delta + (\alpha_{s}-\alpha_{p})(t-t_c)\right]\rho_{12} \\
				i\dot{\rho_{22}} =& \Omega^*_3(t)\rho_{12} - \Omega_3(t)\rho_{21} 
			\end{aligned}
		\end{align}
		The Hamiltonian is now given by:
		\begin{equation}\label{supereff_HAM_FE}
			\small{
				\textbf{H} = \hbar
				\begin{pmatrix}
					\Omega_1(t)	&	\Omega_3(t) \\
					\Omega^*_3(t)	&	\Omega_2(t)-(\delta-(\alpha_s-\alpha_p)(t-t_c))	\\
			\end{pmatrix}}
		\end{equation}
		The final ``super-effective" Hamiltonian can be found by adding $\frac{1}{2}(\delta-(\alpha_s-\alpha_p)(t-t_c)-\Omega_1(t)-\Omega_2(t))$ to both the diagonal:
		\begin{equation}\label{HAM_supereff}
			\mathbf{H}_{se}(t) = 
			\frac{\hbar}{2}
			\begingroup 
			\setlength\arraycolsep{-15pt}
			\begin{pmatrix}
				\small
				\delta-(\alpha_s-\alpha_p)(t-t_{c})+\Omega_{1}(t)-\Omega_{2}(t)  & 2\Omega_{3}(t)\\ 
				2\Omega^*_{3}(t) & -\delta+(\alpha_s-\alpha_p)(t-t_{c})-\Omega_{1}(t)+\Omega_{2}(t)  \\
			\end{pmatrix}\
			\endgroup
		\end{equation}
		\pagebreak
		
		\section*{Appendix B}
		{\bf Derivation of coupled Maxwell - Liouville von Neumann Equations}
		\vskip 0.3in
		Maxwell's equations, with no free currents and charges, read \begin{eqnarray}
			\nabla \cdot (\epsilon_0 E + P)=0 \label{one}\\
			\nabla \times E = - \partial B / \partial t \label{two}\\
			\nabla \times B = \mu_0  \partial (\epsilon_0E + P) / \partial t \label{three}\\
			\nabla \cdot B=0
		\end{eqnarray}
		From Eqs.(\ref{two},\ref{three}) we obtain the wave equation   
		\begin{equation}
			\nabla^2 E - \epsilon_0 \mu_0 \frac{\partial^2 E }{ \partial^2 t} = \nabla (\nabla \cdot E) + \mu_0 \frac{\partial^2 P }{ \partial^2 t }.
		\end{equation}
		It follows from Eq.(\ref{one}) that $\nabla \cdot E = \nabla \cdot P / \epsilon_0 $ in a space free from charges. In a plane wave limit, when 
		the wave length is much less than the beam radius and neglecting any diffraction effects in transverse direction, fields propagate in the $\hat{z}$ direction and have polarization  in the XY plane. Then $\nabla \cdot P$ may be set to zero and the wave equation reads
		\begin{eqnarray}
			\left(\pdv{}{z} + \frac{1}{c}\pdv{}{t} \right) \left( -\pdv{}{z} + \frac{1}{c} \pdv{}{t} \right)E =  - \mu_0 \pdv[2]{P}{t} \label{Sv4}
		\end{eqnarray}
		Assuming the field is $E(z,t) = \frac{1}{2} (E_0(z,t)e^{-i[\omega t - kz - \phi(z,t)]}+c.c) $ and polarization is $P(z,t) = \frac{1}{2} (P_0(z,t)e^{-i[\omega t - kz - \phi(z,t)]}+c.c )$ and considering $E_0(z,t)$ and $\phi(z,t)$ as slowly varying functions of position and time, we write 
		\begin{eqnarray}
			-\pdv{E(z,t)}{z}=-\frac{1}{2} ( e^{-i\omega t} e^{i k z} e^{i \phi(z,t)} \pdv{E_0(z,t)}{z} +
			i k  E_0(z,t) e^{-i\omega t} e^{i k z} e^{i \phi(z,t)}\\ 
			+ i  \pdv{\phi(z,t)}{z} E_0(z,t) e^{-i \omega t} e^{i k z} e^{i \phi(z,t)} + c.c.) \nonumber \\
			\frac{1}{c} \pdv{E(z,t)}{t}=\frac{1}{2c} ( e^{-i\omega t} e^{i k z} e^{i \phi(z,t)} \pdv{E_0(z,t)}{t} -
			i \omega  E_0(z,t) e^{-i\omega t} e^{i k z} e^{i \phi(z,t)}\\ 
			+ i  \pdv{\phi(z,t)}{t} E_0(z,t) e^{-i \omega t} e^{i k z} e^{i \phi(z,t)} + c.c.) \nonumber 
		\end{eqnarray}
		
		Then 
		\begin{eqnarray}
			\left( -\pdv{}{z} + \frac{1}{c}\pdv{}{t} \right)E &=& - \frac{ik}{2} ( E_0(z,t) e^{-i\omega t} e^{i k z} e^{i \phi(z,t)} - c.c.) \nonumber \\
			& &-\frac{i \omega}{2 c} ( E_0(z,t) e^{-i\omega t} e^{i k z} e^{i \phi(z,t)} - c.c.) \label{Sv5} \\
			&=& -ik ( E_0(z,t) e^{-i(\omega t - k z - \phi(z,t))} - c.c.) \\
			&=& -2ik \Im{E}. \nonumber
		\end{eqnarray}
		By substituting Eq.(\ref{Sv5}) to Eq.(\ref{Sv4}) and using $\omega/c = k$ and later assuming real fields we arrive at 
		\begin{eqnarray}
			&&-ik\left(\pdv{}{z} + \frac{1}{c}\pdv{}{t} \right) ( E_0(z,t) e^{-i(\omega t - k z - \phi(z,t))} + c.c.)  \nonumber \\
			&=& -ik \pdv{ E_0(z,t)}{z}e^{-i(\omega t - k z - \phi(z,t))}+ik \pdv{ E_0^*(z,t)}{z}e^{i(\omega t - k z - \phi(z,t))} \nonumber \\
			& &-\frac{ik}{c} \pdv{ E_0(z,t)}{t}e^{-i(\omega t - k z - \phi(z,t))}+\frac{ik}{c} \pdv{ E_0^*(z,t)}{t}e^{i(\omega t - k z - \phi(z,t))}\nonumber \\
			&=&-ik \left[ \left(\pdv{ E_0(z,t)}{z}+ \frac{1}{c}  \pdv{ E_0(z,t)}{t} \right)e^{-i(\omega t - k z - \phi(z,t))} \right] \nonumber \\
			& & +ik \left[\left(\pdv{ E_0^*(z,t)}{z}+ \frac{1}{c}  \pdv{ E_0^*(z,t)}{t} \right)e^{i(\omega t - k z - \phi(z,t))} \right] \nonumber \\
			&=&-2k  \left(\pdv{ E_0(z,t)}{z}+ \frac{1}{c}  \pdv{ E_0(z,t)}{t} \right) \frac{1}{2i} \left( -e^{-i(\omega t - k z - \phi(z,t))} +e^{i(\omega t - k z - \phi(z,t))}\right) \nonumber \\
			&=&-2k \left(\pdv{ E_0(z,t)}{z}+ \frac{1}{c}  \pdv{ E_0(z,t)}{t} \right) \sin{(\omega t - k z - \phi(z,t))} \nonumber\\
			&=& - \mu_0 \pdv[2]{}{t}P(z,t)  \label{eq:1}
		\end{eqnarray} 
		
		For $P(z,t) = \frac{1}{2} (P_0(z,t)e^{-i[\omega t - kz - \phi(z,t)]}+c.c )$, 
		\begin{eqnarray}
			\pdv[2]{}{t}P(z,t)= - \omega^2 \left( \frac{1}{2} (P_0(z,t)e^{-i[\omega t - kz - \phi(z,t)]}+c.c )  \right)=- \omega^2 \Re\left[ P(z,t)\right]
		\end{eqnarray}
		
		Substituting these in Eq.(\ref{eq:1}) gives
		\begin{eqnarray}
			&-2k (\pdv{ E_0(z,t)}{z}+ \frac{1}{c}  \pdv{ E_0(z,t)}{t}) \sin{(\omega t - k z - \phi(z,t))}=\nonumber \\
			&\mu_0 \omega^2 \left( Re\left[ P_0(z,t)\right] \cos{(\omega t-kz)} + Im\left[ P_0(z,t)\right] \sin{(\omega t-kz- \phi(z,t) )} \right),
		\end{eqnarray}
		leading to
		\begin{eqnarray}
			&-2k (\pdv{ E_0(z,t)}{z}+ \frac{1}{c}  \pdv{ E_0(z,t)}{t})=
			&\mu_0 \omega^2  Im\left[ P_0(z,t)\right].\label{slow}
		\end{eqnarray}
		
		In quantum theory, a measurable quantity is the expectation value, which for macroscopic polarization is an expectation value of the electric dipole moment operator $\hat{\mu}$, $ \langle P(z,t) \rangle = N_sTr \{\langle\rho (z,t)\cdot\mu \rangle\} $, where $N_s$ is the atomic density of the target molecules. Applied to the four-level system of CARS, it contains four components corresponding to each of four transitions:

		\begin{equation}
			\begin{aligned}
				P_p(z,t) &= 2 N_s  \Re\left[\mu_{13}\rho_{13}(z,t)e^{i(\omega_pt-k_pz- \phi(z,t))}\right]\\
				P_s (z,t)&= 2 N_s \Re\left[\mu_{23}\rho_{23}(z,t)e^{i(\omega_st-k_sz- \phi(z,t))}\right]\\
				P_{pr} (z,t)&= 2 N_s \Re\left[\mu_{24}\rho_{24}(z,t)e^{i(\omega_{pr}t-k_{pr}z- \phi(z,t))}\right]\\
				P_{as}(z,t) &= 2 N_s \Re\left[\mu_{14}\rho_{14}(z,t)e^{i(\omega_{as}t-k_{as}z- \phi(z,t))}\right],
			\end{aligned}
		\end{equation}
		giving $P_{0p}(z,t)=N_s \mu_{13}\rho_{13}(z,t)$, $P_{0s}(z,t)=N_s \mu_{23}\rho_{23}(z,t)$, $P_{0pr}(z,t)=N_s \mu_{24}\rho_{24}(z,t)$, and $P_{0as}(z,t)=N_s \mu_{14}\rho_{14}(z,t)$.
		
		For four components of propagating fields in CARS, the Eq.(\ref{slow}) reads  as follows
		\begin{eqnarray}
			&\pdv{E_{p}}{z} + \frac{1}{c} \pdv{E_{p}}{t} = - N_s \frac{\mu_0 \mu_{13} \omega_{p}^2}{k_p} \Im {\rho_{13}(z,t)} \label{eq:2}\\
			&\pdv{E_{s}}{z} + \frac{1}{c} \pdv{E_{s}}{t} = -N_s \frac{\mu_0 \mu_{23} \omega_{s}^2}{k_s} \Im {\rho_{23}(z,t)} \nonumber\\
			&\pdv{E_{pr}}{z} + \frac{1}{c} \pdv{E_{pr}}{t} = -N_s \frac{\mu_0 \mu_{24} \omega_{pr}^2}{k_{pr}} \Im {\rho_{24}(z,t)} \nonumber \\
			&\pdv{E_{as}}{z} + \frac{1}{c} \pdv{E_{as}}{t} = - N_s \frac{\mu_0 \mu_{14} \omega_{as}^2}{k_{as}} \Im {\rho_{14}(z,t)}.\nonumber
		\end{eqnarray}
		If $\bar{t} = (t-\frac{z}{c})$, then $ \dv{t}{z} = (\dv{\bar{t}}{z} + \frac{1}{c}), $ which leads to  $\pdv{}{z} = \pdv{}{t}\pdv{t}{z} = \frac{1}{c} \pdv{}{t}$. Taking into account that $k_{q}=\omega_{q}/c$,  and $c \omega_{q} \hbar =E_{q}$, where $q=p, s, pr, as$, the Eq. \eqref{eq:2} becomes
		\begin{equation}
			\frac{1}{c} \pdv{E_{q}}{t} = -N_s \mu_0 \mu_{ij} \frac{E_{q} (t)}{\hbar}  \Im {\rho_{ij}} \label{eq:3}
		\end{equation}

		We find the density matrix elements $\rho_{ij}$ from the Liouville von Neumann equation $i\hbar\dot{\rho} =[H,\rho]$ and using the above Hamiltonian in Eq.(\ref{HAM}). We start by opening the commutator and applying the substitutions 
		\begin{eqnarray}
			\begin{aligned}
				\rho_{12}&=\tilde{\rho}_{12} e^{i ( \alpha_p  -\alpha_s) t^2 /2} \nonumber \\
				\rho_{13}&=\tilde{\rho}_{13} e^{i (\Delta_s t + \alpha_p t^2 /2)} \nonumber \\
				\rho_{14}&=\tilde{\rho}_{14} e^{ i \Delta_{as} t}\nonumber \\
				\rho_{23}&=\tilde{\rho}_{23} e^{i (\Delta_s t + \alpha_s t^2 /2)} \nonumber \\
				\rho_{24}&=\tilde{\rho}_{24} e^{i (\Delta_{as} t + \alpha_{pr} t^2 /2)} \nonumber \\
				\rho_{34}&=\tilde{\rho}_{34} e^{i (\Delta_{as} - \Delta_{s}) t - i\alpha_p t^2 /2}
			\end{aligned}
		\end{eqnarray}
		Next, we apply the rotating wave approximation and use the control condition on the chirp parameters $\alpha_s -\alpha_p=\alpha_{pr}$, and arrive at 
		\begin{equation} \label{densitymrx}
			\begin{aligned}
				\dot{\rho}_{11} &=  -\tfrac{i}{2}\Omega_{p0}(t) \tilde{\rho}_{31}+ \tfrac{i}{2} \Omega^*_{p0}(t) \tilde{\rho}_{13} - \tfrac{i}{2}\Omega_{as0}(t) \tilde{\rho}_{41} + \tfrac{i}{2}  \Omega^*_{as0}(t) \tilde{\rho}_{14} \\
				\dot{\rho}_{22} &= - \tfrac{i}{2}\Omega_{s0}(t) \tilde{\rho}_{32}+ \tfrac{i}{2} \Omega^*_{s0}(t) \tilde{\rho}_{23} - \tfrac{i}{2}\Omega_{pr0}(t) \tilde{\rho}_{42} + \tfrac{i}{2}  \Omega^*_{pr0}(t) \tilde{\rho}_{24}  \\
				\dot{\rho}_{33} &=  \tfrac{i}{2}\Omega_{p0}(t) \tilde{\rho}_{31} - \tfrac{i}{2} \Omega^*_{p0}(t) \tilde{\rho}_{13} + \tfrac{i}{2} \Omega_{s0}(t) \tilde{\rho}_{32} - \tfrac{i}{2}  \Omega^*_{s0}(t) \tilde{\rho}_{23}  \\
				\dot{\rho}_{44} &=  \tfrac{i}{2}\Omega_{as0}(t) \tilde{\rho}_{41}- \tfrac{i}{2} \Omega^*_{as0}(t) \tilde{\rho}_{14} + \tfrac{i}{2} \Omega_{pr0}(t) \tilde{\rho}_{42} - \tfrac{i}{2}  \Omega^*_{pr0}(t) \tilde{\rho}_{24}  \\
				\dot{\tilde{\rho}}_{12} &=  i \alpha_{pr}t \tilde{\rho}_{12}- \tfrac{i}{2}\Omega_{p0}(t) \tilde{\rho}_{32} -\tfrac{i}{2}\Omega_{as0}(t) \tilde{\rho}_{42}+ \tfrac{i}{2} \Omega^*_{s0}(t) \tilde{\rho}_{13} + \tfrac{i}{2} \Omega^*_{pr0}(t) \tilde{\rho}_{14}\\
				\dot{\tilde{\rho}}_{13} &=  -i (\Delta_s+\alpha_p t) \tilde{\rho}_{13}- \tfrac{i}{2}\Omega_{p0}(t) (\rho_{33}-\rho_{11}) -\tfrac{i}{2}\Omega_{as0}(t) \tilde{\rho}_{43}+ \tfrac{i}{2}\Omega_{s0}(t) \tilde{\rho}_{12} \\
				\dot{\tilde{\rho}}_{14} &=  -i \Delta_{as} \tilde{\rho}_{14}- \tfrac{i}{2}\Omega_{p0}(t) \tilde{\rho}_{34}  -\tfrac{i}{2}\Omega_{as0}(t) (\rho_{44}-\rho_{11})+ \tfrac{i}{2} \Omega_{pr0}(t) \tilde{\rho}_{12} \\
				\dot{\tilde{\rho}}_{23} &=  -i (\Delta_s+\alpha_s t) \tilde{\rho}_{23}- \tfrac{i}{2}\Omega_{s0}(t) (\rho_{33}-\rho_{22}) -\tfrac{i}{2}\Omega_{pr0}(t) \tilde{\rho}_{43}+ \tfrac{i}{2} \Omega_{p0}(t) \tilde{\rho}_{21} \\
				\dot{\tilde{\rho}}_{24} &=  -i (\Delta_{as}+\alpha_{pr} t) \tilde{\rho}_{24}- \tfrac{i}{2}\Omega_{pr0}(t) (\rho_{44}-\rho_{22}) -\tfrac{i}{2}\Omega_{s0}(t) \tilde{\rho}_{34}+ \tfrac{i}{2} \Omega_{as0}(t) \tilde{\rho}_{21} \\
				\dot{\tilde{\rho}}_{34} &=  i (\Delta_s-\Delta_{as}+\alpha_{p} t) \tilde{\rho}_{34}- \tfrac{i}{2}\Omega^*_{p0}(t) \tilde{\rho}_{14} -\tfrac{i}{2}\Omega^*_{s0}(t) \tilde{\rho}_{24}+ \tfrac{i}{2} \Omega_{as0}(t)/2 \tilde{\rho}_{31} \\
				& + \tfrac{i}{2}  \Omega_{pr0}(t) \tilde{\rho}_{32}. \nonumber
			\end{aligned}
		\end{equation}
		
		After performing adiabatic elimination of the excited states assuming that $ \dot{\rho}_{13},\dot{\rho}_{14},\dot{\rho}_{23},\dot{\rho}_{24}, \dot{\rho}_{34} \approx 0, \; \rho_{34} \approx 0, \; \rho_{33},\rho_{44} \ll \rho_{11},\rho_{22} $ \quad and \quad $\dot{\rho}_{33},\dot{\rho}_{44} \approx 0 $, the density matrix elements $\rho_{13},\rho_{23},\rho_{14},\rho_{24}$ read in terms of $\rho_{11},\rho_{22}$ and $\rho_{12}$ as follows

		\begin{equation}
			\begin{aligned}
				\rho_{13} &=  \dfrac{1}{2(\Delta_s + \alpha_p t)}\Omega_{p0}(t)\rho_{11} + \dfrac{1}{2(\Delta_s + \alpha_p t)}\Omega_{s0}(t)\rho_{12}\\
				\rho_{23} &=  \dfrac{1}{2(\Delta_s + \alpha_s t)}\Omega_{s0}(t)\rho_{22} + \dfrac{1}{2(\Delta_s + \alpha_s t)}\Omega_{p0}(t)\rho_{21} \\
				\rho_{14} &=  \dfrac{1}{2\Delta_{as}}\Omega_{as0}(t)\rho_{11} + \dfrac{1}{2\Delta_{as}}\Omega_{pr0}(t)\rho_{12} \\
				\rho_{24} &=  \dfrac{1}{2(\Delta_{as} + \alpha_{pr} t)}\Omega_{pr0}(t)\rho_{22} + \dfrac{1}{2(\Delta_{as} + \alpha_{pr} t)}\Omega_{as0}(t)\rho_{21}
			\end{aligned}  \label{eq:4}
		\end{equation}
		
		Substituting Eq.\eqref{eq:4} in Eq.\eqref{eq:3} and rewriting the equations in terms of Rabi frequencies provide the following Maxwell's equations:
		
		\begin{equation}
			\begin{aligned}
				\pdv{\Omega_{p0}}{t} &= c\pdv{\Omega_{p0}}{z} = -\dfrac{\eta}{2(\Delta_s+\alpha_p t)} \kappa_{13} \omega_p \Omega_{s0}(t) \Im[\rho_{12}] \\
				\pdv{\Omega_{s0}}{t} &= c\pdv{\Omega_{s0}}{z} = \dfrac{\eta}{2(\Delta_s+\alpha_s t)}
				\kappa_{23} \omega_s \Omega_{p0}(t) \Im[\rho_{12}] \\
				\pdv{\Omega_{pr0}}{t} &= c\pdv{\Omega_{pr0}}{z} = \dfrac{\eta}{2(\Delta_{as}+\alpha_{pr} t)} \kappa_{24}\omega_{pr} \Omega_{as0}(t) \Im[\rho_{12}] \\
				\pdv{\Omega_{as0}}{t} &= c\pdv{\Omega_{as0}}{z} = -\dfrac{\eta}{2(\Delta_{as})} \kappa_{14} \omega_{as} \Omega_{pr0}(t) \Im[\rho_{12}].
			\end{aligned}
		\end{equation}
		
	\end{document}